\documentclass[english,11pt]{article}
\usepackage{amsmath,amsfonts,amssymb,amsbsy,amstext}
\usepackage[top=3cm, bottom=2cm, left=2cm, right=2cm]{geometry}
\usepackage{graphicx}
\usepackage[english]{babel}
\usepackage{subfigure}
\usepackage[numbers,compress]{natbib}

\newcommand{\beq}{\begin{eqnarray}}
\newcommand{\eeq}{\end{eqnarray}}
\newcommand{\non}{\nonumber\\}

\newcommand{\diag}{{\rm diag}}

\newcommand{\Li}{{\rm Li}}
\newcommand{\h}[2]{h\left(#1,#2\right)}
\newcommand{\U}{{\rm U}}
\newcommand{\SU}{{\rm SU}}

\begin{document}

\begin{titlepage}

\begin{center}
{\Large\bf Mild-split SUSY with flavor}

\bigskip
{\large
Latif Eliaz, Amit Giveon, Sven Bjarke Gudnason and Eitan Tsuk
}

\bigskip
{\it
Racah Institute of Physics, The Hebrew University,
Jerusalem, 91904, Israel
}

\vskip 3cm

{\bf Abstract}
\end {center}

In the framework of a gauge mediated quiver-like model, the standard
model flavor texture can be naturally generated. The model -- like the
MSSM -- has furthermore a region in parameter space where the lightest
Higgs mass is fed by heavy stop loops, which in turn sets the average
squark mass scale near $10-20$ TeV. We perform a careful flavor
analysis to check whether this type of mild-split SUSY passes all
flavor constraints as easily as envisioned in the original type of
split SUSY. Interestingly, it turns out to be on the border of several
constraints, in particular, the branching ratio of $\mu\to e\gamma$
and, if order one complex phases are assumed, also $\epsilon_K$,
neutron and electron EDM. 
Furthermore, we consider unification as well as dark matter
candidates, especially the gravitino.
Finally, we provide a closed-form formula for the soft masses of matter
in arbitrary representations of any of the gauge groups in a generic
quiver-like model with a general messenger sector.

\vfill
\noindent
\rule{5cm}{0.5pt}\\
{\it\footnotesize
 latif.eliaz(at)mail.huji.ac.il \\
 giveon(at)phys.huji.ac.il  \\
 gudnason(at)phys.huji.ac.il \\
 eitan.tsuk(at)mail.huji.ac.il
}

\end{titlepage}

\section{Introduction}

Supersymmetry is an elegant solution to the hierarchy problem of the
standard model (SM). It is clear that some explanation as to why the Higgs
is so light compared to the Planck scale is sought for in a
fundamental theory of particle physics. What is less clear cut is how
much fine-tuning to allow for in practice. Common lore is
that ten percent is not ``fine'' tuning and the one-percent level
is acceptable. The discussion of how big this number may be is
somewhat philosophical and examples in Nature are known where a
fine-tuning of the 0.1 per-mille level is realized; e.g.~the binding
energy of triplet deuteron is about $\sim 2.2$ MeV which is only
slightly above the energy released in neutron beta decay and is
important because it prevented all the neutrons from decaying during
the 
evolution of the Universe. Furthermore, the singlet state of deuteron
does not exist; nevertheless, virtual particle exchange affects the
neutron cross section, since the negative binding energy -- only 60
keV -- is very small. Naive expectations would set the deuteron
binding energy of the order of a hundred MeV, but cancellation in the
effective theory leaves the binding energy of both the singlet and the
triplet states of order one MeV \cite{Kaplan:1998sz}. Since some
degree of fine-tuning has been observed in Nature, we contemplate a
more relaxed attitude towards fine-tuning, in the spirit of
e.g.~\cite{ArkaniHamed:2012gw}.

The hierarchy problem is not the only piece of the puzzle that one
would wish be explained by the theory of Nature. Assuming that general
relativity is correct at certain astrophysical scales (say at the kilo
parsec scale), the existence of dark matter halos in for instance
dwarf spheroidals is necessary for flattening the rotation curves of
satellites. A popular candidate is a sufficiently weakly interacting
massive particle (WIMP) of which cold dark matter (CDM) is made
of. Certain supersymmetric extensions of the SM come with such a
candidate with a mass and abundance compatible with observations. 
Furthermore, the electric charge is observed to be quantized and the
standard model gauge couplings hint at gauge coupling unification,
both of which calls for the possibility of a grand unified theory
(GUT). Supersymmetry typically further enhances the precision to which
this happens under certain conditions.
Finally, the quark and lepton masses and mixing angles
have a very particular form, which calls
for some underlying mechanism.

In a series of works, following
\cite{Csaki:2001em,Cheng:2001an,Craig:2011yk}, we have considered
a gauge mediated supersymmetric extension of the SM in which we double
the SM gauge group and Higgs them back together at low energies by
means of a link field attaining a VEV
\cite{Auzzi:2010mb,Auzzi:2010xc,Auzzi:2011gh,Auzzi:2011eu,Auzzi:2012dv}.
Our focus has been on the natural part of parameter space,
i.e.~keeping the stops as light as possible \cite{Auzzi:2012dv} and
explaining all 18 parameters of the SM \cite{Auzzi:2011eu}, satisfying
the constraints coming from collider data and flavor physics. In the
particular model we studied, we were able to obtain a natural model
(i.e.~fine tuning of parameters at the percent level at worst), which
however has three short comings. The lightest supersymmetric particle
(LSP), being the gravitino, is too light to be a CDM candidate 
and the embedding of the model into a unifying theory is not
straightforward (some ideas regarding an elaboration that could do the
job were put forward in \cite{Auzzi:2012dv}). The final short-coming
is that even though the natural setting can avoid fine-tuning in the
Higgs quartic, some degree of tuning is necessary in order for this
type of model not to be at odds with CP-violating observables, like
the $\epsilon_K$ parameter of the kaon system. 

If on the other hand, we relax our attitude towards the level of
acceptable fine tuning, as in the spirit of split SUSY
\cite{ArkaniHamed:2004fb,Giudice:2004tc,ArkaniHamed:2004yi,ArkaniHamed:2012gw};
say if we allow for a fine tuning at the 0.1--1 per-mille level for
the Higgs quartic, then the same type of model as described above is
able to still explain the 18 parameters of the SM, to provide a WIMP
CDM candidate in terms of the gravitino (being much heavier than in the
other scenario) and finally to unify without any elaborations of the
model. The supersymmetric flavor problem is then addressed here simply
by universality and decoupling; all the squarks will be heavier than
about 10 TeV in order to amplify the Higgs quartic coupling for
obtaining a 125 GeV Higgs.

In this paper, we focus on two branches of the above described
model. In one case, supersymmetry breaking is mediated near
the grand unified (GUT) scale and, in this case, unification as well
as a gravitino dark matter candidate can be contemplated.
This case comes with a long renormalization running, which can
potentially be probed in future experiments by certain flavor 
observables, e.g.~the electric dipole moment (EDM) of the neutron and
the branching ratio of $\mu\to e\gamma$.
In the other case, supersymmetry breaking is mediated at a relatively
low scale, i.e.~$\sim 10^{6-7}$ GeV, and it has a possible embedding
in SQCD~\cite{Green:2010ww} as well as a light gravitino that
could be contemplated as a warm dark matter candidate.
Both scenarios have only gauginos and the gravitino sparticles
at mass scales
below $\sim 1$ TeV, but are interestingly not far from constraints due
to flavor observables, especially, the CP-violating ones. Future
experiments, for e.g.~the EDMs, will be able to probe considerable
parts of their parameter spaces.

The paper is organized as follows. In sec.~\ref{sec:overview} we
present an overview of the model without too many technical
details. The reader can then skip to the discussion if not interested in
further details.
In secs.~\ref{sec:highscale} and \ref{sec:lowscale} we present the two
branches of the model, with high- and low-scale mediation of SUSY
breaking, respectively, and their corresponding spectra. In the
high-scale case, we contemplate unification, which is analyzed in
sec.~\ref{sec:uni}. The flavor constraints for both model types are
studied in detail in sec.~\ref{sec:sflavor}. Then the prospects of
gravitino dark matter is discussed in sec.~\ref{sec:dm} and finally,
sec.~\ref{sec:discussion} concludes with a discussion.
In app.~\ref{app:sfermionmasses}, we provide a closed-form formula for
the soft masses of matter in arbitrary representations of any of the
gauge groups in a generic quiver-like model with a general messenger
sector. App.~\ref{app:montecarlo} contains a Monte Carlo analysis of
the diagonalization matrix elements entering the flavor constraints.

\section{Overview of the model\label{sec:overview}}

\begin{figure}[!ht]
\begin{center}
\includegraphics[width=0.35\linewidth]{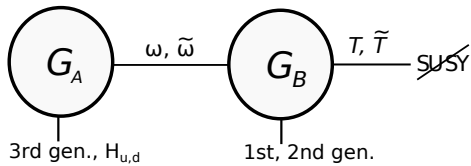}
\caption{A diagram describing the model with gauge groups
  $G_A,G_B=\U(1)\times\SU(2)\times\SU(3)$
  and link fields $\omega,\tilde{\omega}$. SUSY breaking is
  connected via messenger fields $T,\tilde{T}$ only to $G_B$.
  We refer to the model as depicted above as the \emph{normal model}
  while the \emph{inverted model} has the matter content on the two
  nodes swapped, i.e.~the 1st and 2nd generations are charged under
  node $A$ while the 3rd generation and the Higgses are charged under
  node $B$.}
\label{fig:quiver2nodes}
\end{center}
\end{figure}

The model of beyond-SM (BSM) physics we study in this paper is
sketched in fig.~\ref{fig:quiver2nodes} and is generically a
non-flavor-blind extension in the class of gauge mediated
supersymmetric models. The structure of the gauge groups is
used to generate the SM flavor texture, which we will describe
shortly.

The model is characterized by the following scales. Supersymmetry is
broken in a secluded sector and mediated to the node $B$ at the
messenger scale $M$, which is taken roughly an order of magnitude
lower than the Higgsing scale (of the link fields)
$\langle \omega\rangle=\langle\tilde{\omega}\rangle=v$, viz.~the VEV
of the link fields. This means that at the messenger scale the theory
is basically just a single node -- i.e.~an MSSM-like theory -- at
scale $M$ with more structure at scale $v>M$.
Further up in scale, we contemplate a UV completion with dynamics
generating higher-dimension operators suppressed by the scale
$\Lambda_{\rm flavor}$, giving rise to fermion masses and SM flavor
texture.
Hence in this paper we are considering the part of the two-node
parameter space where $\Lambda_{\rm flavor} > v > M$. In
sec.~\ref{sec:highscale} and \ref{sec:lowscale}, we consider the
messenger scale $M$ to be near the GUT scale and near
$\sim 10^{6-7}$ GeV, respectively.

The matter content of the supersymmetric SM (SSM) is split on the two
gauge groups $G_A$ and $G_B$ as follows: the complete third generation
is charged under $G_A$ together with the two Higgs superfields,
$H_u,H_d$, giving a tree-level top-Yukawa of order one, while the
first two generations sit on the other group $G_B$, giving vanishing
Yukawas at tree-level. This explains why the top, bottom and tau have
larger mass than the rest of the SM fermions.
The representation of the link fields $\omega,\tilde{\omega}$
determines the flavor texture of the SM fermions as we
shall review next.
We denote the model as just described by the \emph{normal model},
whereas if we simply swap the matter content of the two gauge groups,
we call it the \emph{inverted model}, see
fig.~\ref{fig:quiver2nodes}; such a swapping does not affect the flavor texture
of the SM particles, though it does affect some aspects of flavor constraints.

\subsection{Flavor texture}

Summarizing the results of \cite{Auzzi:2011eu}, the
Yukawa matrices can be generated via higher-dimension operators like,
for instance
\beq
\frac{\lambda_{ij}^2}{\Lambda_{\rm flavor}^2} Q_i H_u u_j^c \omega_Q
\omega_{u^c} \, , \qquad
i,j=1,2 \ \ (\textrm{generation indices}).
\eeq
Flavor texture is ideal for the choice of
$\omega$ ($\tilde{\omega}$) transforming in the
block-diagonal representation of $({\bf 10},\overline{\bf 10})$
($(\overline{\bf 10},{\bf 10})$). The ${\bf 10}$ decomposes like
$Q\oplus u^c\oplus e^c$ under $\U(1)\times\SU(2)\times\SU(3)$, where
the labels refer to the representations $R$ of the SM fields. Assuming
order-one coefficients of the higher-dimension operators, this
representation gives rise to the following Yukawa textures
\cite{Auzzi:2011eu}
\beq
Y^u \sim
\begin{pmatrix}
\epsilon^2 & \epsilon^2 & \epsilon \\
\epsilon^2 & \epsilon^2 & \epsilon \\
\epsilon & \epsilon & 1 \\
\end{pmatrix} \, , \qquad
Y^d \sim
\begin{pmatrix}
\epsilon^2 & \epsilon^2 & \epsilon \\
\epsilon^2 & \epsilon^2 & \epsilon \\
\epsilon^2 & \epsilon^2 & 1 \\
\end{pmatrix} \, , \qquad
Y^e \sim
\begin{pmatrix}
\epsilon^2 & \epsilon^2 & \epsilon^2 \\
\epsilon^2 & \epsilon^2 & \epsilon^2 \\
\epsilon & \epsilon & 1 \\
\end{pmatrix} \, ,
\label{eq:flavortexture}
\eeq
with $\epsilon=\epsilon_Q=\epsilon_{u^c}=\epsilon_{e^c}$ being the ratio
\beq
\epsilon_R = \frac{\langle\omega_R\rangle}{\Lambda_{\rm flavor}} \sim
\frac{1}{10} \, .
\eeq
The textures roughly make the following prediction
\beq
\frac{m_c}{m_t} \sim \frac{m_s}{m_b} \sim \frac{m_\mu}{m_\tau} \sim
\mathcal{O}(\epsilon^2) \, , \qquad
\frac{m_t}{m_b} \sim \frac{m_t}{m_\tau} \sim \tan\beta \, .
\eeq
The numerical value is hence determined from the observed quark
masses and if $\tan\beta$ is sizable, the top-bottom mass hierarchy is
generated naturally.
The above pattern reproduces the
quark and lepton masses as well as the measured CKM matrix with
coefficients in the range $[0.8$--$1.1]$ for $\tan\beta=40$, see
\cite{Auzzi:2011eu}.
Here, however, we will not insist on such a high degree of precision.

The off-diagonal elements of the
Yukawas are crucial in order to produce a sufficient amount of quark
mixing, and therefore $\epsilon$ of order $1/10$ is preferred.
This fixes the Higgsing scale in terms of the flavor scale:
$v=\langle\omega\rangle\sim \Lambda_{\rm flavor}/10$.

\subsection{The Higgs and gravitino masses}

The Higgs boson has been found at the LHC with a mass of 125--126 GeV
\cite{ATLAS-CONF-2012-093,CMS-PAS-HIG-12-020} and hence we need to
accommodate such a ``large'' Higgs quartic in the model.
In the part of parameter space chosen in the present model
(viz.~$v>M$), the D-terms associated with the enhanced gauge symmetry
are decoupled and do not give any
observable contribution to the Higgs quartic coupling. Also, because we
are using gauge mediation, the trilinears vanish at the messenger
scale and are nowhere near sizable enough for increasing the Higgs
mass with light squarks.
Hence, the simplest possibility, which we utilize in this model, is
having very heavy sfermions, in particular, we need the stops of order
$\sim 10$ TeV, which will feed mass at one loop to the Higgs as
\cite{Martin:1997ns}
\beq
\delta m_{h_0}^2 = \frac{3}{4\pi^2} \cos^2(\alpha) Y_t^2 m_t^2
\log\left(\frac{m_{\tilde{t}_1} m_{\tilde{t}_2}}{m_t^2}\right) \, ,
\eeq
where $\alpha$ is the Higgs mixing angle \cite{Martin:1997ns},
$Y_t$ is the top Yukawa, $m_t$ is the top mass and finally
$m_{\tilde{t}_{1,2}}$ are the stop masses.

This means that the scale of the soft masses is
$\sqrt{2}\alpha F/(4\pi M)\sim 10$ TeV.
Hence, in the high-scale mediation case, where $M\sim 10^{15}$ GeV, we
have roughly
\beq
\sqrt{F} \sim 3 \times 10^{10} \; {\rm GeV} \, ,
\eeq
and in turn $x\equiv F/M^2 \sim 8 \times 10^{-11}$ giving a gravitino
mass of roughly
\beq
m_{3/2} \gtrsim 20 \; {\rm GeV} \, ,
\eeq
which is suitable as a cold dark matter candidate.
In the low-scale mediation case on the other hand, the gravitino will
be much lighter and can at best be a warm dark matter candidate, see
sec.~\ref{sec:dm_ls}.
The above calculated value assumes $k\equiv F/F_0=1$, where $\sqrt{F}$
is the SUSY-breaking scale felt by the messenger field whereas
$\sqrt{F_0}$ is the SUSY-breaking scale determining the gravitino
mass (though $k\leq 1$ and could be $\ll 1$
\cite{Gherghetta:1998tq,Giudice:1998bp}).

\subsection{Messenger sector, soft masses and the sparticle spectrum}

In order to get reasonably light gaugino masses compared to the
necessarily very heavy sfermion masses, we choose to work with a
messenger sector having more than one pair of messengers.
For concreteness, we choose a messenger sector having two
messengers (times an integer $p$, which has a trivial impact on the
gaugino mass to sfermion mass ratio)
\beq
\int d^4\theta \left(T_i^\dag T_i + \widetilde{T}_i^\dag \widetilde{T}_i\right)
+\int d^2\theta \; \widetilde{T}_i \widetilde{\mathcal{M}}_{ij} T_j
+ {\rm c.c.} \, , \qquad
\widetilde{\mathcal{M}} = \mathbf{1}_p \otimes \mathcal{M} \, , \qquad
\mathcal{M}_{ab} = m_{ab} + S \lambda_{ab} \, ,
\eeq
where $i,j=1,\ldots,2p$, $p\in\mathbb{Z}_{>0}$, $a,b=1,2$ and the
SUSY-breaking spurion attains an F-term VEV
\beq
\langle S\rangle = \theta^2 F \, ,
\eeq
and we assume messenger parity as well as CP conservation in the
messenger sector.
The above is a messenger sector characterized by a two-by-two matrix
$\mathcal{M}$ whose determinant specifies whether the gaugino masses
vanish two leading order or not \cite{Komargodski:2009jf}. Namely, if
$\det\mathcal{M}$ is independent of $S$, then the gaugino masses
vanish to leading order in SUSY breaking. We consider such a case in
sec.~\ref{sec:lowscale} while in sec.~\ref{sec:highscale} we study a
case where the gaugino mass does not vanish to leading order in
SUSY-breaking and thus allowing for a high messenger scale $M$,
suitable for a single scale unification scenario.
The sparticle spectrum, obtained via RG evolution down to the weak
scale, is presented for two corresponding benchmark points in
figs.~\ref{fig:hs} and \ref{fig:ls}. 

\subsection{Sflavor constraints}

The way the supersymmetric flavor problem is tackled in this type of
mild-split models is by degeneracy. In the limit of the Higgsing
scale, $m_v^2\equiv 2(g_A^2+g_B^2)v$, being much larger than the
messenger scale, $m_v\gg M$, universality of the squark masses holds
true. 
On the other hand, we would like to have as few scales in the model as
possible, that is, if $m_v$ would be of the order of $M$ (but still
larger), we would think of this as being ``one scale.''
Furthermore, the scale $M$ determines the gravitino mass, which in
turn determines whether the model can have a neutralino LSP or only a
gravitino LSP. This is important for the dark matter candidate in
question.
In order to make a quantitative assessment of the necessary separation
of scales $m_v,M$, we perform an extensive analysis of flavor
constraints in sec.~\ref{sec:sflavor}. The result of many sflavor
checks is that the $K-\bar{K}$ meson mixing with double insertion in
the mass insertion (MI) approximation is the most important of the
meson mixings with respect to the mass splitting induced at the
messenger scale. The $D-\bar{D}$ meson mixing, however, at double
insertions is sensitive to the top-Yukawa induced splitting for large
RG evolution. The bottom meson mixings are subdominant to the
mentioned ones. The branching ratio for $b\to s\gamma$ is
potentially important. However, due to vanishing $A$-terms,
$\delta^{\rm LR}$ is not inducing any sizable flavor changing effects.
The branching ratio of $\mu\to e\gamma$ is one of the major
constraints
and with a future experimental upgrade, it has potential to
probe quite far in the parameter space of the model.
The EDM coming both from gluino/squark diagrams and from the slepton
sector are important if no assumptions are made about complex
phases. Again with future experimental limits, the EDM of the electron
will be able to probe much farther in parameter space.
Finally, let us mention that the constraints due to the CP-violating
observable $\epsilon_K$ can also be satisfied with no assumption of
alignment in the high-scale model of sec.~\ref{sec:highscale} with
inverted matter content.

\subsection{Results}

Setting the gluino mass near 1150 GeV
-- its present bound \cite{ATLAS:2013ama},
the sparticle spectrum that we found then
depends mainly on the choice of $\tan\beta$ and the messenger scale. 
The average squark masses weigh in at
about 12+ TeV and the slepton masses at about 7+ TeV.
The wino is near 400 GeV and the bino sits near 200
GeV. The gravitino is almost always the LSP, although the bino can be
lighter in a corner of parameter space, see shortly.
Two benchmark points are presented in figs.~\ref{fig:hs} and
\ref{fig:ls}, for the high-scale and low-scale models, respectively. 

We have made a simple one-loop estimate to see how well the high-scale
model unifies. It turns out to match the measured value of the strong 
coupling only at the $2-3\sigma$ level. However, two-loop effects,
threshold effects and more importantly, matter from the link sector
has not been taken into account, which for just a slight splitting
could alter this substantially.

The flavor constraints can be satisfied, although some of the
constraints are on the border of probing the model, depending on
whether the normal or the inverted quiver model is chosen.
Even with order one complex phases, the models pass more or less the
limit on the $\epsilon_K$ CP-violating parameter of the kaon system as
well as constraints from the electric dipole moments of the neutron
and electron. 

The high-scale model has a gravitino dark matter candidate in most of
the parameter space, although a bino LSP is possible in a corner. The
gravitino is a cold dark matter candidate and can account for all the
measured dark matter abundance in accord with the recent observation
of Planck with a not-too-low reheating temperature to even be
compatible with a leptogenesis scenario. A potential problem, however,
is due to the NLSP -- the bino, decaying along with the emitted
photons potentially destroying light nuclei, synthesized during
BBN. There are some assumptions built into such cosmological
calculations and we have quoted a couple of ways out in
sec.~\ref{sec:dm}.

In the low-scale mediation case,
which can be embedded in a dynamical model,
the gravitino could potentially be a
warm dark matter candidate. We leave the verdict of the validity of
such a possibility to the astrophysics community.

In conclusion, we have presented a model with two different
incarnations, that address {\it all} the SM parameters,
with the measured Higgs mass.
It is able to pass sflavor constraints,
it may have perturbative unification,
and it has possibilities for
providing a dark matter candidate.

\subsection{Reading on}

In the following sections we will go into detail with the two
different model choices, their parameter spaces
(secs.~\ref{sec:highscale} and \ref{sec:lowscale}), and then in turn
their flavor constraints (sec.~\ref{sec:sflavor}) and dark matter
prospectives (sec.~\ref{sec:dm}). For the
high-scale case we contemplate also the quality of unification in
sec.~\ref{sec:uni}. In app.~\ref{app:sfermionmasses} we give two-loop
mass formulae for the scalars in a generic quiver with a general
messenger sector. The reader not interested in further details can
take a look at the spectra of figs.~\ref{fig:hs} and \ref{fig:ls} as
well as at the summary plot \ref{fig:sflavor_summary} for the sflavor
constraints and then jump to the discussion
(sec.~\ref{sec:discussion}).

\section{A high-scale model\label{sec:highscale}}

This model is chosen as an example of mild-split SUSY which
enjoys gauge coupling unification and a gravitino dark matter candidate,
though it does not come from a manifest dynamical embedding
(but perhaps an embedding in some `uplifted vacuum' exists).
The reason is the following.
Generic dynamical embeddings have vanishing gaugino masses
to leading order in SUSY breaking~\cite{Komargodski:2009jf},
unless the theory sits in an `uplifted vacuum' of the type studied
e.g.~in \cite{Giveon:2009yu,Auzzi:2010wm}.
Here, since we take the messenger scale $M$ to be near the GUT scale, a
messenger sector having vanishing gaugino masses to leading order
cannot produce a viable phenomenology. The leading order gaugino
mass goes like $M x$, where $x\equiv F/M^2$, while the next-to-leading
order contribution can be shown to go like $M x^3$, which for $x\sim
10^{-10}$ is completely negligible even for $M$ at the GUT scale.
Therefore, we consider here a different messenger sector which
interpolates that of minimal gauge mediation (MGM) (with two
messengers) and that of the dynamical embedding by a single real
parameter $\alpha$ 
\beq
\mathcal{M} = M
\begin{pmatrix}
1-2\alpha & 0 \\
0 & 1
\end{pmatrix}
+ S
\begin{pmatrix}
1 & \alpha \\
\alpha & 1
\end{pmatrix}
\, .
\eeq
For $\alpha=1$ the determinant is $-M^2$, i.e.~independent of $S$ and
hence due to the results of \cite{Komargodski:2009jf}, the gaugino
masses vanish to leading order in SUSY breaking. For vanishing
$\alpha$, on the other hand, the messenger sector is that of MGM with
two messenger fields. This interpolation is very simple, but has a
sick region, namely $\alpha$ should not be taken to be near $1/2$ as
one of the fermionic messengers is massless (or very light) and hence
the phenomenology is not viable in that region.
Since we are interested in a mild-split SUSY scenario, we consider
$\alpha$ less than, but close to, unity. \footnote{This type of
  mild-split SUSY spectrum comes also naturally in axion mediation
  models, see e.g.~\cite{Baryakhtar:2013wy}. }

The gaugino masses for this particular messenger sector are given by
\beq
m_{\tilde{g},k} = \frac{\alpha_k}{4\pi} M x \; 2p \;
\frac{1-\alpha}{2\alpha-1}
+ \mathcal{O}(\alpha_k M x^3) \, , \qquad
\alpha\in [0,1] \, ,
\eeq
where $\alpha_k\equiv g_k^2/(4\pi)$ are the gauge couplings with
$k=1,2,3$ corresponding to $\U(1)_Y,\SU(2)_L,\SU(3)_c$, respectively,
and $x\equiv F/M^2$.
Notice that for $\alpha=1$ the above expression vanishes and for
$\alpha=0$ the standard MGM formula is formally recovered (up to the
minus sign) with Dynkin index $2p$, corresponding to $2p$
messengers. Note also that the expression is not valid for
$\alpha=1/2$.

The sfermion masses are given by
\beq
m_{\tilde{f}}^2 = 2 \sum_{k=1}^3 \left(\frac{\alpha_k}{4\pi}\right)^2
C_{\tilde{f},k} M^2 x^2 p
\left(1 + \frac{1}{(1-2\alpha)^2}
+\frac{\alpha \log|1-2\alpha|}{\alpha-1} \right)
+\mathcal{O}(\alpha_k^2 M^2 x^4) \, ,
\eeq
where $C_{\tilde{f},k}$ is the quadratic Casimir of the sfermion
$\tilde{f}$ with respect to the gauge group $k$.
In the limit of $\alpha\to 1$, the above expression is formally equal
to that of MGM with Dynkin index $4p$, whereas for $\alpha=0$ it
recovers MGM with Dynkin index $2p$, corresponding to $2p$ messengers.

The given masses are all calculated at the messenger scale $M$ and
need to be RG evolved down to the electroweak scale for the physical
low-energy spectrum from which we can understand the phenomenology of
the model. Since we work in the part of parameter space where the
two-nodes quiver-like model effectively is MSSM-like, we can directly use
the spectrum calculator SOFTSUSY 3.3.4 \cite{Allanach:2001kg}.

In our model, the gluino mass is a(n) (almost) free parameter, so we
set it at 1150 GeV, which is close to current exclusion limits from
the LHC, see e.g.~\cite{ATLAS:2013ama}. The other gaugino masses
follow approximately from the GUT relation. 
If the bino were lighter than about 100 GeV, the charged wino would be
excluded up to $\sim 315$ GeV by ATLAS for decoupled sleptons 
\cite{ATLAS-CONF-2013-035}. 
However, for a bino heavier than around 120 GeV, there is practically
no bound from the LHC, although Tevatron data still excludes such a
charged wino below 270 GeV \cite{Meade:2009qv}. 
Due to the GUT relation among the gaugino masses, the exclusion limits
on the charged wino and on the bino are automatically satisfied when
the bound on the gluino is satisfied. 

\begin{figure}[!tp]
\begin{minipage}[t]{0.48\linewidth}
\begin{center}
\includegraphics[width=\linewidth]{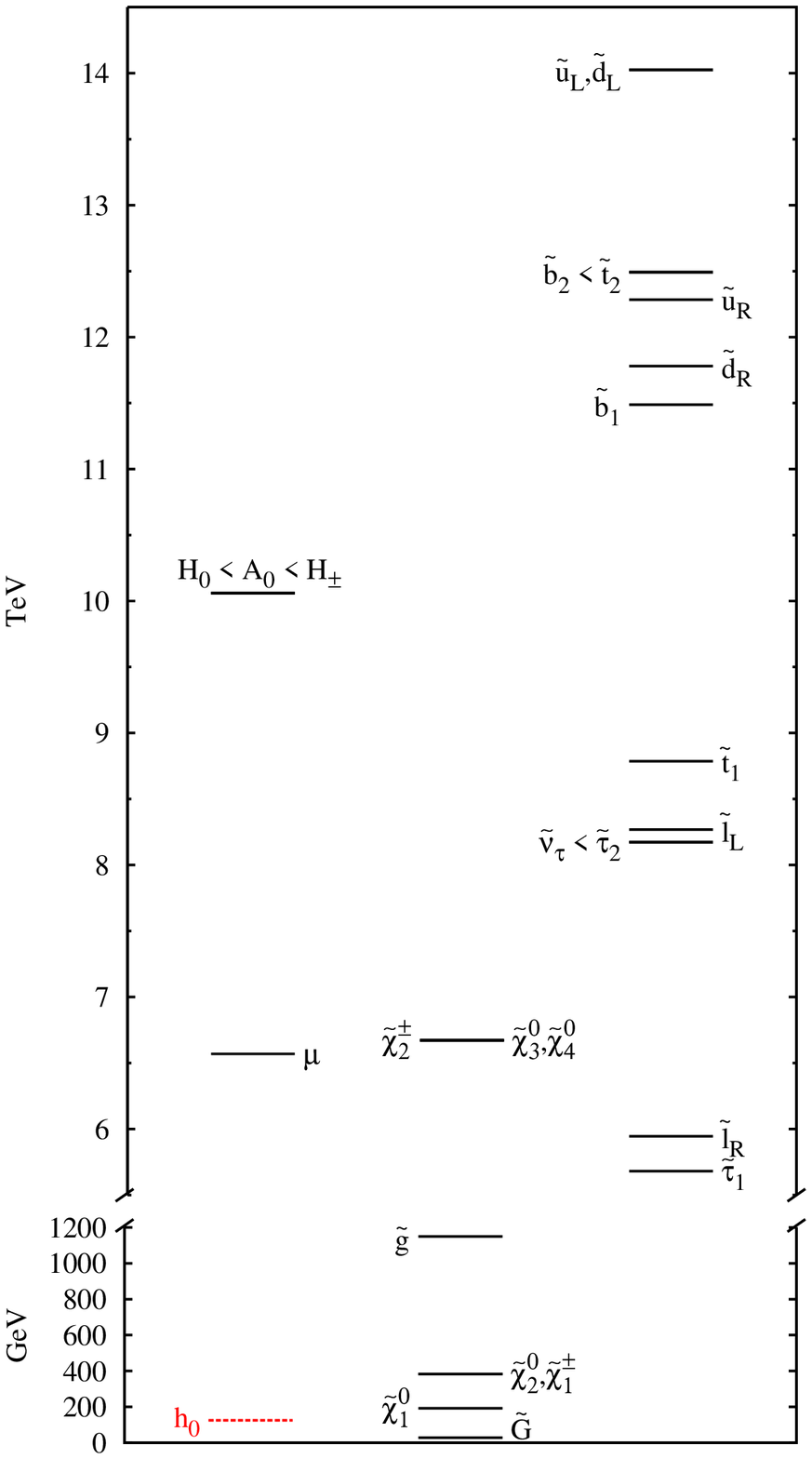}
\caption{Benchmark point for the high-scale model with
  $M=1.15\times 10^{14}$ GeV, $x=8.97\times 10^{-9}$, $p=1$, $y=35$,
  $\alpha=0.941$, $\tan\beta=20$, $\Delta_\mu^{-1}=9.6\times 10^{-5}$. }
\label{fig:hs}
\end{center}
\end{minipage}\ \
\begin{minipage}[t]{0.48\linewidth}
\begin{center}
\includegraphics[width=\linewidth]{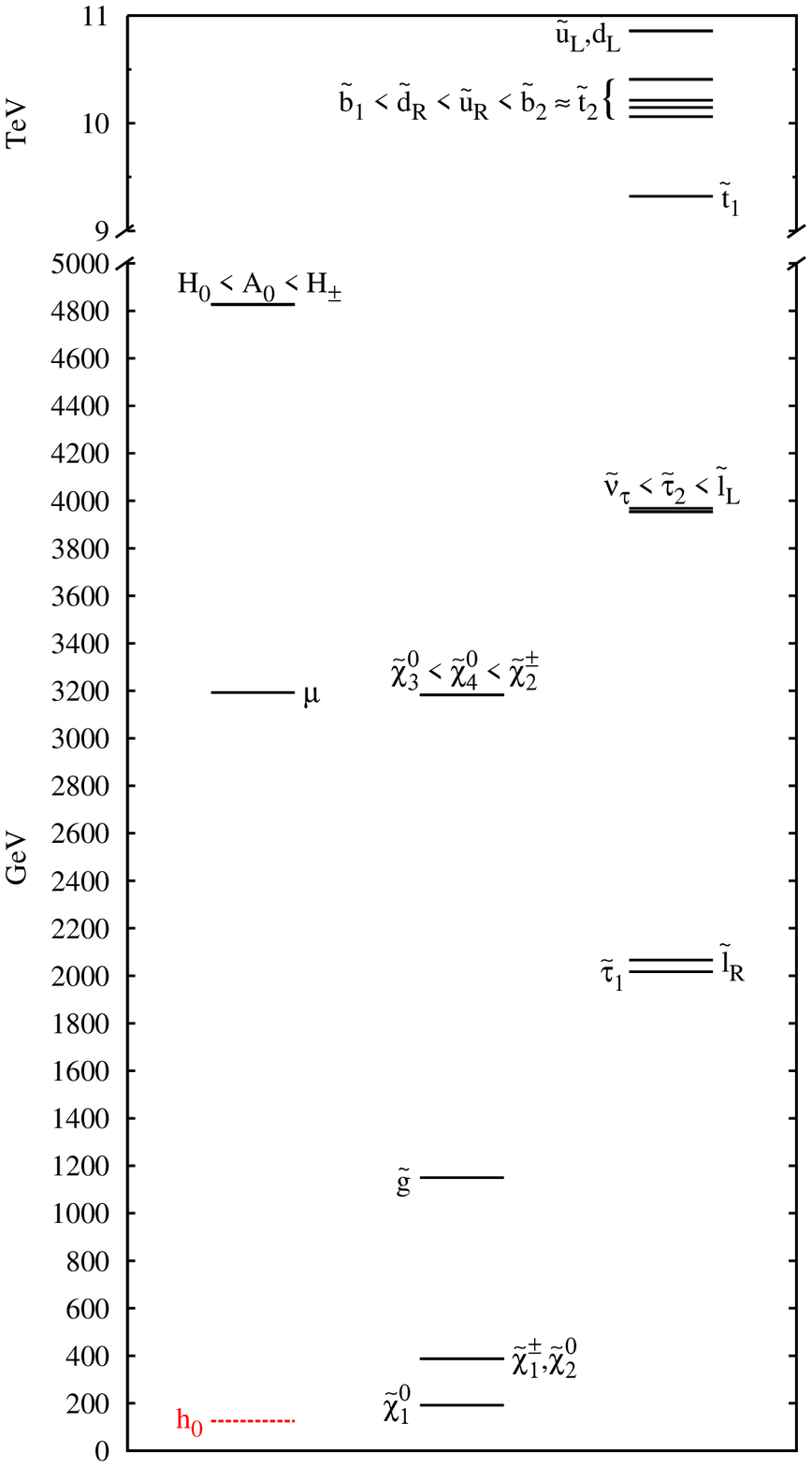}
\caption{Benchmark point for the low-scale model with
  $M=1.61\times 10^{6}$ GeV, $x=0.850$, $p=1$, $z=1$, $\tan\beta=20$,
  $\Delta_\mu^{-1}=4.1\times 10^{-4}$. }
\label{fig:ls}
\end{center}
\end{minipage}
\end{figure}

We show a benchmark point in fig.~\ref{fig:hs}.
The shown low-energy spectrum consists basically of the lightest
CP-even Higgs at 125.5 GeV \cite{ATLAS-CONF-2013-014}\footnote{For
  recent fits to the Higgs mass, see e.g.~\cite{Giardino:2013bma}
  in which a lower face-value is obtained. In order to be
  conservative, we choose to stick with a higher Higgs mass as a
  worst-case-scenario. } and the bino,
wino and gluino around 192 GeV, 383 GeV and 1150 GeV, respectively. The
LSP is the gravitino with a mass bigger than 28 GeV and the
fine-tuning according to the Barbieri-Giudice measure
\cite{Barbieri:1987fn}
\beq
\Delta_\mu \equiv \frac{2|\mu|^2}{m_Z^2} \, , 
\eeq
is roughly $\Delta_\mu^{-1}\sim 0.1$ per-mille.
In the chosen benchmark point, we have set $\tan\beta=20$ in order to
naturally produce a top-bottom hierarchy in the SM fermion mass sector
and the lightest CP-even Higgs is set at 125.5 GeV which then fixes
the stop masses and by means of the chosen parameter space also the
rest of the sfermions. 

The scales of the average squark masses and the gravitino mass are 
\beq
\tilde{m}_L\sim 13.5{}_{-2.4}^{+3.4}{}_{-3.1}^{+5.2}{}_{-0.93}^{+1.1}
\ {\rm TeV} \, , \qquad
\tilde{m}_R\sim 11.4{}_{-2.1}^{+2.9}{}_{-2.6}^{+4.4}{}_{-0.80}^{+0.94}
\ {\rm TeV} \, , \qquad
m_{3/2} \gtrsim 28.4{}_{-5.4}^{+7.6}{}_{-6.8}^{+11.5}{}_{-2.0}^{+2.3}
\ {\rm GeV} \, , \nonumber
\eeq
where the first uncertainty is estimated from changing the lightest
Higgs mass by $\pm 0.585$ GeV which corresponds to $\pm 1\sigma$,
statistical and systematical combined, and the second is due to the
uncertainty in the measurement of the top mass, i.e.~$\mp 1$ GeV
corresponding to $\mp 1\sigma$, combined. The last uncertainty is
estimated by changing the strong coupling $\alpha_3$ by $\pm 1.1\times
10^{-3}$ corresponding to $\pm 1\sigma$, experimentally.

\begin{figure}[!tp]
\begin{center}
\mbox{\subfigure[LSP/NLSP]{\includegraphics[width=0.49\linewidth]{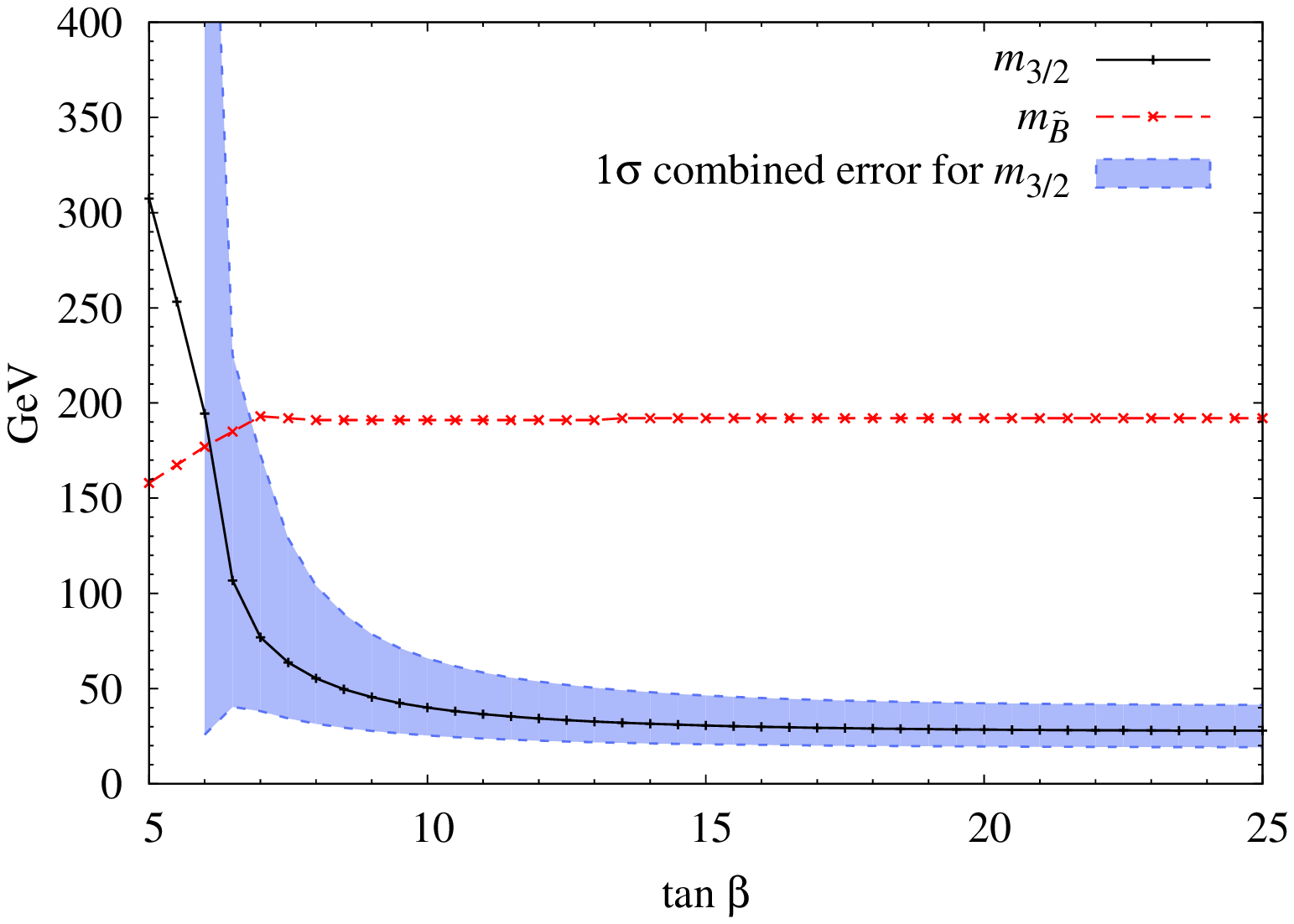}}
\subfigure[Average squark
  mass]{\includegraphics[width=0.49\linewidth]{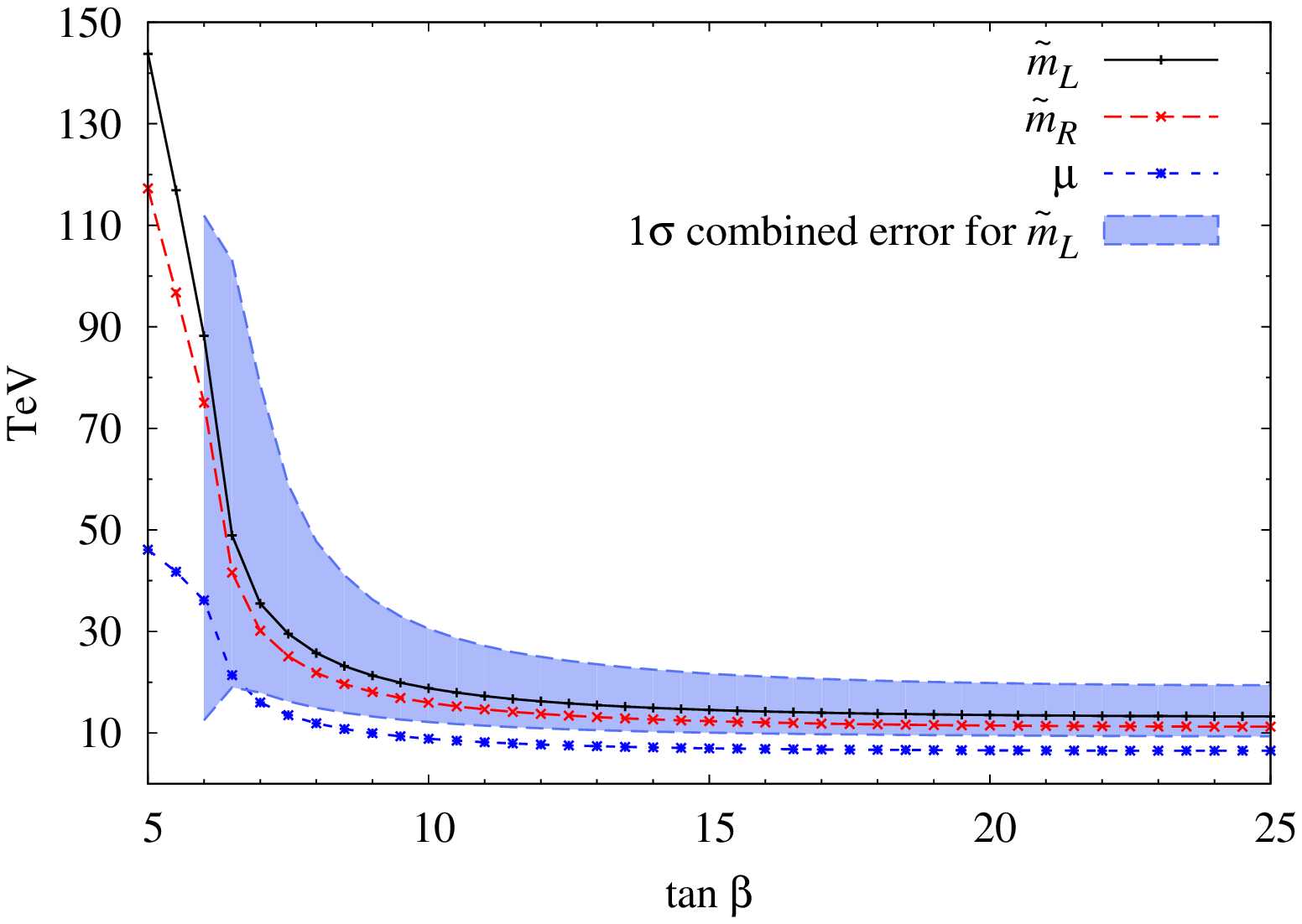}}}
\caption{a) The LSP/NLSP and b) the average squark masses as functions
  of $\tan\beta$ for the high-scale model. The plots are made keeping
  the Higgs mass fixed at 
  125.5 GeV and the gluino mass at 1150 GeV. 
  The relation between $\tilde{m}$ and $\tan\beta$ is
  sensitive to the uncertainties in $m_t,m_h$ and $\alpha_3$. The
  $1\sigma$ areas show the combined errors on the gravitino
  mass and on the left-handed average squark mass, respectively. The
  combined error is calculated by adding each error in quadrature,
  i.e.~$\pm\sqrt{\Delta_{\mp t}^2+\Delta_{\pm h}^2+\Delta_{\pm
      \alpha_3}^2}$. The errors are very large for $\tan\beta<6$ and
  are not shown here.}
\label{fig:tanbeta}
\end{center}
\end{figure}

One can however ask what happens to the spectrum by lowering
$\tan\beta$, since some tuning
of the top-bottom hierarchy is acceptable. What happens when
keeping the Higgs fixed at 125.5 GeV and the gluino at 1150 GeV (near
the bound), is that the squarks become heavier. With these
constraints and of course asserting electroweak symmetry breaking, the
mean squark masses raise to $\tilde{m}_L\sim 36$ TeV, $\tilde{m}_R\sim
31$ TeV, $m_{3/2}\sim 79$ GeV for $\tan\beta=7$ and to
$\tilde{m}_L\sim 144$ TeV, $\tilde{m_R}\sim 117$ TeV, $m_{3/2}\sim
307$ GeV, for $\tan\beta=5$, see fig.~\ref{fig:tanbeta}. 
We have not been able to find a spectrum with a heavy enough Higgs for
$\tan\beta\lesssim 5$ for fixed gluino mass, due to problems of
convergence of the numerical calculation. 
We have been able to find spectra with $\tan\beta\lesssim 5$ by
raising the gluino mass, but as we are investigating the scenario with
a gluino as close to discovery as possible we do not consider further
such possibility.

We have calculated the branching ratios of the spectrum in
fig.~\ref{fig:hs} using SUSY-HIT \cite{Djouadi:2006bz} and found the
following SUSY decays shown in tab.~\ref{tab:brs}.
\begin{table}[!htp]
\begin{center}
\begin{tabular}{l|l|l}
\hline\hline
gluino $\to$ & Br (high-scale) & Br (low-scale)\\
\hline
\phantom{.}&\\[-11pt]
$\widetilde{B}t\bar{t}$ & $18.7\%$ & $5.5\%$\\
$\widetilde{B}q\bar{q}$, $q=u,c$ & $7.9\%$ & $6.2\%$\\
$\widetilde{W}^+b\bar{t}$ or c.c. & $7.2\%$ & $7.9\%$\\
$\widetilde{W}^+q\bar{p}$, $(q,p)=(d,u),(s,c)$ or c.c. & $6.0\%$ & $8.8\%$\\
$\widetilde{W}^0b\bar{b}$ & $4.7\%$ & $5.2\%$\\
$\widetilde{W}^0q\bar{q}$, $q=u,d,c,s$ & $3.0\%$ & $4.4\%$\\
$\widetilde{B}b\bar{b}$ & $2.9\%$ & $1.9\%$\\
$\widetilde{B}q\bar{q}$, $q=d,s$ & $2.5\%$ & $1.8\%$\\
$\widetilde{W}^0t\bar{t}$ & $2.5\%$ & $2.7\%$\\
$\widetilde{B}g$ & $0.013\%$ & $1.1\times 10^{-5}$\\
$\widetilde{W}^0g$ & $8.3\times 10^{-8}$ & $9.5\times 10^{-7}$\\
$\widetilde{G}g$ & $2.0\times 10^{-21}$ & $2.2\times 10^{-6}$\\
\hline\hline
charged wino $\to$ & Br & Br\\
\hline
\phantom{.}&\\[-11pt]
$\widetilde{B}W^+$ & $100\%$ & $100\%$\\
$\widetilde{G}W^+$ & $2.4\times 10^{-24}$ & $1.0\times 10^{-9}$\\
\hline\hline
neutral wino $\to$ & Br & Br\\
\hline
\phantom{.}&\\[-11pt]
$h\widetilde{B}$ & $99\%$ & $96\%$\\
$Z\widetilde{B}$ & $1.4\%$ & $3.8\%$\\
$Z\widetilde{G}$ & $1.1\times 10^{-24}$ & $6.5\times 10^{-10}$\\
$\gamma\widetilde{G}$ & $7.4\times 10^{-25}$ & $2.9\times 10^{-10}$\\
$h\widetilde{G}$ & $2.8\times 10^{-31}$ & $3.0\times 10^{-15}$\\
\hline\hline
bino $\to$ & Br & Br\\
\hline
\phantom{.}&\\[-11pt]
$\gamma\widetilde{G}$ & $84\%$ & $89\%$\\
$Z\widetilde{G}$ & $16\%$ & $11\%$\\
$h\widetilde{G}$ & $2.6\times 10^{-9}$ & $4.8\times 10^{-8}$
\end{tabular}
\caption{Branching ratios of selected SUSY decays calculated using
  SUSY-HIT \cite{Djouadi:2006bz}. The above estimates for the
  branching ratios do not include the flavor violating decays that are
  present in our model, as they in turn depend on the parameter space
  of the model.}
\label{tab:brs}
\end{center}
\end{table}
A typical split-SUSY decay is the gluino decaying into a gravitino and
a gluon \cite{ArkaniHamed:2004yi}. This is, however, too suppressed in
this type of mild-split model to have any phenomenological
consequence. In this particular spectrum the branching ratio for such
a decay is about $2\times 10^{-21}$.
The production modes in this model will be via Drell-Yan production of
either $\widetilde{W}^+\widetilde{W}^-$ or
$\widetilde{W}^0\widetilde{W}^{\pm}$ \cite{Arvanitaki:2012ps}. 
As can be read off from tab.~\ref{tab:brs}, the neutral wino decays
almost exclusively to a bino via Higgs emission while the bino decays
dominantly into a photon and a gravitino. 
Due to the mentioned Higgs emission, one should consider search
strategies of \cite{Howe:2012xe} and due to the charged wino decays,
also searches for opposite-sign dilepton+missing transverse energy
are potentially important \cite{ATLAS-CONF-2013-049}.

\section{A low-scale model\label{sec:lowscale}}

Here we study a low-scale model which does not allow for conventional
gauge coupling unification, but instead it has a known dynamical
embedding in a deformed $\SU(N)$ SQCD \cite{Green:2010ww,Auzzi:2011gh}
with an appropriate number of flavors.
In particular, the messenger sector is a specific outcome of the above
mentioned scenario \cite{Green:2010ww,Auzzi:2011gh}
\beq
\mathcal{M} = M
\begin{pmatrix}
z & 1 \\
1 & 0
\end{pmatrix}
+ S
\begin{pmatrix}
1 & 0 \\
0 & 0
\end{pmatrix}
\, .
\eeq
The explicit formulae for the gaugino masses and the sfermion masses
are given in \cite{Marques:2009yu,Dumitrescu:2010ha,Auzzi:2011gh};
specifically we will use the parametrization given in
\cite{Auzzi:2011gh}.

\begin{figure}[!tp]
\begin{center}
\includegraphics[width=0.48\linewidth]{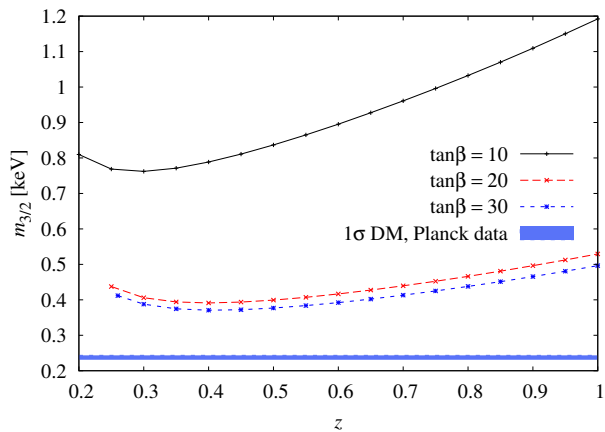}
\caption{Gravitino mass in the low-scale model as function of the
  parameter $z$. The messenger scale $M$ and $x$ are fixed by setting
  the Higgs and gluino masses equal to 125.5 GeV and 1150 GeV,
  respectively. The left-most end-point of each curve is approximately
  where $x$ approaches unity (recall that $x\leq 1$ in order not to
  have tachyonic messengers). We do not present the according change in
  the average mass of the squarks as it varies only about $1.5\%$ with
  $z$ in the range of the graph. }
\label{fig:lsgrav}
\end{center}
\end{figure}

A benchmark point for this model is shown in fig.~\ref{fig:ls}.
The low-energy spectrum is similar to the high-scale model and it
consists of the lightest CP-even Higgs at 125.5 GeV and the bino, wino
and gluino around 192 GeV, 387 GeV and 1150 GeV, respectively. The LSP
is the gravitino with a mass bigger than 0.53 keV, which is slightly
too large with respect to the bound from overclosure of the Universe,
see fig.~\ref{fig:lsgrav}.
The fine-tuning, however, according to the Barbieri-Giudice measure is
slightly better than in the high-scale case, viz.~0.4 per-mille.
As in the high-scale case, we have chosen $\tan\beta=20$ in order to
produce a top-bottom hierarchy in the SM fermion mass sector and the
lightest CP-even Higgs is set at 125.5 GeV which then fixes the stop
masses and by means of the chosen parameter space also the rest of the
sfermions. The main difference in this spectrum with respect to that
of the high-scale model, is the mass of gravitino and that the
sfermions, in particular the Higgsini, are somewhat lighter than in
the high-scale case. The branching ratios for the spectrum in
fig.~\ref{fig:ls} are calculated using SUSY-HIT \cite{Djouadi:2006bz}
and shown in tab.~\ref{tab:brs}.

\section{Unification\label{sec:uni}}

In this section we will briefly discuss the degree to which gauge
coupling unification works out in the high-scale mediation case.
We will just make a one-loop estimate of the state of
affairs and we will not incorporate a possible splitting in the link
sector here. Defining the unification scale by the intersection of the 
$\U(1)$ and $\SU(2)$ gauge couplings, we can trace back the gauge
coupling of $\SU(3)$ and compare it to the experimentally measured
value at the scale of the $Z$ mass. The expression at one-loop is
independent of complete $\SU(5)$ multiplets and reads
\begin{align}
\alpha_3^{-1}(m_Z) &\simeq \alpha_1^{-1}(m_Z)
-\frac{b_3^g}{2\pi}\log m_Z
-\frac{b_3^{\tilde{g}}}{2\pi}\log m_{\tilde{g}}
+\frac{b_1^h}{2\pi}\log m_H
+\frac{b_1^{\tilde{h}}}{2\pi}\log\mu
\non &\phantom{=\ }
+ \frac{b_3^g + b_3^{\tilde{g}} - b_1^h - b_1^{\tilde{h}}}{b_2^g +
  b_2^{\tilde{g}} + b_2^h + b_2^{\tilde{h}} - b_1^h - b_1^{\tilde{h}}}
\bigg(\alpha_2^{-1}(m_Z) - \alpha_1^{-1}(m_Z)
  + \frac{b_2^g}{2\pi}\log m_Z
  + \frac{b_2^{\tilde{g}}}{2\pi}\log m_{\tilde{W}}
\non &\phantom{\simeq + \frac{b_3^g + b_3^{\tilde{g}} - b_1^h -
    b_1^{\tilde{h}}}{b_2^g +
  b_2^{\tilde{g}} + b_2^h + b_2^{\tilde{h}} - b_1^h -
  b_1^{\tilde{h}}}\bigg(\ }
  + \frac{b_2^h - b_1^h}{2\pi}\log m_H
  + \frac{b_2^{\tilde{h}} - b_1^{\tilde{h}}}{2\pi}\log\mu
\bigg) \, ,
\end{align}
where $m_Z$, $m_{\tilde{g}}$, $m_{\tilde{W}}$, $m_H$ and $\mu$ are the
$Z$ mass, the gluino mass, the wino mass, the mass of the heavy Higgses
and the higgsino mass, respectively, and $b_3^g=-11$,
$b_3^{\tilde{g}}=2$, $b_2^g=-22/3$, $b_2^{\tilde{g}}=4/3$,
$b_2^h=1/3$, $b_2^{\tilde{h}}=2/3$, $b_1^h=1/5$ and
$b_1^{\tilde{h}}=2/5$.
\begin{figure}[!tp]
\begin{center}
\includegraphics[width=0.5\linewidth]{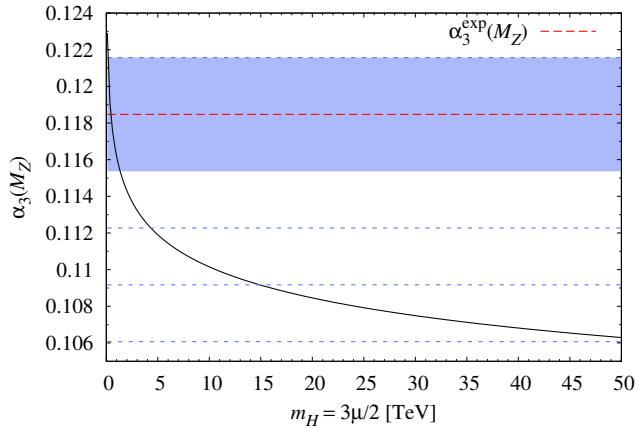}
\caption{Unification at one loop is estimated by the intersection of
  the $\U(1)$ and $\SU(2)$ gauge couplings and then the $\SU(3)$ gauge
  coupling is traced back to the $m_Z$-scale where it is compared
  to the experimental value of \cite{Beringer:1900zz}. The blue strip
  denotes the $1\sigma$ band while blue-dashed lines show the $2-4$
  standard deviations from top to bottom. We use 0.0031 as the
  standard deviation, which is a combined value. We have set
  $\mu=\tfrac{2}{3}m_H$, $m_{\tilde{g}}=1150$ GeV and
  $m_{\widetilde{W}}=383$ GeV as a representative example using the
  spectrum of fig.~\ref{fig:hs}. }
\label{fig:uni}
\end{center}
\end{figure}
In fig.~\ref{fig:uni} is shown the degree to which gauge coupling
unification works in terms of the strong gauge coupling matching with
the experimental measured value. If, for instance, the strong gauge
coupling should match up with its experimentally measured value at
$m_Z$ to within $2\sigma$s, the heavy Higgses should weigh less than
about $4.7$ TeV. Comparing with fig.~\ref{fig:tanbeta},
$\alpha_3(m_Z)$ matches its experimental value between $2$ and
$4\sigma$s, depending on the value of $\tan\beta$ and hence the scale
of Higgses and higgsinos.

The gauge coupling unification described here is solely that of gauge
group $G_A$ of fig.~\ref{fig:quiver2nodes} and it is not spoiled by
the (un)Higgsing with the gauge group $G_B$ as long as the gauge
couplings on $G_B$ are kept $\SU(5)$ invariant and the messenger
and link fields transform in complete $\SU(5)$ representations.

\section{Superpartner flavor and CP phases\label{sec:sflavor}}

In this section, we will explain the sflavor constraints in
detail. Many constraints were checked that were subdominant and those
are just mentioned with references to the literature whilst the
important ones for the model are explained here (for reviews, see
e.g.~\cite{Gedalia:2010rj,Jager:2008fc}).
As the model has near-flavor universality, it is adequate to use the
Mass Insertion (MI) approximation \cite{Hall:1985dx}.
Because the first two generations of squarks are on the same node,
one could naively expect the $K-\bar{K}$ and $D-\bar{D}$
constraints to be automatically satisfied, whilst the constraints from
$B_d-\bar{B}_d$ (and $B_s-\bar{B_s}$ to a lesser extent) to be
important.
It turns out, however, that the gluino box diagrams contributing to
the mass difference in the neutral kaon system -- even though having
vanishing contributions from single flavor-flip insertions -- have
dominating/competitive contributions from double flavor-flip mass
insertions $(2\to 3)\times (3\to 1)$ \cite{Giudice:2008uk}. This type
of effective mass-insertion has recently received attention due to
much interest in natural SUSY models with a hierarchy in the soft
masses
\cite{Dimopoulos:1995mi,Cohen:1996vb,Barbieri:2010pd,Barbieri:2011ci,Brust:2011tb,Papucci:2011wy}. 
Even though we do not have a hierarchy in the soft masses, only an
$\SU(2)$ flavor symmetry is preserved at the messenger scale in the
two-nodes model and thus the mass differences -- although small --
resides between the first two and the third generation of
squarks. This explains the importance of the double flavor-flip
effective mass insertions. The reason why the kaon system is
competitive with the $B$ meson system, is due to the tighter
experimental limit.

The meson mixing itself does not pose the strongest limit on the model
at hand, but the CP-violating parameter $\epsilon_K$ in the kaon
system provides one of the toughest constraints.

Flavor violation is possible also in the leptonic
sector where the branching ratio of $\mu\to e\gamma$ gives rise to
strong constraints, especially for large $\tan\beta\gtrsim 20-30$. If
furthermore order one complex phases -- 
of which our model has two of -- are not tuned away somehow, then the
electric dipole moment of the electron sets an even stronger
constraint. Finally, allowing for order one complex phases, also the
squark sector induces electric dipole moments, in this case, affecting
that of the neutron.

We will start by explaining the sflavor constraint calculation for the
$K-\bar{K}$ meson mixing and CP-violation, in order to set a
limit on how small $y=m_v/M$ can be; $m_v^2\equiv 2(g_A^2+g_B^2)v^2$ is
the Higgsing scale of the link fields
$\omega,\tilde{\omega}$.\footnote{Strictly speaking there are three
  different values of the Higgsing scale: $m_{v_k}^2 =
  2(g_{A_k}^2+g_{B_k}^2)v^2$, but when we omit the index $k$, a
  geometric average value of the three scales is understood. } 
We work in the framework of the effective Hamiltonian
\beq
\mathcal{H}_{\rm eff} = C_1 O_1 + \tilde{C}_1 \tilde{O}_1
+ C_4 O_4 + C_5 O_5 \, ,
\eeq
where the operators are defined as
\beq
O_1 = (\bar{d}_L^\alpha\gamma_\mu s_L^\alpha)
(\bar{d}_L^\beta\gamma^\mu s_L^\beta) \, , \quad
O_4 = (\bar{d}_R^\alpha s_L^\alpha)(\bar{d}_L^\beta s_R^\beta) \, ,
\quad
O_5 = (\bar{d}_R^\alpha s_L^\beta)(\bar{d}_L^\beta s_R^\alpha) \, ,
\eeq
with $\alpha,\beta$ being color indices and $\tilde{O}_1$ is given by
$O_1$ with $L\to R$. Since we are working in a gauge-mediated
SUSY-breaking model with negligible $A$-terms, the operators $O_{2,3}$
and their corresponding tilded ones are subdominant.
The double flavor-flip mass insertion \cite{Giudice:2008uk}
contributions from gluino box diagrams have Wilson coefficients
\cite{Hall:1985dx,Hagelin:1992tc,Gabbiani:1996hi} given by
\cite{Altmannshofer:2009ne}
\begin{align}
C_1^{\tilde{g}} &\simeq -\frac{\alpha_3^2}{\tilde{m}^2}
\left[(\delta_d^{LL})_{23}(\delta_d^{LL})_{31}\right]^2 g_1^{(3)}(x_{\tilde{g}})
\, , \non
C_4^{\tilde{g}} &\simeq -\frac{\alpha_3^2}{\tilde{m}^2}
\left[(\delta_d^{LL})_{23}(\delta_d^{LL})_{31}
(\delta_d^{RR})_{23}(\delta_d^{RR})_{31}\right] g_4^{(3)}(x_{\tilde{g}}) \, ,
\label{eq:WilsonKKbar}\\
C_5^{\tilde{g}} &\simeq -\frac{\alpha_3^2}{\tilde{m}^2}
\left[(\delta_d^{LL})_{23}(\delta_d^{LL})_{31}
(\delta_d^{RR})_{23}(\delta_d^{RR})_{31}\right] g_5^{(3)}(x_{\tilde{g}}) \, ,
\nonumber
\end{align}
where $x_{\tilde{g}}\equiv m_{\tilde{g}}^2/\tilde{m}^2$, $\tilde{m}$
being the average squark mass, $m_{\tilde{g}}$ the gluino mass, and
the loop functions are given in app.~A of
\cite{Altmannshofer:2009ne}. 

Another contribution to the $K-\bar{K}$ mixing comes from double
neutral Higgs penguin diagrams, again at fourth order in squark MIs
\cite{Altmannshofer:2009ne}
\beq
C_4^H \simeq -\frac{\alpha_3^2\alpha_2}{4\pi}
\frac{m_b^2}{2m_W}
\frac{\tan^4\beta}{(1+\epsilon_{\tilde{g}}\tan\beta)^4}
\frac{|\mu|^2 m_{\tilde{g}}^2}{m_A^2\tilde{m}^4}
(\delta_d^{LL})_{23}(\delta_d^{LL})_{31}
(\delta_d^{RR})_{23}(\delta_d^{RR})_{31}
h_2^2(x_{\tilde{g}}) \, ,
\eeq
where $m_b$, $m_W$, $\mu$, $m_A$ and $\tilde{m}$ are the masses of the
bottom quark, the $W$-bosons, the supersymmetric Higgs mass, the
CP-odd Higgs state and the average squark mass, respectively; the loop
function $h_2(x_{\tilde{g}})$ can be found in the app.~A of
\cite{Altmannshofer:2009ne} while
\beq
\epsilon_{\tilde{g}} \simeq \frac{2\alpha_3}{3\pi}
\frac{\mu m_{\tilde{g}}}{\tilde{m}^2}
\left(\frac{1}{1-x_{\tilde{g}}} + \frac{x_{\tilde{g}}}{(1-x_{\tilde{g}})^2}\log x_{\tilde{g}}\right) \, .
\eeq
The neutral Higgs contribution becomes important and hence competitive
with the gluino box contribution for $\tan\beta\gtrsim 30-50$. For
$\tan\beta\lesssim 20$ it is negligible compared to that of the gluino
box.

The $K_L-K_S$ mass difference and the CP-violating parameter of the
kaon system are then calculated by
\begin{align}
\Delta m_K &= 2\Re\langle K^0|\mathcal{H}_{\rm eff}|\bar{K}^0\rangle \,
, \\
\epsilon_K &= \frac{1}{\sqrt{2}\Delta m_K} \Im\langle
K^0|\mathcal{H}_{\rm eff}|\bar{K}^0\rangle \, .
\end{align}
The matrix elements can be found in
\cite{Bagger:1997gg,Ciuchini:1998ix,Contino:1998nw} and recent results
for the bag parameters in \cite{Boyle:2012qb,Bertone:2012cu}, while
the RG evolution of the matrix elements is done using the magic
numbers of \cite{Ciuchini:1998ix}.

The mass insertions are given by the off-diagonal elements of the
(here down-type) squark mass-squared matrix in the super-CKM basis
\cite{Misiak:1997ei} divided by the average squark mass-squared
\beq
(\delta_{d}^{MN})_{ij} =
\frac{(\mathcal{M}_{\tilde{d}}^{MN})_{ij}^2}{\tilde{m}^2} =
\frac{\left(V_M^d
  \diag(m_{\tilde{d},M}^2,m_{\tilde{s},M}^2,m_{\tilde{b},M}^2)
  (V_N^d)^\dag\right)_{ij} \delta^{MN}}{\tilde{m}^2} \, ,
\label{eq:delta}
\eeq
where $M,N=L,R$, and $i,j=1,2,3$. We are neglecting the $LR,RL$
elements due to vanishing $A$-terms.
The matrices $V_{M,N}^d$ are the bi-unitary rotation matrices used to
diagonalize the down-type fermions. Since we work with the two-nodes
model of fig.~\ref{fig:quiver2nodes}, the first two generations of
squarks are mass-degenerate at the messenger scale $M$. Defining
$\Delta m_{\tilde{d}}^2\equiv m_{\tilde{b}}^2 - m_{\tilde{d}}^2 =
m_{\tilde{b}}^2 - m_{\tilde{s}}^2$, we can write
\beq
(\delta_d^{MM})_{ij} = (\eta_d^{MM})_{ij}
\frac{\Delta m_{\tilde{d}}^2}{\tilde{m}^2} \, , \qquad
(\eta_d^{MM})_{ij} \equiv (V_M^d)_{i3} (V_M^{d\dag})_{3j} \, ,
\label{eq:etadef}
\eeq
where $(\eta_d^{MM})_{ij}\leq 1/2$ are matrix elements that depend on
the basis of the Yukawa matrices. The reason why the coefficients are
smaller than $1/2$ is geometric. Let us consider an $\SU(2)$ subgroup,
for which we can write e.g.~the element
\beq
\left|(\eta_d^{MM})_{13}\right| = \sqrt{x(1-x)} \leq \frac{1}{2} \, ,
\qquad
[0,1]\ni x = (n_1^2+n_2^2)\sin^2\alpha \, ,
\eeq
where $n_{1,2,3}$ is a (real) three-component unit vector and
$\alpha\in\mathbb{R}$ is a real number. We would like to present a
conservative flavor analysis, i.e.~not assuming any alignments or
tuning of complex phases. The most conservative choice would thus be
to set $(\eta_d^{MM})_{13,23}=1/2$. However, in order to see if
such a choice is by any means realistic in the model at hand, we
perform a Monte Carlo analysis, presented in
app.~\ref{app:montecarlo}. It is done setting $\tan\beta=20$ and
assuming the Yukawa texture \eqref{eq:flavortexture} with coefficients
in the range $0.1-2$. All the random matrices generated have
coefficients producing the measured quark (lepton) masses as well as
the best-fit CKM matrix. For each such generated Yukawa matrix, we
calculate the diagonalization matrices which by eq.~\eqref{eq:etadef}
gives us $(\eta_{u,d}^{MM})_{ij}$. The analysis shows that in the
squark sector the largest values of $(\eta_d^{MM})_{13,23}$ are about
$\sim 1/4$ and $\sim 1/2$ in the slepton sector. Notice that the
specific choice of texture, $\tan\beta$ and the coefficients are such that
the $(\eta_{d}^{RR})_{13,23}\lesssim 0.08$. We will not incorporate
this into the presented flavor constraints, as it depends on choices
made in the Monte Carlo, but we will use this information in the
summary plots presented in the end of this section.

That is, we would like the analysis to represent
the generic constraints for the model, also if we change $\tan\beta$
or the link field representation \cite{Auzzi:2011eu}. Due to above
discussion, we choose to present the flavor constraints with
\beq
(\eta_u^{MM})_{13,23}=\frac{1}{4} \, , \qquad
(\eta_d^{MM})_{13,23}=\frac{1}{4} \, , \qquad
(\eta_\ell^{MM})_{13,23}=\frac{1}{2} \, , \qquad
(\eta_e^{MM})_{13,23}=\frac{1}{2} \, .
\eeq

The second ingredient in the $\delta$s is the actual mass-splitting,
which is intrinsic and is generated at the 
messenger scale $M$ due to the fact that the model is a two-nodes
model (see fig.~\ref{fig:quiver2nodes}).
The mass-splitting generated at the messenger scale is calculated
exactly in app.~\ref{app:sfermionmasses}. The exact expressions are
given by eq.~\eqref{eq:general_massfunction} with the form factors in
app.~\ref{app:twonodes}, while an expansion to second order in $1/y$
is given by eq.~\eqref{eq:massfunction_largeyexpansion}.
Using eq.~\eqref{eq:massfunction_largeyexpansion}, we find the
mass splitting at the messenger scale
\beq
\frac{\Delta m_{L,R}^2}{\tilde{m}^2} \sim
\pm c \times \frac{2\lambda_{\rm av}^{LL,RR} \log^2 y^2}{y^2} \, , \qquad
\lambda_{\rm av}^{LL} \equiv
\frac{1}{3}\sum_{k=1}^3\frac{g_B^2}{g_{{\rm eff},k}^2} \, , \qquad
\lambda_{\rm av}^{RR} \equiv
\frac{1}{2}\sum_{k=1,3}\frac{g_B^2}{g_{{\rm eff},k}^2} \, ,
\eeq
where $+$ ($-$) is for the normal (inverted) model and $c$ is an order 
one constant determined by the messenger sector matrix $\mathcal{M}$
while $y\equiv m_v/M$. In the case of the high-scale mediation model
of sec.~\ref{sec:highscale}, $c\sim 1$ will be a function of $\alpha$
but having only a mild dependence for $\alpha\lesssim 1$ in the
relevant parameter space. 

The contribution to the $\delta$s described so far is the intrinsic
one due to the mass splitting at the messenger scale. RG evolution
\cite{Martin:1993zk} will also induce a contribution to the $\delta$s
which can be split into a MFV (minimal flavor violation) and a non-MFV
part
\begin{align}
\big(\delta_u^{LL}\big)_{i3}^{\rm RG} &\sim 
\frac{t}{8\pi^2}\left(2+\frac{m_{H_d}^2}{\tilde{m}^2}\right) 
|Y_b|^2K_{i3}^\dag K_{33}
+\frac{t}{16\pi^2}\frac{\Delta m_L^2}{\tilde{m}^2}\bigg[
|Y_t|^2 \big(\eta_u^{LL}\big)_{i3} 
+ |Y_b|^2 \big(\eta_u^{LL}\big)_{ik} K_{k3}^\dag K_{33}
\non &\phantom{=\ }
+ |Y_b|^2 K_{i3}^\dag K_{3k} \big(\eta_u^{LL}\big)_{k3}
+ 2|Y_b|^2\frac{\Delta m_R^2}{\Delta m_L^2} K_{i3}^\dag
\big(\eta_d^{RR}\big)_{33} K_{33}\bigg] \,
, \label{eq:deltauLLsimp} \\ 
\big(\delta_u^{RR}\big)_{i3}^{\rm RG} &\sim
\frac{t}{8\pi^2}\frac{\Delta m_R^2}{\tilde{m}^2}
|Y_t|^2 \big(\eta_u^{RR}\big)_{i3} \, , \label{eq:deltauRRsimp} \\
\big(\delta_d^{LL}\big)_{i3}^{\rm RG} &\sim 
\frac{t}{8\pi^2}\left(2+\frac{m_{H_u}^2}{\tilde{m}^2}\right) 
|Y_t|^2 K_{i3} K_{33}^\dag
+\frac{t}{16\pi^2}\frac{\Delta m_L^2}{\tilde{m}^2} \bigg[
|Y_b|^2 \big(\eta_d^{LL}\big)_{i3}
+ |Y_t|^2 \big(\eta_d^{LL}\big)_{ik} K_{k3} K_{33}^\dag
\non &\phantom{=\ }
+ |Y_t|^2 K_{i3} K_{3k}^\dag \big(\eta_d^{LL}\big)_{k3}
+ 2|y_t|^2 \frac{\Delta m_R^2}{\Delta m_L^2} K_{i3}
\big(\eta_u^{RR}\big)_{33} K_{33}^\dag\bigg] \,
, \label{eq:deltadLLsimp} \\ 
\big(\delta_d^{RR}\big)_{i3} &\sim 
\frac{t}{8\pi^2} \frac{\Delta m_R^2}{\tilde{m}^2}
|Y_b|^2 \big(\eta_d^{\rm RR}\big)_{i3} \, , \label{eq:deltadRRsimp}
\\
\big(\delta_\ell^{LL}\big)_{i3} &\sim 
\frac{t}{8\pi^2} \frac{\Delta m_{\ell,L}^2}{\tilde{m}_{\ell}^2}
|Y_\tau|^2 \big(\eta_\ell^{\rm LL}\big)_{i3} \, , \label{eq:deltaeLLsimp}
\\
\big(\delta_e^{RR}\big)_{i3} &\sim 
\frac{t}{8\pi^2} \frac{\Delta m_{e,R}^2}{\tilde{m}_{\ell}^2}
|Y_\tau|^2 \big(\eta_e^{\rm RR}\big)_{i3} \, , \label{eq:deltaeRRsimp}
\end{align}
where $i=1,2$, $t = \log(m_t/M)$ is the range of the RG running and $K^{\rm T}$
is the CKM matrix. 
The first terms in eqs.~\eqref{eq:deltauLLsimp} and
\eqref{eq:deltadLLsimp} are of MFV type while all the other terms are
not, but are proportional to the intrinsic mass splitting $\Delta m^2$
due to the model being a two-nodes quiver, generated at the messenger
scale. 

\begin{figure}[!tp]
\begin{center}
\mbox{\subfigure[Two-node
    quiver]{\includegraphics[width=0.49\linewidth]{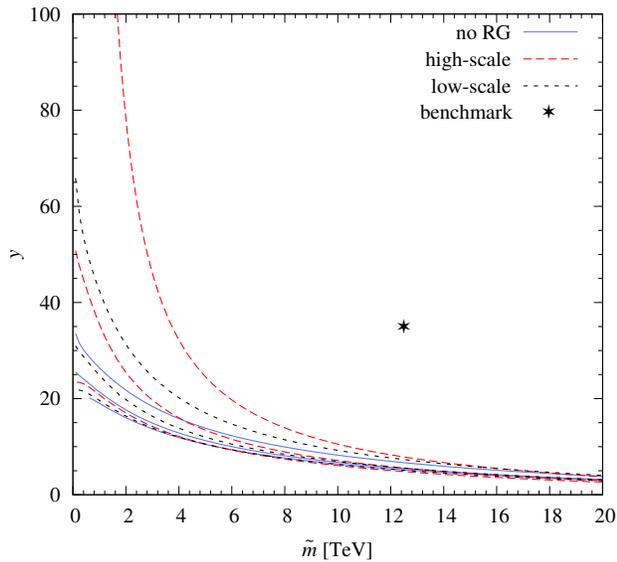}}
\subfigure[Inverted two-node
    quiver]{\includegraphics[width=0.49\linewidth]{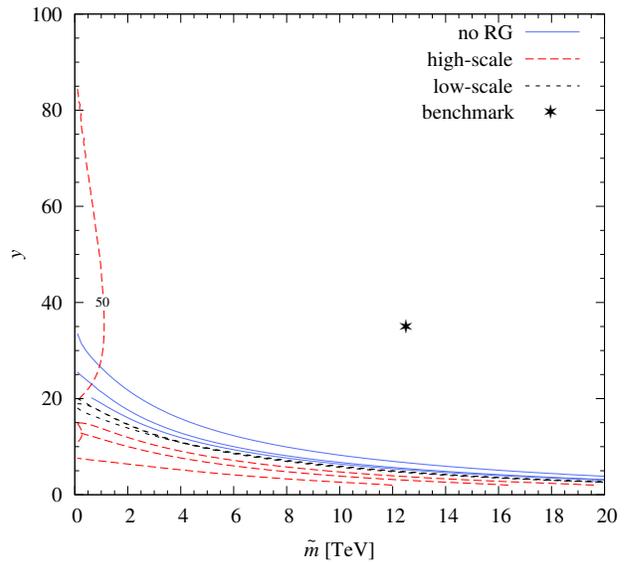}}}
\caption{Constraints on $\Delta m_K$ from gluino box diagrams and
  double Higgs penguin diagrams in the ($\tilde{m},y$)-plane for a)
  the normal and b) the inverted quiver model.
  The three lines (from above) represent $\tan\beta=50,30,10$.
  The gluino mass is set to 1150 GeV, $\mu=\tilde{m}/2$ and
  $m_H=2\tilde{m}/3$. The Higgs contributions kick in at
  $\tan\beta\gtrsim 30$. 
  The relation between the average squark mass, $\tilde{m}$, and
  $\tan\beta$ is sensitive to the properties of the messenger sector
  (e.g.~the values of $M,\alpha,z$) as well as the precise value of the SM
  parameters (e.g.~a slight change of $m_t,m_h$ and $\alpha_3$ feeds
  large modifications to $\tilde{m}$ via RG evolution); this and the
  following figures thus span the full relevant parameter space. }
\label{fig:deltamk}
\end{center}
\end{figure}

Now we are ready to present the flavor constraints for the models at
hand.
In fig.~\ref{fig:deltamk} we show the results of the constraints due
to $K-\bar{K}$ mixing coming from both gluino box diagrams at fourth
order in the MI approximation, i.e.~two MIs with double flavor-flip
($2\to 3$ and $3\to 1$) in the super-CKM basis (recall that the flavor
changing $2\to 1$ process is negligible as the first two generations
are degenerate at the messenger scale). We have set the gluino mass to
be 1150 GeV for all the constraints. The experimental limits used are
summarized in tab.~\ref{tab:explimits}. The benchmark point shown in
fig.~\ref{fig:hs} is indicated in all the flavor constraints with a
black star.

\begin{table}[!ht]
\begin{center}
\begin{tabular}{l|l}
Observable & Limit\\
\hline
$\Delta m_K$ & $3.484\times 10^{-15}$ GeV \cite{Beringer:1900zz}\\
$\Delta m_D$ & $1.30\times 10^{-14}$ GeV \cite{Amhis:2012bh}\\
$\Delta m_{B_d}$ & $3.337\times 10^{-13}$ GeV \cite{Beringer:1900zz}\\
$|\epsilon_K|$ & $2.228\times 10^{-3}$ \cite{Beringer:1900zz}\\
${\rm Br}(\mu\to e\gamma)$ & $5.7\times 10^{-13}$ \cite{Adam:2013mnn}\\
$|d_e|$ & $1.05\times 10^{-27}$ e cm (90\% CL.) \cite{Hudson:2011zz}\\
$|d_n|$ & $2.9\times 10^{-26}$ e cm (90\% CL.) \cite{Baker:2006ts}
\end{tabular}
\caption{Experimental limits used in the presented plots.}
\label{tab:explimits}
\end{center}
\end{table}

\begin{figure}[!tbp]
\begin{center}
\mbox{\subfigure[Two-node
    quiver]{\includegraphics[width=0.49\linewidth]{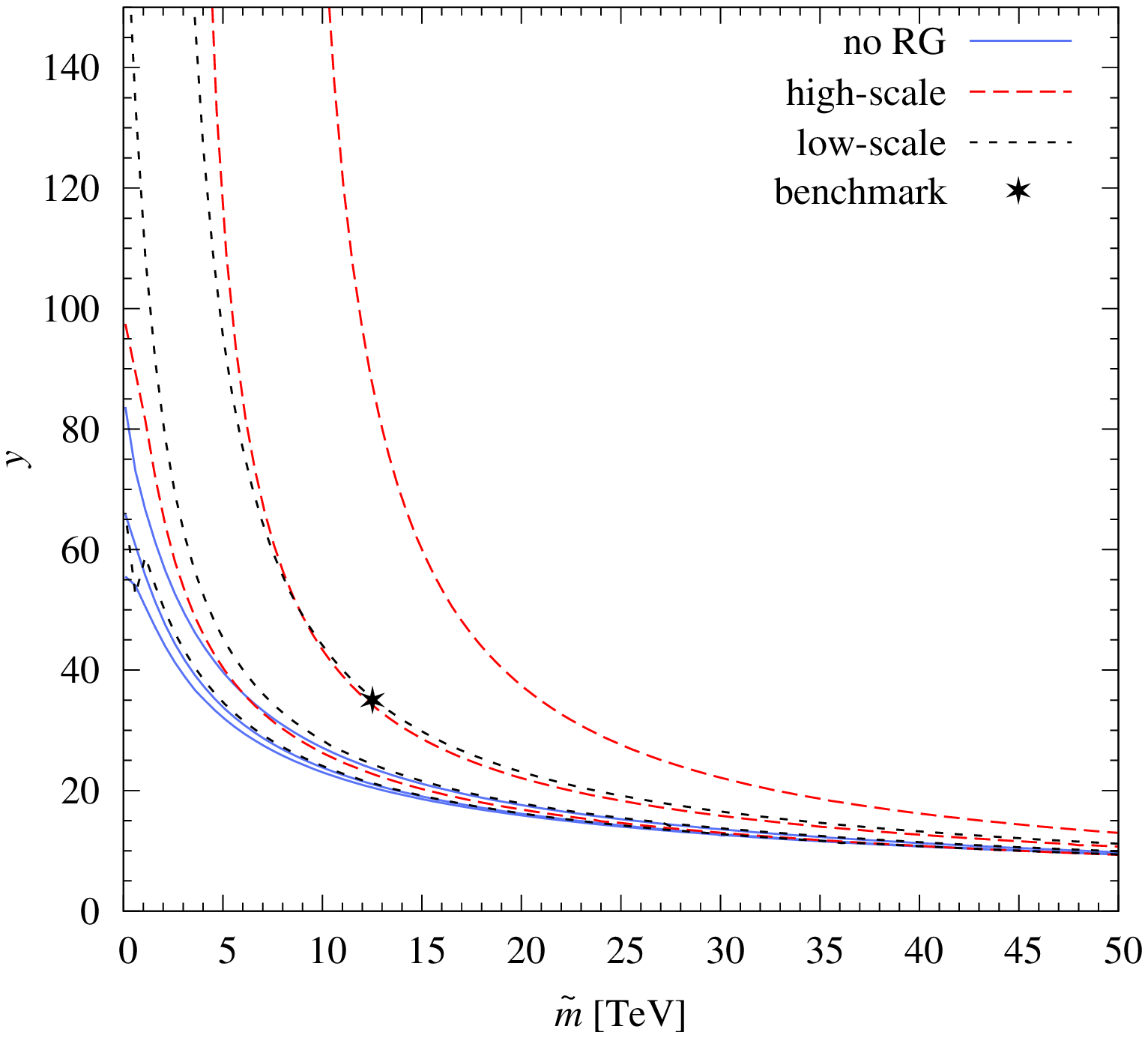}}
\subfigure[Inverted two-node
  quiver]{\includegraphics[width=0.49\linewidth]{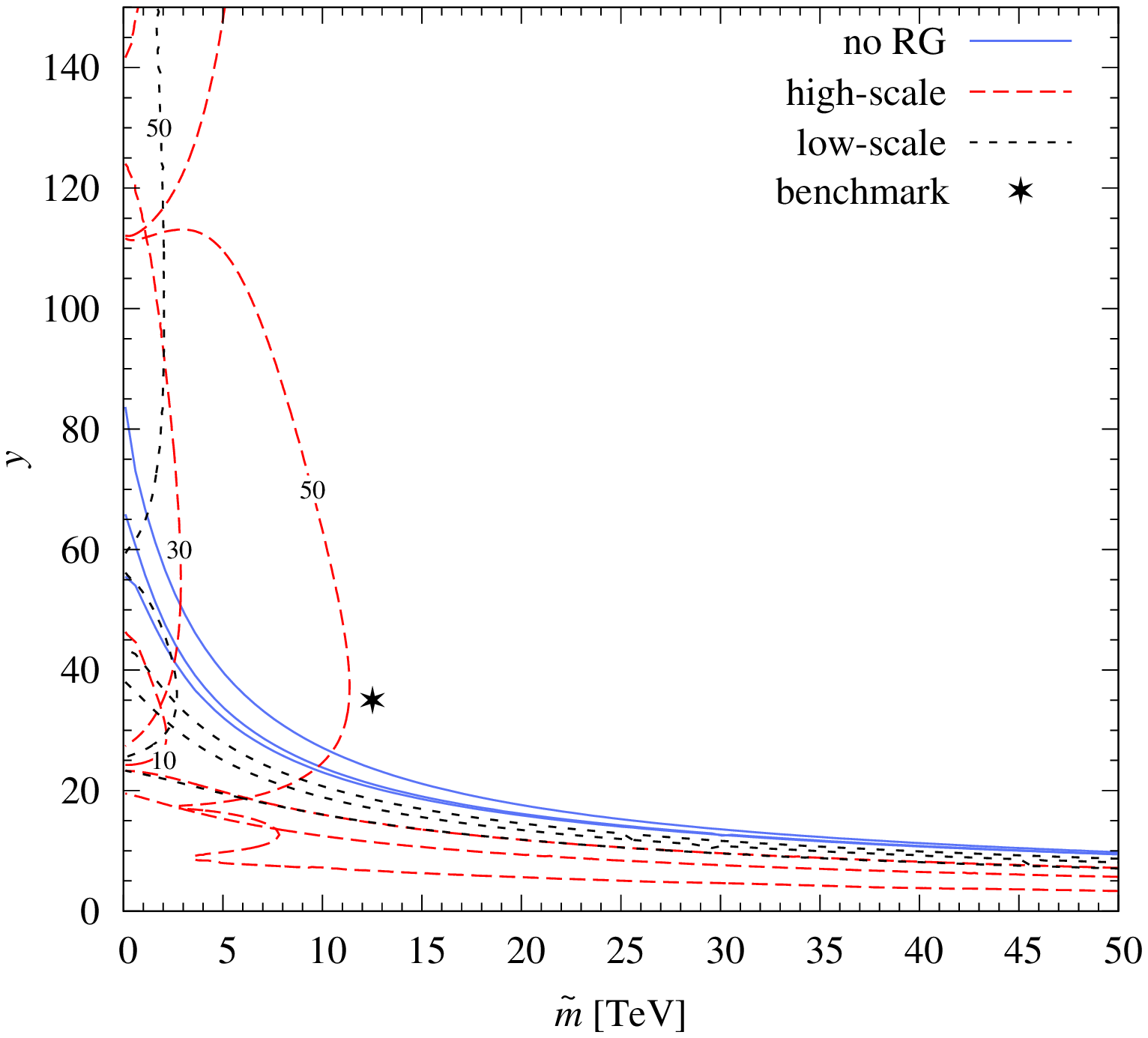}}}
\caption{Constraints on $|\epsilon_K|$ from gluino box diagrams as
  well as double Higgs penguin diagrams in the
  ($\tilde{m},y$)-plane.
  The three lines (from above) represent $\tan\beta=50,30,10$.
  The gluino mass is set to 1150 GeV, $\mu=\tilde{m}/2$ and
  $m_H=2\tilde{m}/3$. The Higgs contributions kick in at
  $\tan\beta\gtrsim 30$.}
\label{fig:epsilonk}
\end{center}
\end{figure}

Using the same diagrams, i.e.~gluino boxes and Higgs penguins, we
calculate the constraints on the CP-violating parameter $\epsilon_K$
which are shown in fig.~\ref{fig:epsilonk}. For the plots, $\Delta
m_K$ is set to the experimental bound, and we assume order one complex 
phases in the matrix elements, hence this is a conservative
estimate. Notice that in the model at hand, we have potentially two
complex phases that cannot be set to zero \cite{Auzzi:2011eu} and
hence can induce CP-violating effects as measured by $\epsilon_K$ (as
well as the EDMs to be discussed shortly).

In the case of $B_d-\bar{B}_d$, the dominant Wilson coefficients are
given by single flavor-flipped second order MIs, see
\cite{Altmannshofer:2009ne}, and the magic numbers and matrix
elements can be found in \cite{Becirevic:2001jj}.
Even though the boxes are only at second order in the MI expansion, the
$B_q-\bar{B}_q$ contributions are only comparable to the $K-\bar{K}$
ones (even though at fourth order in MI). By explicit calculations we
find that the $B_d-\bar{B}_d$ constraints are down with respect to the
$K-\bar{K}$ ones by about a factor of two (the $B_s-\bar{B}_s$ are
even more subdominant).
We have also checked the $D-\bar{D}$ mixing which like in the kaon
case needs the double flavor-flipped MI to kick in (the Wilson
coefficients are given by eq.~\eqref{eq:WilsonKKbar} with
$\delta_d\to\delta_u$). The matrix elements, bag parameters and magic
numbers are given in \cite{Bona:2007vi}. Due to the experimental
limits (see tab.~\ref{tab:explimits}), these constraints are less
severe than the kaon ones. We find that due to the RG effects on the
mass-splitting -- because of the large top-Yukawa -- the $D-\bar{D}$
meson mixing constraint is of the same order as the $B_d-\bar{B}_d$
constraint; both about a factor of two less important than the kaon
ones. The $B_d-\bar{B}_d$ constraints are stronger than the
$D-\bar{D}$ ones for the average squark mass less than about 6 TeV,
which however is not possible in the model and part of parameter space
that we are investigating here.

\begin{figure}[!tp]
\begin{center}
\includegraphics[width=0.49\linewidth]{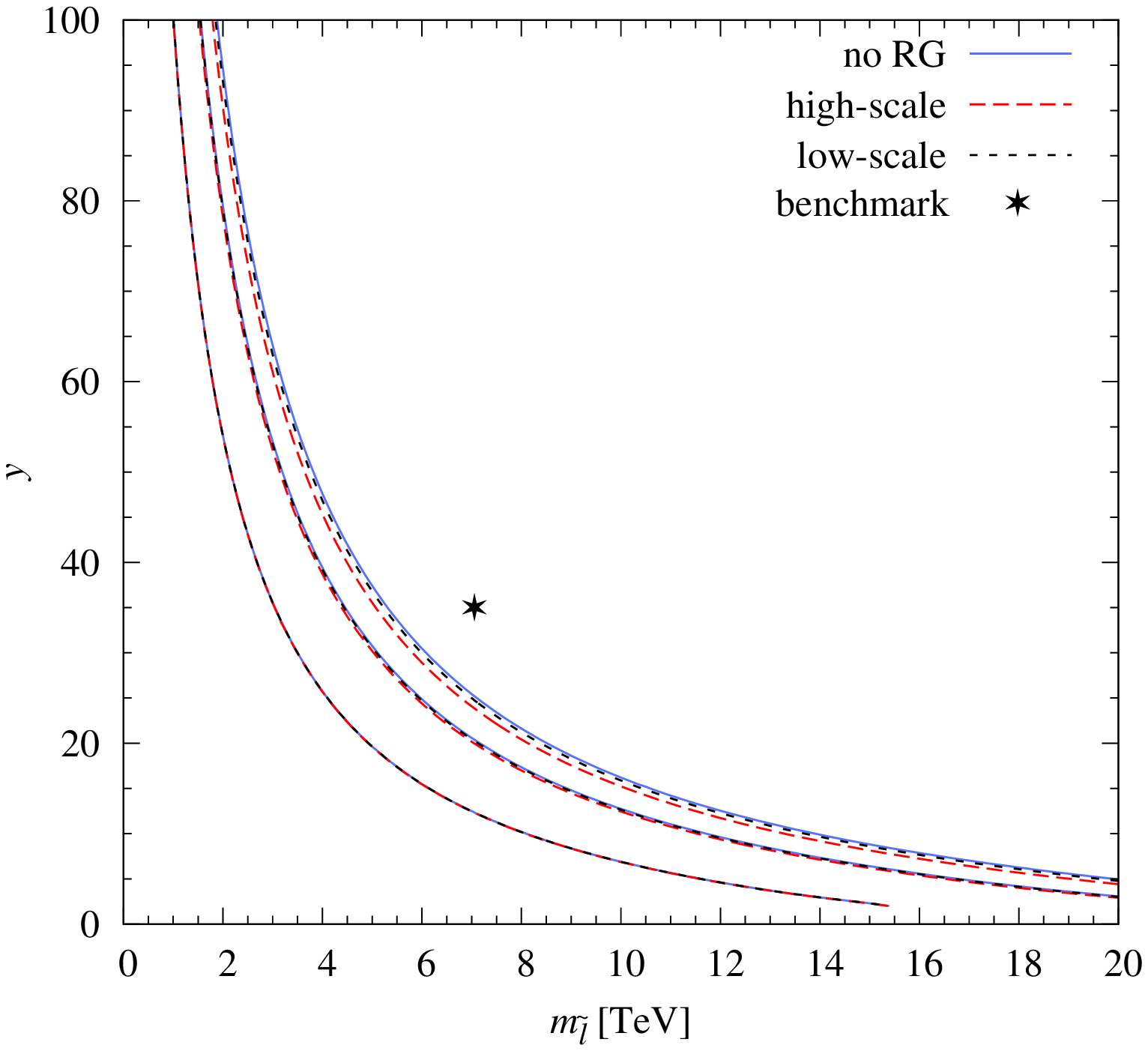}
\caption{Constraints on ${\rm Br}(\mu\to e\gamma)$ from gaugino and
  slepton mediated decay in the ($m_{\tilde{\ell}},y$)-plane, where
  $m_{\tilde{\ell}}$ is the average slepton mass. The
  three lines (from above) represent $\tan\beta=50,30,10$.
  We took $\mu=m_{\tilde{\ell}}$ and $M_1=192$ GeV. The constraints
  are equal for the normal and the inverted model. }
\label{fig:brmuegamma}
\end{center}
\end{figure}

The next flavor observables we check are $\Delta F=1$ processes.
We find that the most important one is due to the gaugino and slepton
mediated decay, $\mu\to e\gamma$. The important amplitude again has a
double flavor-flipped MI and reads
\cite{Paradisi:2005fk,Altmannshofer:2009ne}
\beq
A_{L,R}^{21} \simeq \frac{\alpha_1}{4\pi}
\frac{m_\tau}{m_\mu}
\frac{\mu M_1 \tan\beta}{m_{\tilde{\ell}}^4}
(\delta_e^{RR})_{23}(\delta_\ell^{LL})_{31} f_{4n}(x_1) \, ,
\eeq
where $m_\tau$, $m_\mu$, $\mu$, $M_1$ and $m_{\tilde{\ell}}$ are the
tau mass, the muon mass, the supersymmetric Higgs mass, the bino mass
and the average slepton mass, respectively. $x_1\equiv
M_1^2/m_{\tilde{\ell}}^2$. The loop function $f_{4n}(x_1)$ can be
found in app.~A of \cite{Altmannshofer:2009ne}.
This amplitude is especially important due to an enhancement
factor of $m_\tau/m_\mu$ with respect to that of
\cite{Gabbiani:1988rb,Hagelin:1992tc}. \footnote{Note that the Monte
  Carlo analysis of app.~\ref{app:montecarlo} finds relatively large
  maximum values for $(\delta_\ell^{RR})_{12}$ which could potentially
  be important. We checked however the single MI contribution and it
  is much less important than the double MI contribution discussed
  here. }
The branching ratio can then be expressed as
\beq
\frac{{\rm BR}(\mu\to e\gamma)}{{\rm BR}(\mu\to e\nu_\mu\bar{\nu}_e)}
= \frac{48\pi^3\alpha}{G_F^2}
\left(|A_L^{21}|^2+|A_R^{21}|^2\right) \, .
\eeq
The constraints are shown in fig.~\ref{fig:brmuegamma}.
Let us mention that the MEG corporation plans a possible upgrade which
could increase the experimental limit by one order of magnitude
\cite{Baldini:2013ke}, potentially probing very far with this flavor
observable; see also \cite{Moroi:2013sfa,Moroi:2013vya}.

We checked also other $\Delta F=1$ processes such as the gluino
mediated contribution to $b\to s\gamma$ (the charged Higgses are too
heavy in our model to be of any importance \cite{Hermann:2012fc}) as
well as the Higgs mediated penguin contributions to
$B_s\to\mu^+\mu^-$ \cite{Altmannshofer:2009ne}. These constraints were
subdominant with respect to the above discussed processes.

\begin{figure}[!tp]
\begin{center}
\mbox{\subfigure[Two-node
    quiver]{\includegraphics[width=0.49\linewidth]{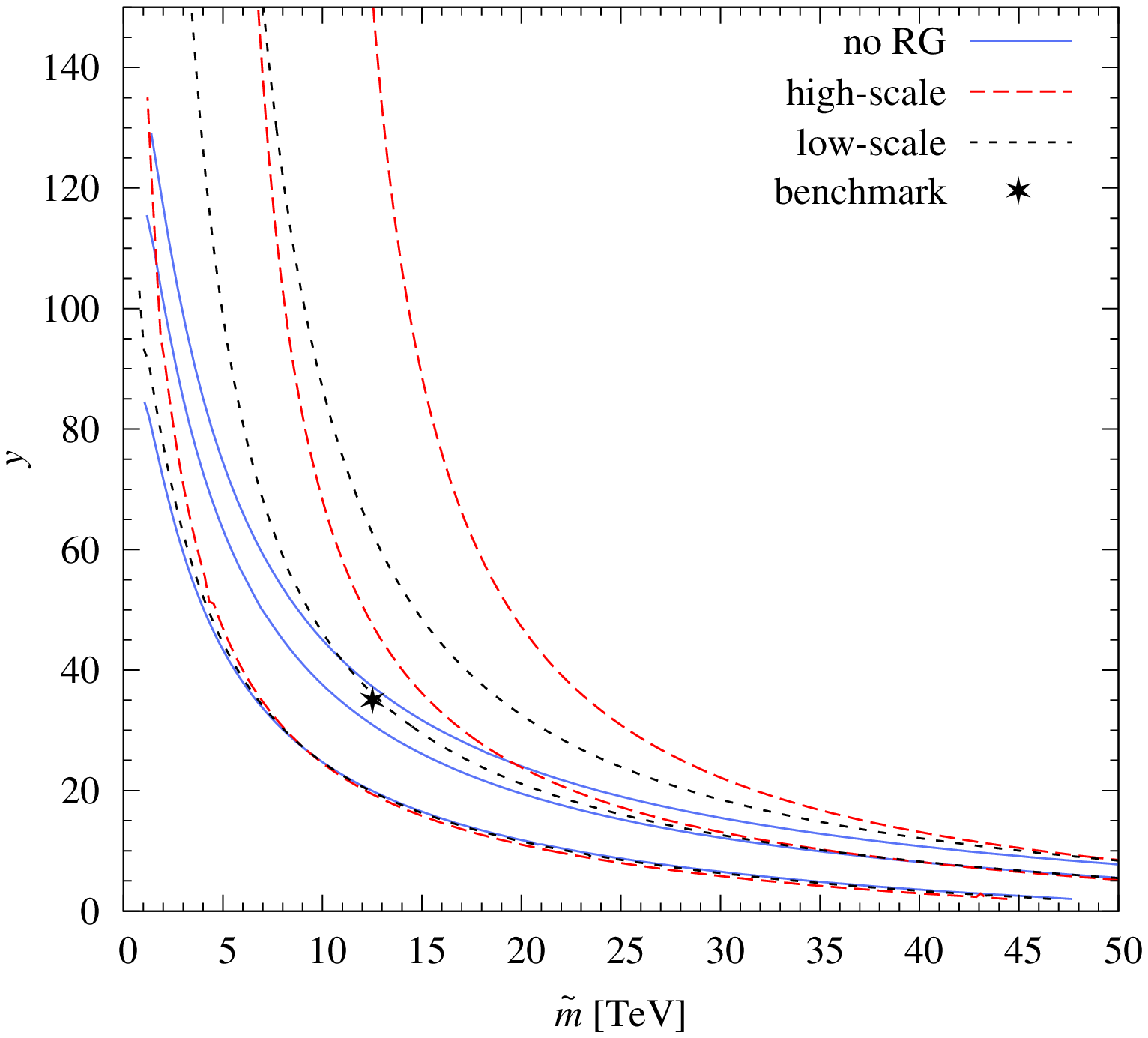}}
\subfigure[Inverted two-node
  quiver]{\includegraphics[width=0.49\linewidth]{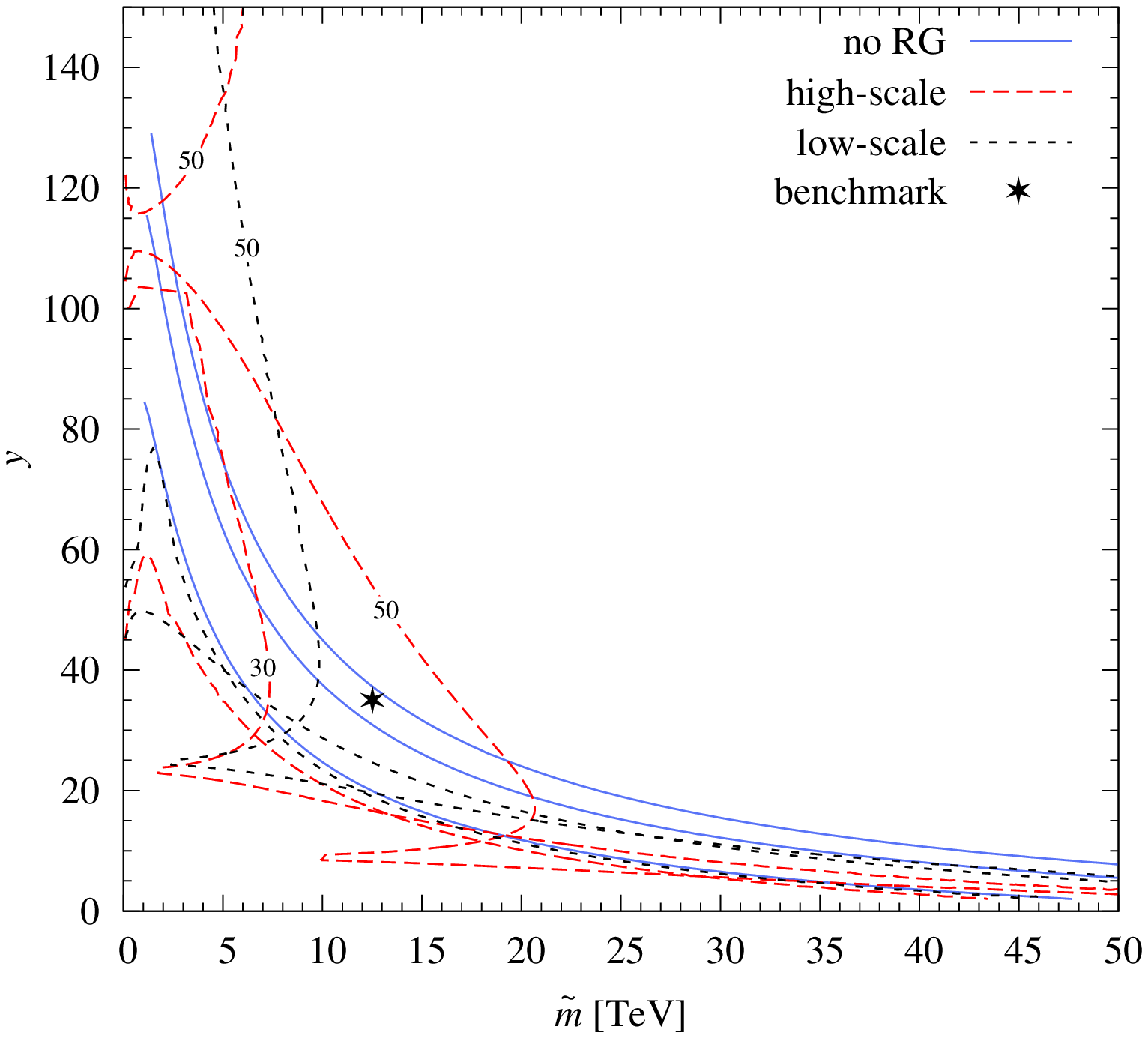}}}
\caption{Constraints on $|d_n|$ from gluino/squark diagrams in the
  ($\tilde{m},y$)-plane. The three lines (from above) represent
  $\tan\beta=50,30,10$.
  We took $\mu=\tilde{m}/2$ and $m_{\tilde{g}}=1150$ GeV.}
\label{fig:edmneutron}
\end{center}
\end{figure}

The next checks we perform are on the electric dipole moments
(EDMs). Although they typically are classified as $\Delta F=0$
observables, the dominant ones that we check here are of the so-called
``flavored'' type, i.e.~they consist of two $\Delta F=1$
transitions. Since we do not have sizable $A$-terms, the dominant
contribution is to the down-quark chromo-EDM and reads
\cite{Hisano:2006mj,Altmannshofer:2009ne}
\beq
\left\{d_d/e, d_d^c\right\} \simeq
-\frac{\alpha_3}{4\pi}\frac{m_b}{\tilde{m}^2}
\frac{m_{\tilde{g}}\mu}{\tilde{m}^2}
\frac{\tan\beta}{1+\epsilon_{\tilde{g}}\tan\beta}
\Im\left[(\delta_d^{LL})_{13}(\delta_d^{RR})_{31}\right]
f_{\tilde{g}}^d(x_{\tilde{g}}) \, ,
\eeq
where the loop functions can be found in app.~A of
\cite{Altmannshofer:2009ne}.
Although the quark EDMs have not been measured, they are related to
that of the neutron by a QCD sum rule estimate \cite{Pospelov:2000bw}
\beq
d_n = (1\pm 0.5)\left[1.4(d_d-0.25d_u)+1.1e(d_d^c+0.5d_u^c)\right] \, ,
\eeq
which is limited by experiments on ultra-cold neutrons, see
tab.~\ref{tab:explimits}. As mentioned, we have two complex phases
that we cannot eliminate and that can be of order one. Here we assume
order one phases, although accidentally smaller phases would reduce
the present constraints.
The constraints we find are shown in fig.~\ref{fig:edmneutron}.

\begin{figure}[!tp]
\begin{center}
\includegraphics[width=0.49\linewidth]{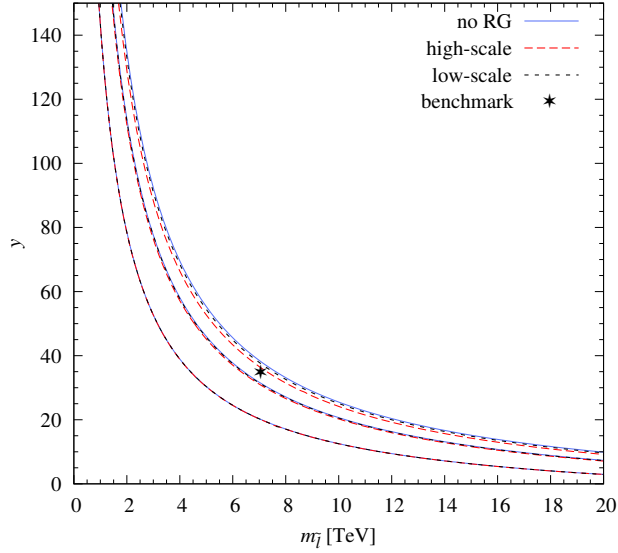}
\caption{Constraints on $|d_e|$ from gluino/squark diagrams in the
  ($m_{\tilde{\ell}},y$)-plane, with $m_{\tilde{\ell}}$ being the
  average slepton mass. The three lines (from above) represent
  $\tan\beta=50,30,10$.
  We took $\mu=m_{\tilde{\ell}}$ and $M_1=192$ GeV. The constraints
  are equal for the normal and the inverted model.}
\label{fig:edmelectron}
\end{center}
\end{figure}

We also check the EDM of the electron which has the leading
contribution at double flavor-flipped MI (second order)
\cite{Hisano:2008hn,Altmannshofer:2009ne}
\beq
\frac{d_e}{e} \simeq \frac{\alpha_1}{4\pi}\frac{M_1}{m_{\tilde{\ell}}^2}
\frac{m_\tau\tan\beta}{m_{\tilde{\ell}}^2}
\Im\left[\mu(\delta_\ell^{LL})_{13}(\delta_e^{RR})_{31}\right] f_{4n}(x_1) \, .
\eeq
The constraints are displayed in fig.~\ref{fig:edmelectron}.
Experimentally, the limit comes from measuring Ytterbium-Fluoride
(YbF) and is given in tab.~\ref{tab:explimits}. Future EDM
measurements could conceivably push up the limit by almost two orders
of magnitude \cite{Vutha:2009ux}, which could make this flavor
observable one of the tightest.

A comment in store is due to the conservative choice of setting
$(\eta_{u,d}^{MM})_{13,23}=1/4$ and $(\eta_{\ell,e}^{MM})_{13,23}=1/2$
(see eq.~\eqref{eq:etadef}) in the shown flavor constraints. This is
the worst-case scenario for the model at hand. This conservative
value is chosen as to display the farthest possible sensitivity for
the respective constraint. If the $\eta$s are changed by a factor of
order one, we can crudely ignore the loop factor and thus for fixed
$y$ on the graphs, the $x$-axis is simply rescaled by a power law, see
tab.~\ref{tab:mtildescaling}.
\begin{table}[!tbp]
\begin{center}
\begin{tabular}{l|l|l|l}
Constraint & Scaling formula & max$(\eta)$ &
$\sqrt{\langle\eta^2\rangle}$\\
\hline
$\Delta m_K$, $|\epsilon_K|$ ($C_4$) &
$\tilde{m}\to
16\sqrt{(\eta_d^{LL})_{23}(\eta_d^{LL})_{31}(\eta_d^{RR})_{23}(\eta_d^{RR})_{31}}\ \tilde{m}$
& $\tilde{m}\to 0.29\tilde{m}$ & $\tilde{m}\to 0.081\tilde{m}$\\
$\Delta m_D$ &
$\tilde{m}\to
16\sqrt{(\eta_u^{LL})_{23}(\eta_u^{LL})_{31}(\eta_u^{RR})_{23}(\eta_u^{RR})_{31}}\ \tilde{m}$
& $\tilde{m}\to 0.71\tilde{m}$ & $\tilde{m}\to 0.20\tilde{m}$\\
$|d_n|$ &
$\tilde{m}\to\left(16(\eta_d^{LL})_{13}(\eta_d^{RR})_{31}\right)^{1/3}\tilde{m}$
& $\tilde{m}\to 0.66\tilde{m}$ & $\tilde{m}\to 0.43\tilde{m}$\\
${\rm Br}(\mu\to e\gamma)$ &
$m_{\tilde{\ell}}\to\left(8(\eta_\ell^{LL})_{31}^2(\eta_e^{RR})_{23}^2+8(\eta_\ell^{RR})_{31}^2(\eta_e^{LL})_{23}^2\right)^{1/6}
m_{\tilde{\ell}}$ & $m_{\tilde{\ell}}\to 0.77m_{\tilde{\ell}}$ & $m_{\tilde{\ell}}\to 0.48m_{\tilde{\ell}}$\\
$|d_e|$ &
$m_{\tilde{\ell}}\to\left(4(\eta_\ell^{LL})_{13}(\eta_e^{RR})_{31}\right)^{1/3}
m_{\tilde{\ell}}$
& $m_{\tilde{\ell}}\to 0.77m_{\tilde{\ell}}$ & $m_{\tilde{\ell}}\to 0.48m_{\tilde{\ell}}$
\end{tabular}
\caption{Scaling of the average mass in the presented graphs for fixed
  $y$. $\eta$ in the 3rd and 4th column refers to the value of the
  elements of eq.~\eqref{eq:etadef} which are estimated by the Monte
  Carlo method in app.~\ref{app:montecarlo}. }
\label{tab:mtildescaling}
\end{center}
\end{table}

Finally, we also checked whether the model is able to produce a
measured discrepancy \cite{Roberts:2010cj} between the SM prediction
and the current measurements of the $(g-2)_\mu$ using the formulae of
\cite{Moroi:1995yh}. However, due to the heavy sleptons and large
value of $\mu$, the SUSY contribution is typically two orders of
magnitude too small to explain the $3\sigma$ anomaly
\cite{Endo:2013bba}.

\begin{figure}[!htp]
\begin{center}
\mbox{\subfigure[High-scale -- two-node
    quiver]{\includegraphics[width=0.49\linewidth]{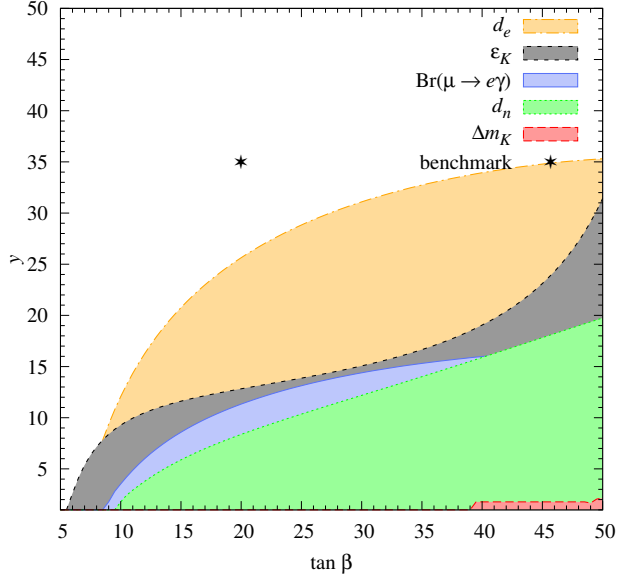}}
\subfigure[High-scale -- inverted two-node
  quiver]{\includegraphics[width=0.49\linewidth]{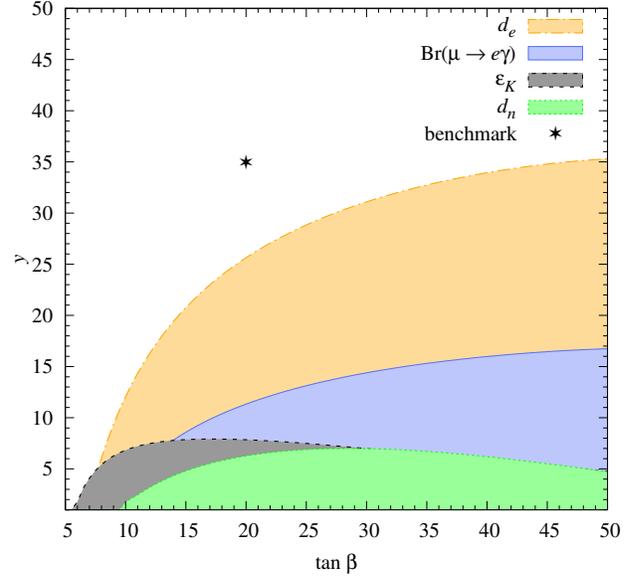}}}
\mbox{\subfigure[Low-scale -- two-node
    quiver]{\includegraphics[width=0.49\linewidth]{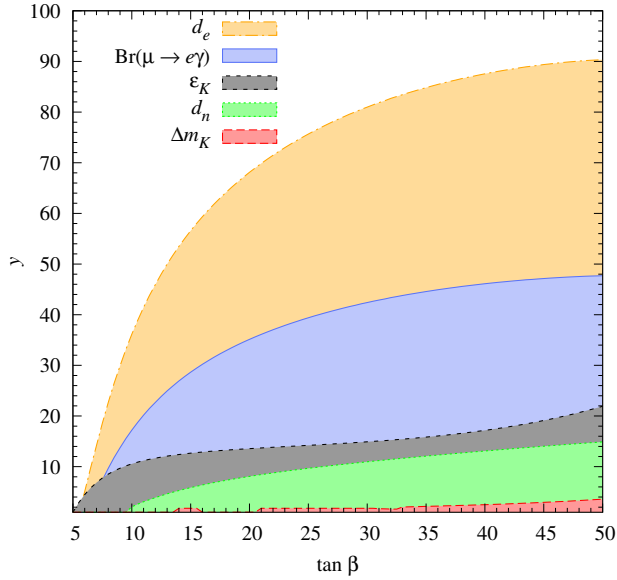}}
\subfigure[Low-scale -- inverted two-node
  quiver]{\includegraphics[width=0.49\linewidth]{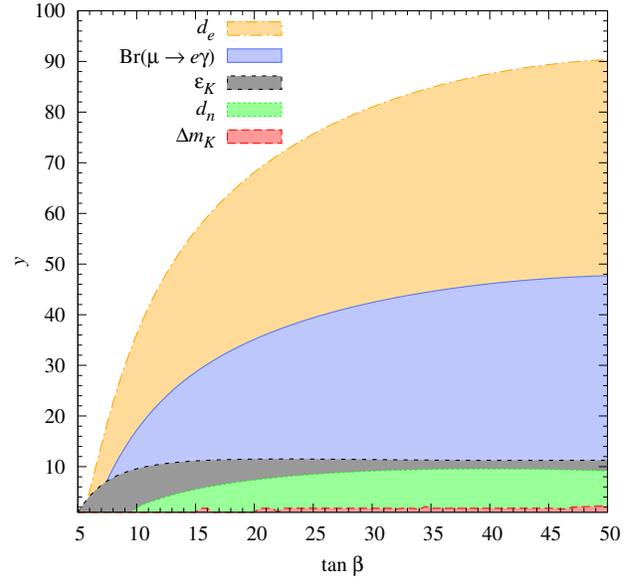}}}
\caption{Summary of all the presented flavor constraints in the
  ($\tan\beta,y$)-plane. The $\eta$s have
  been set to their maximal values according to the Monte Carlo
  analysis of app.~\ref{app:montecarlo}, see
  tab.~\ref{tab:etavalues}. 
  We have set $M_1=192$ GeV and $m_{\tilde{g}}=1150$ GeV. 
  Notice that the constraints shown are sensitive to the model
  point (and thus the messenger sector).
  Note also that
  the average squark mass, $\tilde{m}$, is calculated for each value of
  $\tan\beta$ using the face-values of the SM parameters, $m_t,m_h$
  and $\alpha_3$; this plot is therefore sensitive to variations of
  the latter parameters as well as of the $\eta$s. }
\label{fig:sflavor_summary}
\end{center}
\end{figure}

To recapitulate, we present a summary plot for both the high-scale
and the low-scale models in fig.~\ref{fig:sflavor_summary}, using the
maximal values of the $\eta$s from the Monte Carlo analysis of
app.~\ref{app:montecarlo}. 

\section{Dark matter\label{sec:dm}}

\subsection{High-scale mediation case}

In the high-scale mediation case there are two possibilities, either
the gravitino or the mostly-bino neutralino is the LSP and can
potentially make it as a dark matter candidate.
For sizable $\tan\beta\gtrsim 6$ the gravitino is the LSP with the
bino being the NLSP, while for $\tan\beta \lesssim 6$ the bino becomes
the LSP with the gravitino the NLSP.

The corner where the neutralino becomes the LSP -- i.e.~when
$\tan\beta\lesssim 6$ -- is not a favorable situation for dark matter
as the $\mu$-term rises sharply into the multi-TeV regime, see
fig.~\ref{fig:tanbeta}. Hence the neutralino which is almost purely
bino has a mass $\sim 160$ GeV $\ll \mu\sim 35-50$ TeV and therefore
the bino is nearly decoupled and annihilates too weakly in the early
Universe \cite{Giudice:2004tc}.

For most values of $\tan\beta$, the gravitino is the LSP and since we
are working in a gauge-mediated model, it is stable.
Due to the reheating temperature being higher than the average
squark-mass scale in the model, the dominant production of gravitinos
comes from high-energy SUSY scattering processes and the contribution
to the present energy density is \cite{Bolz:2000fu,ArkaniHamed:2004yi} 
\beq
\Omega_{3/2} h^2 \simeq
\left(\frac{{\rm GeV}}{m_{3/2}}\right)
\left(\frac{m_{\tilde{g}}}{1150 \; {\rm GeV}}\right)^2
\left(\frac{T_R}{10^{10} \; {\rm GeV}}\right)
\left(\frac{228.75}{g_*(T_R)}\right)^{3/2} 26.5 \, .
\eeq
Using the recent Planck result, $\Omega_{\rm DM} h^2=0.1186\pm 0.0031$
(including lensing) \cite{Ade:2013zuv} which is a larger amount of dark
matter compared to results for instance from WMAP, we obtain an upper
bound (from overclosure of the Universe) and a lower bound (from
contributing all the dark matter of the Plank result) on the reheating
temperature
\beq
1.2 \times 10^9 \; {\rm GeV}
\lesssim T_R
\lesssim 1.3 \times 10^9 \; {\rm GeV} \, ,
\eeq
for a 28 GeV gravitino in the $1\sigma$ window of the Planck
measurement or less if the gravitinos do not make up all the observed
dark matter.
This reheating temperature is just near the lower bound for a
successful thermal leptogenesis
\cite{Fukugita:1986hr,Davidson:2002qv,Rychkov:2007uq}.
For big bang nucleosynthesis (BBN) much lower reheating temperatures
are allowed.

A more severe constraint is due to the bino decaying into gravitinos
and photons, where the photons are potentially damaging, depending on
the life time of the bino, which reads \cite{ArkaniHamed:2004yi}
\beq
\tau_{\chi_1^0} \simeq
\left(\frac{m_{3/2}}{\rm GeV}\right)^2
\bigg(\frac{\rm GeV}{m_{\chi^0_1}}\bigg)^5
6 \times 10^{14} \; {\rm sec} \, .
\label{eq:tauNLSPtoLSP}
\eeq
For the life time of the bino in the range $10^2-10^7$ sec (hadron
decays) and $10^7-10^{10}$ sec (electromagnetic decays), there are
constraints due to the photo destruction of light nuclei that are
synthesized during BBN
\cite{Moroi:1993mb,Gherghetta:1998tq,Kawasaki:2008qe}.
For the life time in the range $10^{10}-10^{13}$ sec, there are
constraints from spectral distortions of the cosmic microwave
background radiation
\cite{Ellis:1984eq,Hu:1993gc,Fixsen:1996nj,Feng:2003uy}. Finally, for
the life time in the range $10^{13}-10^{18}$ sec, diffuse gamma-ray
observations put limits on the decays \cite{Yuksel:2007dr}. Life times
even longer yield practically stable NLSPs, which means that the only
constraint is due to the total energy density of the dark matter.

The life time given in eq.~\eqref{eq:tauNLSPtoLSP} for a 28 GeV
gravitino and 192 GeV bino yields $\tau_{\chi_1^0}\sim 1.8 \times
10^6$ sec, which is in conflict with the above mentioned bounds from
BBN \cite{Feng:2004mt,Kawasaki:2008qe}\footnote{Notice that
  assumptions about the bino yield as well as its scaling with the
  bino mass have been made in the calculation of
  \cite{Kawasaki:2008qe}. }. This kind of analysis typically makes
assumptions about the NLSP yield and that the released energy is of
the order of the NLSP mass. Taking a conservative attitude in view of
the BBN constraints \cite{Feng:2004mt,Kawasaki:2008qe}, the life time
should be smaller than roughly $120$ sec in order to avoid the
overproduction of deuterium or ${}^4$He.
This means that for having a stable gravitino, it should either be
lighter than $\sim 230$ MeV (for a 192 GeV bino) or the bino should be
heavier than $\sim 1310$ GeV (for a 28 GeV gravitino). This would
lower the reheating temperature by two orders of magnitude and hence
rule out the leptogenesis scenario\footnote{Note however that some
  regions of parameter space of e.g.~the CMSSM with a gravitino and a
  bino with the life time in the window of $10^4-10^6$ sec, have been
  found to be consistent with BBN constraints in
  \cite{Ellis:2003dn}. }.

There are several ways to modify the properties of the benchmark
points chosen here,
to allow viable gravitino DM within our class of models, e.g.:
\begin{itemize}
\item The simplest way, suggested above, is to lower the messenger
  scale $M$ and hence the gravitino mass $m_{3/2}$ so that the
  conservative bounds from BBN are satisfied. The rest of the spectrum
  remains almost unaltered.
\item One could contemplate the possibility that $k\equiv F/F_0 =
  \mathcal{O}(10^{-1})$, which would make the gravitino of the benchmark
  point of fig.~\ref{fig:hs} as heavy as the bino. In the case that it
  is nearly degenerate (the possibility of the nearly degenerate NLSP
  gravitino and LSP neutralino is also viable), the BBN constraints
  are drastically weakened \cite{Boubekeur:2010nt}.
\end{itemize}

\subsection{Low-scale mediation case\label{sec:dm_ls}}
In the low-scale mediation case, the gravitino is always the LSP and
extremely light -- of order of keV's -- and hence it would make up
warm dark matter whose mass is constrained by the Lyman-alpha forest
\cite{Boyarsky:2008xj}, gamma-ray bursts \cite{deSouza:2013wsa} and
galaxy formation \cite{Maccio:2012qf,Kang:2012up}. Since in this case
the average mass of the superpartners is much larger than that of the
gravitino and if so is the reheating temperature, then the gravitinos
go in equilibrium giving the density \cite{ArkaniHamed:2004yi}
\beq
\Omega_{3/2} h^2 \simeq \left(\frac{m_{3/2}}{\rm keV}\right)
\left(\frac{228.75}{g_*}\right) 0.5 \, .
\eeq
If the gravitino mass is $0.237$ keV, this yields a warm DM candidate
with the observed relic abundance. The question of whether warm DM is
a viable candidate is currently being reconsidered, see
e.g.~\cite{Angulo:2013sza,deVega:2013ysa,Menci:2013ght}.

Alternatively, in this scenario, one could contemplate the possibility
of pseudomoduli DM particles from the secluded sector
\cite{Shih:2009he,KerenZur:2009cv} or the lightest messenger field if
a conserved quantum number in the messenger sector is assumed
\cite{Dimopoulos:1996gy}.
We will not elaborate on these possibilities here.

\section{Discussion\label{sec:discussion}}

In this paper we have studied a mild-split SUSY part of a parameter
space present in a class of quiver-like models. The simplest type is a
two-nodes quiver model which is able to produce the SM flavor texture,
namely, the quark and lepton masses as well as the CKM matrix,
and the newly found Higgs mass.
In the model as studied in this paper, the Higgs mass is simply fed by
top loops up till about 12 TeV, where the stops eventually kick in.

We made a simple one-loop calculation to estimate to which degree the
model unifies. Its simplest version matches the measured value of the
strong coupling at the $2-3\sigma$ level. There are, however, further 
effects to consider such as two-loop effects, threshold effects and more
importantly, matter from the link sector has not been taken into
account. Since the amount of matter in the link field sector is very
large, a tiny splitting can modify the unification substantially.

The supersymmetric flavor problem -- although simplistically dealt with
using near-universality -- turns out to be quite near to present
flavor constraints, if order one complex phases are taken into
account. In this model, two such complex phases cannot be set to zero
and can be expected to be of order one. Among the important flavor
constraints are $\epsilon_K$, the branching ratio of $\mu\to e\gamma$,
the neutron EDM as well as the electron EDM, at least two of whose
limits are expected to be upgraded in the near future.

We have furthermore
contemplated the possibility of the gravitino making up the observed
dark matter component of the energy balance of the presently observed
Universe. There are two possibilities in the model as it stands. In
the high-scale mediation case, where standard unification is possible,
the gravitino is a cold dark matter
component while in the low-scale mediation case,
where dynamical embedding is possible,
it could plausibly be
a warm dark matter component. The cold dark matter component is
unlikely to be detected at direct detection experiments and although
it does not pose any problems itself, the decays of the NLSP, namely
the bino, could make considerable damage to the concordance of
cosmology by decaying together with photons that would destroy
light nuclei.

We will conclude by tying up loose ends and discussing various
differences between the model we have studied in the present paper and
other types of split-SUSY models in the literature.
The type of split SUSY theories in
\cite{ArkaniHamed:2004fb,Giudice:2004tc,ArkaniHamed:2004yi,ArkaniHamed:2012gw}
has a different set of ground rules, i.e.~the mass scale of the
fermions is set by the dark matter and the requirement of unification
keeps also the higgsini light by means of e.g.~some
symmetry. Therefore, in those theories it is possible to have squarks
heavy enough for having displaced vertices by means of the longevity
of the gluino \cite{ArkaniHamed:2004fb,Gambino:2005eh}. In our model,
on the other hand, the higgsini and scalar masses are tied together by
the nature of gauge mediation, even though we are able to have
electroweak scale gaugini. A further restriction is coming from
keeping the gluino mass near the present experimental bound while
producing electroweak symmetry breaking. 
Although theoretically conceivable, we have not been able to find
numerical spectra for values $\tan\beta$ less than roughly 5 in our
model, due to numerical problems with the precision (the convergence)
of the code. In this work, we have focused on the mild-split part of
the parameter space, however, for $\tan\beta$ of order unity, there is
possibly a corner with ``more split'' supersymmetry than what we have
studied here. We leave such an option as a future study. 
In our model for $\tan\beta \geq 5$, we have left-handed squarks
weighing less than around 140 TeV, which in turn makes the gluino
lifetime smaller than about $6\times 10^{-18}$ sec for a 1.15 TeV
gluino. This is $\sim 2$ nm and thus not long enough for detecting
displaced vertices.

As discussed in more detail in \cite{Auzzi:2011eu}, the two-nodes
quiver model does in fact only naturally produce the hierarchy between
the first two and the third generation fermions. In order to have a
naturally generated hierarchy also between the first two generations
of fermions, namely, between the up and charm (or down and strange)
quarks, a three-nodes quiver can be considered
\cite{Auzzi:2011eu}. This requires a sufficiently large
VEV of the link field connecting the two nodes hosting the first
generations, and requiring this VEV to be below or near the GUT scale
can thus further lower the messenger scale $M$.
This in turn yields a lighter gravitino than in the
example we have considered here. The full investigation of such a
scenario is out of scope of the present paper, but all the necessary
formulae for the sfermion masses are given in
app.~\ref{app:sfermionmasses} upon insertion of the appropriate form
factors of \cite{Auzzi:2011wt}. 
Finally, future investigations could pursue the integration of
neutrino masses into the model and estimate what impact it would have
on the model with respect to the sflavor constraints.

\subsubsection*{Acknowledgments}

We thank Roberto Auzzi for discussions. 
This work is supported in part by the I-CORE Program of the Planning
and Budgeting Committee and the Israel Science Foundation (Center
No. 1937/12), by the BSF -- American-Israel Bi-National Science
Foundation, and by a center of excellence supported by the Israel
Science Foundation (grant number 1665/10).

\appendix

\section{Sfermion masses\label{app:sfermionmasses}}

In this section, we provide two-loop formulae for the soft masses for
a field in an arbitrary representation of any gauge group of a generic
quiver-like model with a general messenger sector.
The mass formula is used to calculate the mass-splitting in the
two-nodes model at the messenger scale, which is the essential
ingredient in the flavor analysis presented in
sec.~\ref{sec:sflavor}.

\subsection{Generic messenger sector in a generic quiver}
Let us consider a generic messenger sector following
\cite{Dumitrescu:2010ha} given by
\beq
\int d^4 \theta \left[
T_i^\dag(\delta_{ij}+V\widetilde{\lambda}_{ij})T_j
+\widetilde{T}_i^\dag(\delta_{ij}+V\widetilde{\lambda}_{ij})\widetilde{T}_j
\right]
+ \int d^2 \theta \; \widetilde{T}_i \widetilde{\mathcal{M}}_{ij} T_j
+ {\rm c.c.} \, ,
\eeq
where $i,j=1,\ldots,2p$ with $p$ a non-negative integer specifying the
number of copies of two-by-two messenger matrices.
Now, passing to a basis where the fermionic messengers have a diagonal
and real mass matrix $m_i$, the complex scalars have, by means of a
unitary transformation, the mass-squared matrix
\beq
\widetilde{\mathcal{M}}_\pm = \mathbf{1}_p \otimes
\left(m \pm F\lambda - D \tilde{\lambda}\right) \, ,
\eeq
whose unitary diagonalization matrices are needed in the calculation
of the soft masses and are denoted by $U_\pm^\dag
\widetilde{\mathcal{M}}_\pm U_\pm = \diag(m_{\pm 1}^2,\cdots,m_{\pm
  2p}^2)$.
The gaugino masses are unaltered compared to \cite{Dumitrescu:2010ha}
and the sfermion masses are given by
\begin{align}
m_{\tilde{f}}^2 &= 2 \sum_{k=1}^3 \left(\frac{\alpha_k}{4\pi}\right)^2
C_{\tilde{f},k} n_k \; \mathcal{E}[f_k(p^2)] \, ,
\end{align}
where the mass function is defined as
\begin{align}
\mathcal{E}[f(p^2)] \equiv \int \frac{d^4p d^4q}{\pi^4} \;
f(p^2) &\Bigg[
-\frac{1}{4}\sum_{\pm,i,j} \big(U_\pm^\dag U_\mp\big)_{ij}
\big(U_\mp^\dag U_\pm)_{ji}
\frac{1}{p^2[q^2+m_{\mp j}^2][(p+q)^2+m_{\pm i}^2]}
\\ & \phantom{=\ }
+\sum_{\pm,i,j} \big(U_\pm^\dag\big)_{ij} \big(U_\pm\big)_{ji}
\frac{p^2+m_{\pm i}^2-m_j^2}{p^4[q^2+m_j^2][(p+q)^2+m_{\pm i}^2]}
\non & \phantom{=\ }
-\sum_{\pm,i}\frac{\tfrac{1}{4}p^2+m_{\pm i}^2}{p^4[q^2+m_{\pm
      i}^2][(p+q)^2+m_{\pm i}^2]}
-\sum_{i}\frac{p^2-2m_i^2}{p^4[q^2+m_i^2][(p+q)^2+m_i^2]} \Bigg] \, .
\nonumber
\end{align}
Using the method of \cite{Auzzi:2011wt}, we can calculate the
integrals for an arbitrary quiver theory in terms of the following
coefficients
\begin{align}
\frac{f(p^2)}{p^2} &= \frac{a_0}{p^2}
+\sum_\ell \frac{a_{1,\ell}}{p^2+m_\ell^2}
+\sum_\ell \frac{a_{2,\ell}}{(p^2+m_\ell^2)^2} \, , \\
\frac{f(p^2)}{p^4} &= \frac{b_{-1}}{p^2} + \frac{b_0}{p^4}
+ \sum_\ell \frac{b_{1,\ell}}{p^2+m_\ell^2}
+\sum_\ell \frac{b_{2,\ell}}{(p^2+m_\ell^2)^2} \, , \nonumber
\end{align}
given by the above partial fractions.
The result is
\begin{align}
\label{eq:general_massfunction}
&\mathcal{E}[f(p^2)] \non
&=
\sum_{\pm,i,j} \big(U_\pm^\dag U_\mp\big)_{ij}
\big(U_\mp^\dag U_\pm)_{ji}
\bigg(a_0 \alpha_0^a(m_{\mp j},m_{\pm i})
+\sum_\ell a_{1,\ell} \alpha_1^a(m_\ell,m_{\mp j},m_{\pm i})
+\sum_\ell a_{2,\ell} \alpha_2^a(m_\ell,m_{\mp j},m_{\pm i}) \bigg)
\non &\phantom{=\ }
+\sum_{\pm,i,j} \big(U_\pm^\dag\big)_{ij} \big(U_\pm\big)_{ji}
\bigg(a_0 \alpha_0^b(m_j,m_{\pm i})
+\sum_\ell a_{1,\ell} \alpha_1^b(m_\ell,m_j,m_{\pm i})
+\sum_\ell a_{2,\ell} \alpha_2^b(m_\ell,m_j,m_{\pm i})
\non &\phantom{=\ } \qquad
+b_{-1} \beta_{-1}^b(m_j,m_{\pm i})
+b_0 \beta_0^b(m_j,m_{\pm i})
+\sum_\ell b_{1,\ell} \beta_1^b(m_\ell,m_j,m_{\pm i})
+\sum_\ell b_{2,\ell} \beta_2^b(m_\ell,m_j,m_{\pm i}) \bigg)
\non &\phantom{=\ }
+\sum_{\pm,i} \bigg(
\sum_\ell a_{1,\ell} \alpha_1^c(m_\ell,m_i,m_{\pm i})
+\sum_\ell a_{2,\ell} \alpha_2^c(m_\ell,m_i,m_{\pm i})
\\ &\phantom{=\ } \qquad
+b_{-1} \beta_{-1}^c(m_i,m_{\pm i})
+b_0 \beta_0^c(m_i,m_{\pm i})
+\sum_\ell b_{1,\ell} \beta_1^c(m_\ell,m_i,m_{\pm i})
+\sum_\ell b_{2,\ell} \beta_2^c(m_\ell,m_i,m_{\pm i}) \bigg) \, ,
\nonumber
\end{align}
where the functions are given by
\begin{align}
\alpha_0^a(m_{\mp j},m_{\pm i}) &=
\frac{1}{2}m_{\pm i}^2 \Li_2\left(1-\frac{m_{\mp j}^2}{m_{\pm
    i}^2}\right) , \label{eq:alphafunctions} \\
\alpha_1^a(m_\ell,m_{\mp j},m_{\pm i}) &=
\frac{1}{4} m_\ell^2
  \h{\frac{m_{\pm i}^2}{m_\ell^2}}{\frac{m_{\mp j}^2}{m_\ell^2}}
+\frac{1}{2} m_{\pm i}^2
  \h{\frac{m_\ell^2}{m_{\pm i}^2}}{\frac{m_{\mp j}^2}{m_{\pm i}^2}} ,
  \non
\alpha_2^a(m_\ell,m_{\mp j},m_{\pm i}) &=
-\frac{1}{4}
  \h{\frac{m_{\pm i}^2}{m_\ell^2}}{\frac{m_{\mp j}^2}{m_\ell^2}} ,
  \non
\alpha_0^b(m_j,m_{\pm i}) &=
-m_{\pm i}^2 \Li_2\left(1-\frac{m_j^2}{m_{\pm i}^2}\right)
-m_j^2 \Li_2\left(1-\frac{m_{\pm i}^2}{m_j^2}\right) , \non
\alpha_1^b(m_\ell,m_j,m_{\pm i}) &=
-m_\ell^2 \h{\frac{m_{\pm i}^2}{m_\ell^2}}{\frac{m_j^2}{m_\ell^2}}
-m_{\pm i}^2
  \h{\frac{m_\ell^2}{m_{\pm i}^2}}{\frac{m_j^2}{m_{\pm i}^2}}
-m_j^2 \h{\frac{m_{\pm i}^2}{m_j^2}}{\frac{m_\ell^2}{m_j^2}} , \non
\alpha_2^b(m_\ell,m_j,m_{\pm i}) &=
\h{\frac{m_{\pm i}^2}{m_\ell^2}}{\frac{m_j^2}{m_\ell^2}} , \non
\alpha_1^c(m_\ell,m_i,m_{\pm i}) &=
\frac{1}{2} m_{\pm i}^2 \h{\frac{m_\ell^2}{m_{\pm i}^2}}{1}
+m_i^2 \h{\frac{m_\ell^2}{m_i^2}}{1}
+\frac{1}{4}m_\ell^2
  \h{\frac{m_{\pm i}^2}{m_\ell^2}}{\frac{m_{\pm i}^2}{m_\ell^2}}
+\frac{1}{2}m_\ell^2
  \h{\frac{m_i^2}{m_\ell^2}}{\frac{m_i^2}{m_\ell^2}} , \non
\alpha_2^c(m_\ell,m_i,m_{\pm i}) &=
-\frac{1}{4}
  \h{\frac{m_{\pm i}^2}{m_\ell^2}}{\frac{m_{\pm i}^2}{m_\ell^2}}
-\frac{1}{2}
  \h{\frac{m_i^2}{m_\ell^2}}{\frac{m_i^2}{m_\ell^2}} , \nonumber
\end{align}
\begin{align}
\beta_{-1}^b(m_j,m_{\pm i}) &=
-m_j^2\big(m_{\pm i}^2-m_j^2\big)
  \Li_2\left(1-\frac{m_{\pm i}^2}{m_j^2}\right)
-m_{\pm i}^2\big(m_{\pm i}^2-m_j^2\big)
  \Li_2\left(1-\frac{m_j^2}{m_{\pm i}^2}\right)
  , \label{eq:betafunctions} \\
\beta_0^b(m_j,m_{\pm i}) &=
m_j^2 \Li_2\left(1-\frac{m_{\pm i}^2}{m_j^2}\right)
-m_{\pm i}^2 \Li_2\left(1-\frac{m_j^2}{m_{\pm i}^2}\right) , \non
\beta_1^b(m_\ell,m_j,m_{\pm i}) &=
-m_\ell^2\big(m_{\pm i}^2-m_j^2\big)
  \h{\frac{m_{\pm i}^2}{m_\ell^2}}{\frac{m_j^2}{m_\ell^2}}
-m_{\pm i}^2\big(m_{\pm i}^2-m_j^2\big)
  \h{\frac{m_j^2}{m_{\pm i}^2}}{\frac{m_\ell^2}{m_{\pm i}^2}} \non
 &\phantom{=\ }
-m_j^2\big(m_{\pm i}^2-m_j^2\big)
  \h{\frac{m_{\pm i}^2}{m_j^2}}{\frac{m_\ell^2}{m_j^2}} , \non
\beta_2^b(m_\ell,m_j,m_{\pm i}) &=
\big(m_{\pm i}^2-m_j^2\big)
  \h{\frac{m_{\pm i}^2}{m_\ell^2}}{\frac{m_j^2}{m_\ell^2}} , \non
\beta_{-1}^c(m_i,m_{\pm i}) &= m_{\pm i}^4 - m_i^4 , \non
\beta_0^c(m_i,m_{\pm i}) &=
m_{\pm i}^2 \log m_{\pm i}^2 - m_i^2 \log m_i^2 , \non
\beta_1^c(m_\ell,m_i,m_{\pm i}) &=
2m_{\pm i}^4 \h{\frac{m_\ell^2}{m_{\pm i}^2}}{1}
-2m_i^4 \h{\frac{m_\ell^2}{m_i^2}}{1}
+m_\ell^2m_{\pm i}^2
  \h{\frac{m_{\pm i}^2}{m_\ell^2}}{\frac{m_{\pm i}^2}{m_\ell^2}}
-m_\ell^2m_i^2 \h{\frac{m_i^2}{m_\ell^2}}{\frac{m_i^2}{m_\ell^2}} ,
\non
\beta_2^c(m_\ell,m_i,m_{\pm i}) &=
-m_{\pm i}^2
  \h{\frac{m_{\pm i}^2}{m_\ell^2}}{\frac{m_{\pm i}^2}{m_\ell^2}}
+m_i^2 \h{\frac{m_i^2}{m_\ell^2}}{\frac{m_i^2}{m_\ell^2}} . \nonumber
\end{align}

\subsection{One-node quiver}

The quiver with just a single node reproduces the result of
\cite{Marques:2009yu,Dumitrescu:2010ha}. Since the form factor is
trivial, i.e.~$f(p^2)=1$, the non-zero coefficients are $a_0=b_0=1$,
see \cite{Auzzi:2011wt}. Hence, the mass function reads
\begin{align}
\mathcal{E}_{\rm GMGM} &=
\sum_{\pm,i,j} \big(U_\pm^\dag U_\mp\big)_{ij} \big(U_\mp^\dag U_\pm)_{ji}
\alpha_0^a(m_{\mp j},m_{\pm i})
+\sum_{\pm,i,j} \big(U_\pm^\dag\big)_{ij} \big(U_\pm\big)_{ji}
\left(\alpha_0^b(m_j,m_{\pm i})+\beta_0^b(m_j,m_{\pm i})\right)
\non &\phantom{=\ }
+\sum_{\pm,i} \beta_0^c(m_i,m_{\pm i}) \, .
\label{eq:GMGM_massfunction}
\end{align}

\subsection{Two-nodes quiver\label{app:twonodes}}

In this section we will calculate the example of the sfermion masses
in two-nodes quiver-like models. In this model, there is only a single mass
of the heavy vector bosons
\beq
\{m_\ell\} = m_v = \sqrt{2(g_A^2 +g_B^2)} v \, .
\eeq
For the node $A$, the form factor is given
by
\cite{McGarrie:2010qr,Auzzi:2010mb,Sudano:2010vt,Auzzi:2011wt}\footnote{For 
a $5d$ model from which this form factor can be deconstructed, see
also \cite{McGarrie:2010kh}. }
\beq
f_A(p^2) = \left(\frac{m_v^2}{p^2-m_v^2}\right)^2 \, ,
\eeq
which gives the coefficients
\beq
a_0 = 1 \, , \quad
a_1 = -1 \, , \quad
a_2 = -m_v^2 \, , \quad
b_{-1} = -\frac{2}{m_v^2} \, , \quad
b_0 = 1 \, , \quad
b_1 = \frac{2}{m_v^2} \, , \quad
b_2 = 1 \, ,
\eeq
while for the node $B$, the form factor is
\beq
f_B(p^2) = \left(\frac{\lambda p^2 - m_v^2}{p^2 - m_v^2}\right)^2 \ ,
\eeq
where $\lambda = (g_A^2 + g_B^2)/g_A^2$ and the coefficients read
\cite{Auzzi:2011wt}
\begin{align}
a_0 &= 1 \, , \quad
a_1 = -(1-\lambda^2) \, , \quad
a_2 = -(1-\lambda)^2 m_v^2 \, , \quad
\\
b_{-1} &= -\frac{2(1-\lambda)}{m_v^2} \, , \quad
b_0 = 1 \, , \quad
b_1 = \frac{2(1-\lambda)}{m_v^2} \, , \quad
b_2 = (1-\lambda)^2 \, .
\nonumber
\end{align}
The mass function is thus given by eq.~\eqref{eq:general_massfunction}
with the above coefficients.

\subsubsection{Large $m_v/M$ limit}

{}From physical considerations, it is clear that in the limit of the
Higgsing scale of the quiver, $m_v$, being much larger than the
messenger scale, $M$, both of the above mass functions will reduce to
that of the single node theory, eq.~\eqref{eq:GMGM_massfunction}
\beq
\lim_{\frac{m_v}{M}\to \infty} \mathcal{E}[f_A(p^2)] =
\lim_{\frac{m_v}{M}\to \infty} \mathcal{E}[f_B(p^2)] =
\mathcal{E}_{\rm GMGM} \, .
\eeq
However, in order to estimate how large $m_v/M$ should be in order
that the model is not at odds with meson mixings, the leading order in
$1/m_v^2$ correction is very useful and by expanding the functions
(\ref{eq:alphafunctions}-\ref{eq:betafunctions}), we obtain
\begin{align}
&\mathcal{E}[f_B(p^2)] = \mathcal{E}_{\rm
    GMGM} \label{eq:massfunction_largeyexpansion}\\
&\phantom{=\ }
+ \sum_{\pm,i,j} \big(U_\pm^\dag U_\mp\big)_{ij}
\big(U_\mp^\dag U_\pm)_{ji}
\frac{(1-\lambda^2)m_{\mp j}^2 m_{\pm i}^2}{2m_v^2}
\left[
\log\left(\frac{m_v^2}{m_{\pm i}^2}\right)
\log\left(\frac{m_v^2}{m_{\mp j}^2}\right)
- \log\left(\frac{m_v^2}{m_{\mp j}^2}\right) -1 + \frac{\pi^2}{3}
\right]
\non &\phantom{=\ }
+ \sum_{\pm,i,j} \big(U_\pm^\dag\big)_{ij} \big(U_\pm)_{ji}
\frac{(1-\lambda)m_j^2m_{\pm i}^2}{m_v^2}
\Bigg[
-(2+\lambda)\log^2\left(\frac{m_v^2}{m_{\pm i}^2}\right)
-\lambda\log^2\left(\frac{m_v^2}{m_j^2}\right)
+2(1+\lambda)\log\left(\frac{m_v^2}{m_j^2}\right)
\non &\phantom{=+ \sum_{\pm,i,j} \big(U_\pm^\dag\big)_{ij} \big(U_\pm)_{ji}
\frac{(1-\lambda)m_j^2m_{\pm i}^2}{m_v^2}
\Bigg[\ }
+(1+\lambda)\log^2\left(\frac{m_{\pm i}^2}{m_j^2}\right)
+2(1+\lambda)\left(1-\frac{\pi^2}{3}\right)
\Bigg]
\non &\phantom{=\ }
-\sum_{\pm,i,j} \big(U_\pm^\dag\big)_{ij} \big(U_\pm)_{ji}
\frac{2(1-\lambda)(m_j^2-m_{\pm i}^2)}{m_v^2}\left[
  m_{\pm i}^2\Li_2\left(1-\frac{m_j^2}{m_{\pm i}^2}\right)
  +m_j^2\Li_2\left(1-\frac{m_{\pm i}^2}{m_j^2}\right)\right]
\non &\phantom{=\ }
+\sum_{\pm, i}\bigg[
-\frac{(1-\lambda)^2 m_{\pm i}^4}{2m_v^2}
  \log^2\left(\frac{m_v^2}{m_{\pm i}^2}\right)
+\frac{(2+\lambda)(1-\lambda)m_i^4}{m_v^2}
  \log^2\left(\frac{m_v^2}{m_i^2}\right)
-\frac{3(1-\lambda^2)m_{\pm i}^4}{2m_v^2}
  \log\left(\frac{m_v^2}{m_{\pm i}^2}\right)
\non &\phantom{=+\sum_{\pm, i}\bigg[\ }
-\frac{m_{\pm i}^4}{2m_v^2}
  \left[3(1-\lambda^2)+\frac{\pi^2}{3}(1-\lambda)^2\right]
+\frac{(2+\lambda)(1-\lambda)m_i^4}{m_v^2}\frac{\pi^2}{3}\bigg]
+\mathcal{O}(m_v^{-4}) \, .
\nonumber
\end{align}
The result for the node $A$ is formally
\beq
\mathcal{E}[f_A(p^2)] = \left.\mathcal{E}[f_B(p^2)]\right|_{\lambda=0}
\, .
\eeq

\section{Monte Carlo analysis of diagonalization
  matrices\label{app:montecarlo}}

In this appendix we will use the flavor texture of
eq.~\eqref{eq:flavortexture} with random coefficients in the range
$[0.1,2]$ to generate realistic Yukawa matrices, i.e.~reproducing the
measured quark and lepton masses as well as the closest fit to the CKM
matrix. Since there is remaining freedom in the basis of the Yukawas,
we generate 100,000 random Yukawa matrices with coefficients in the
above mentioned range $[0.1,2]$ multiplying the texture
\eqref{eq:flavortexture} -- all giving physical Yukawas.
Then we use the actual diagonalization matrices to calculate
$(\eta_{u,d}^{MM})_{ij}=(V_M^{u,d})_{i3}(V_M^{u,d\dag})_{3j}$, see
eq.~\eqref{eq:etadef}. In this analysis we have chosen
$\tan\beta=20$.

\begin{figure}[!tp]
\begin{center}
\mbox{
\subfigure{\includegraphics[width=0.3\linewidth]{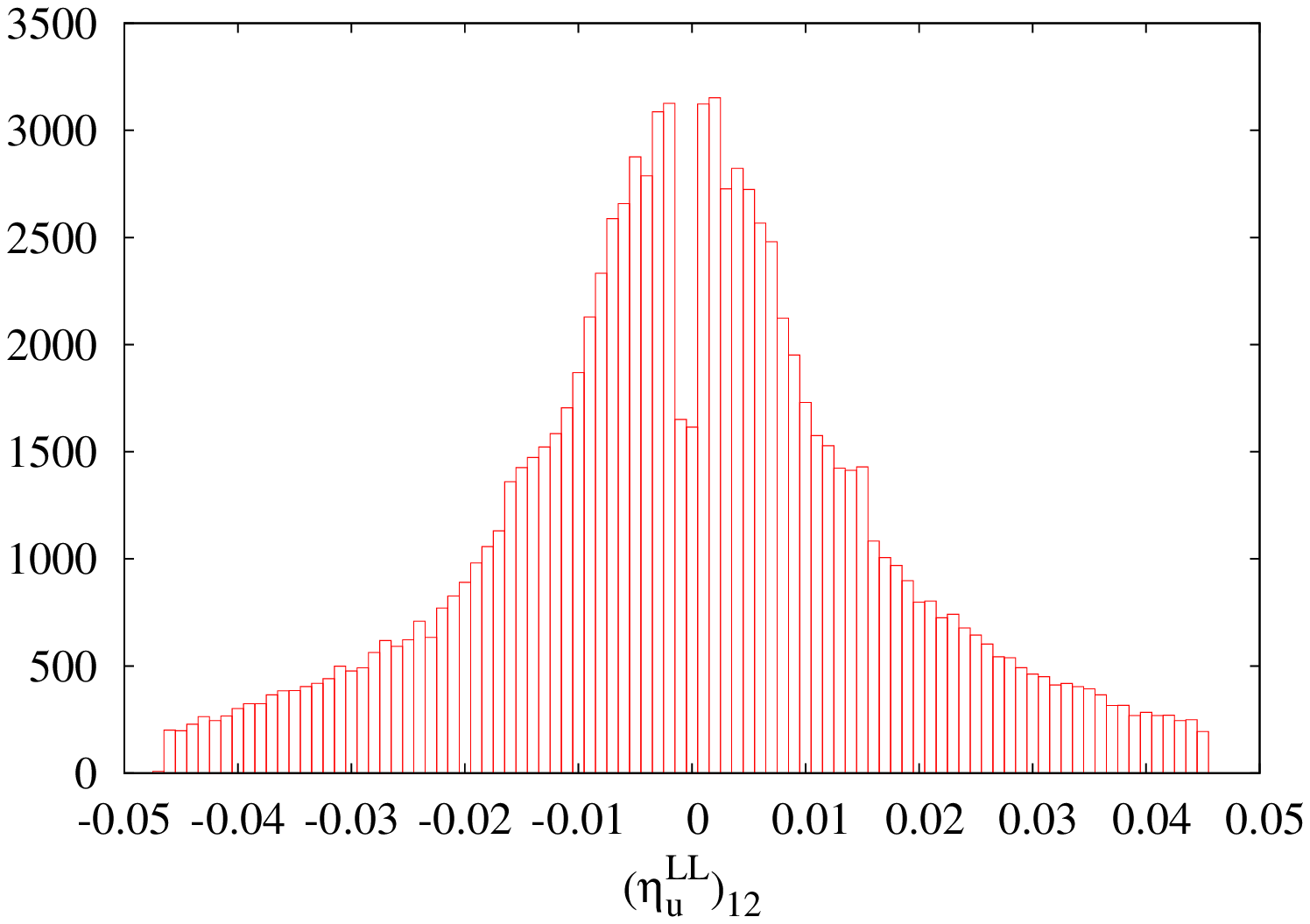}}\
\subfigure{\includegraphics[width=0.3\linewidth]{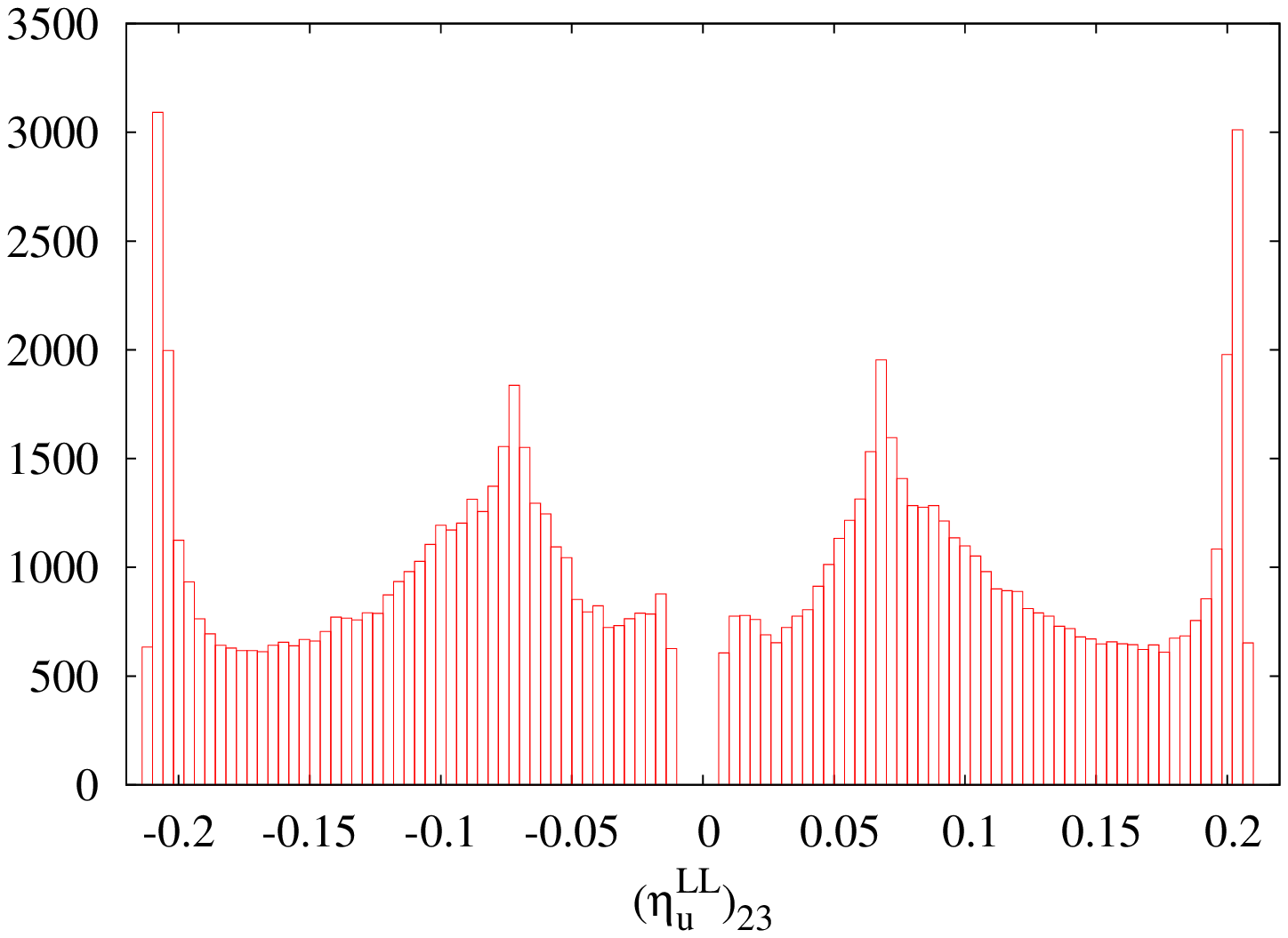}}\
\subfigure{\includegraphics[width=0.3\linewidth]{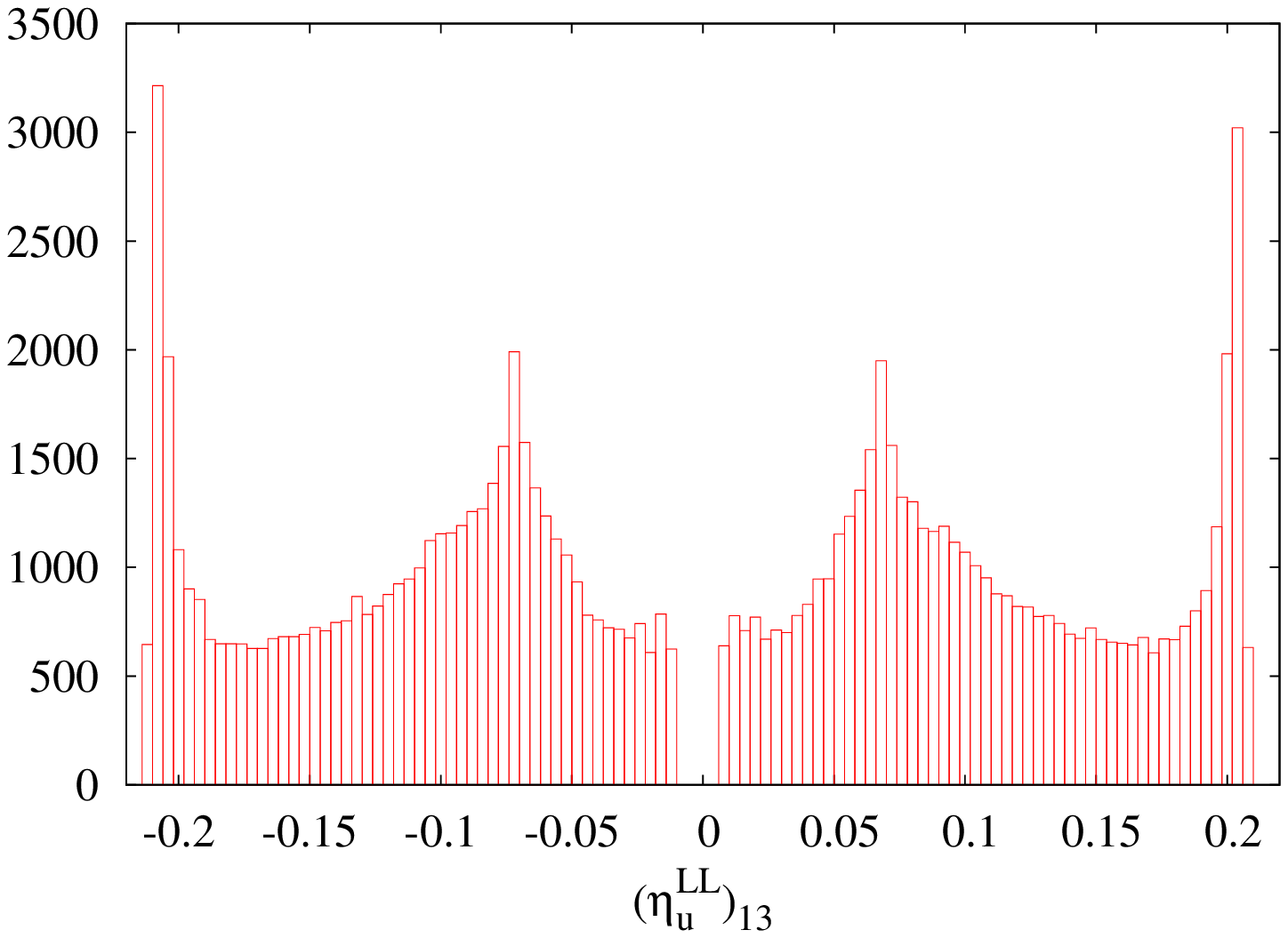}}}
\mbox{
\subfigure{\includegraphics[width=0.3\linewidth]{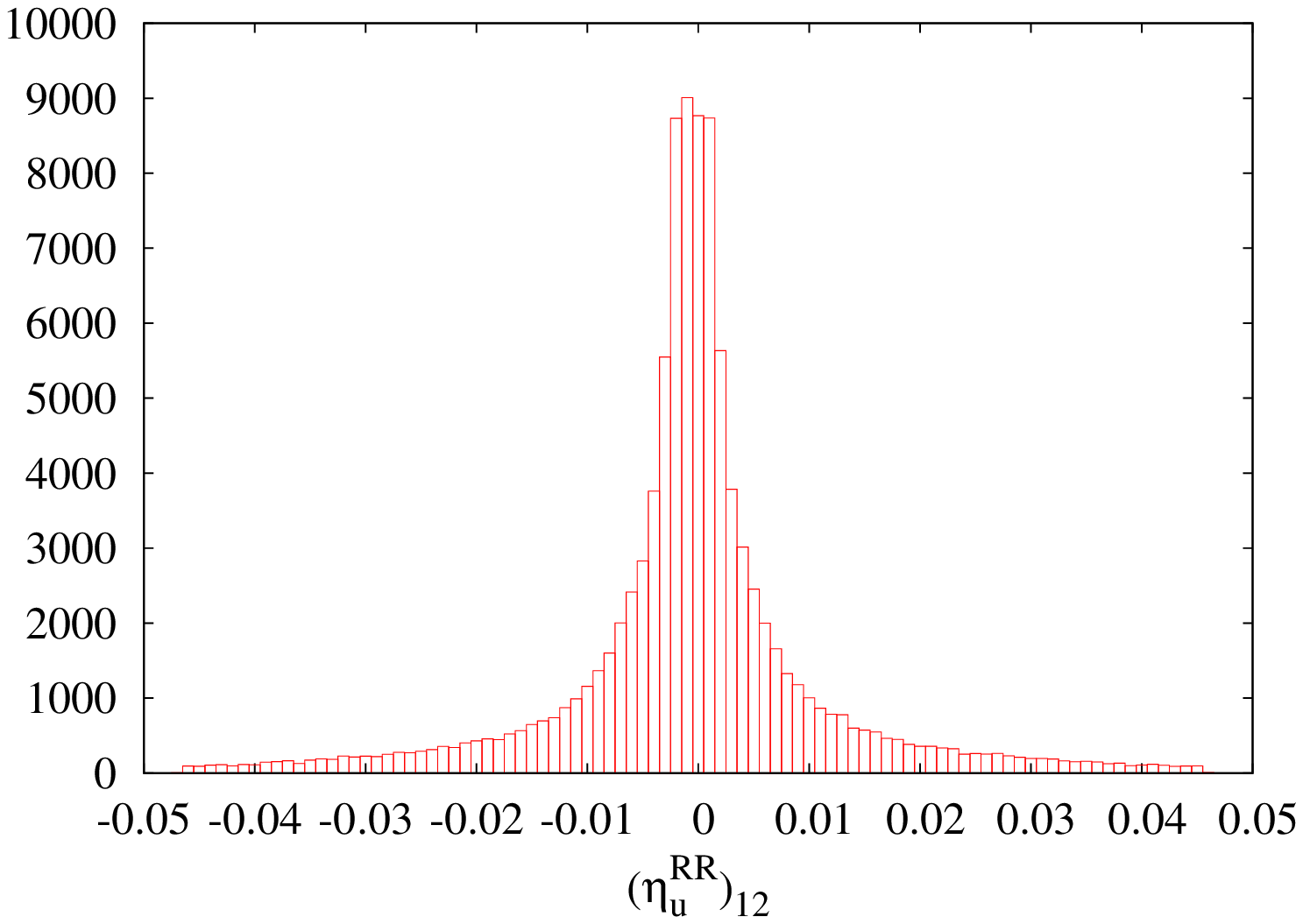}}\
\subfigure{\includegraphics[width=0.3\linewidth]{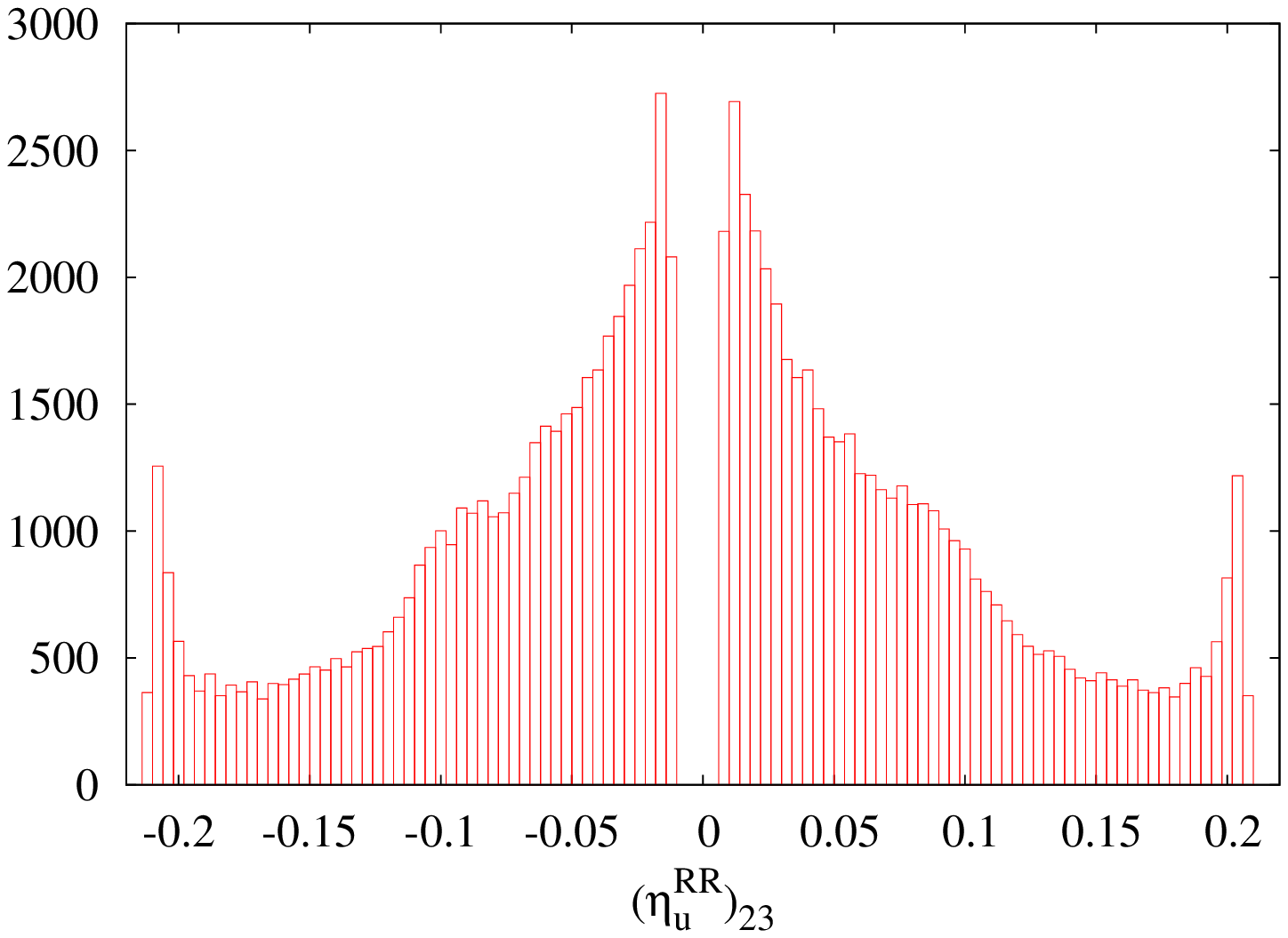}}\
\subfigure{\includegraphics[width=0.3\linewidth]{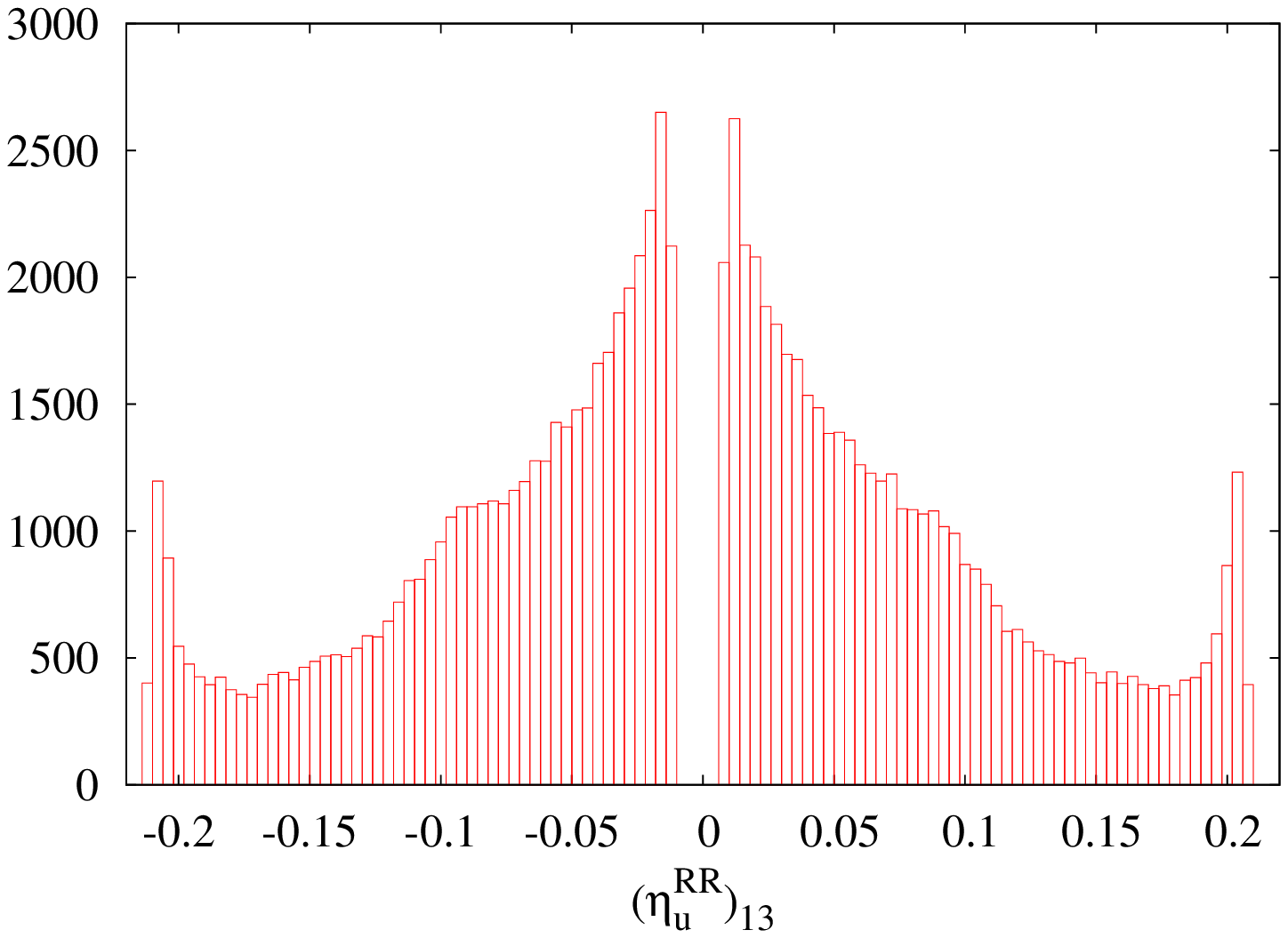}}}
\caption{Monte Carlo analysis using the model flavor texture to obtain
via the diagonalization matrices the elements
$(\eta_u^{LL,RR})_{ij}$.}
\label{fig:deltasu}
\end{center}
\end{figure}
\begin{figure}[!tp]
\begin{center}
\mbox{
\subfigure{\includegraphics[width=0.3\linewidth]{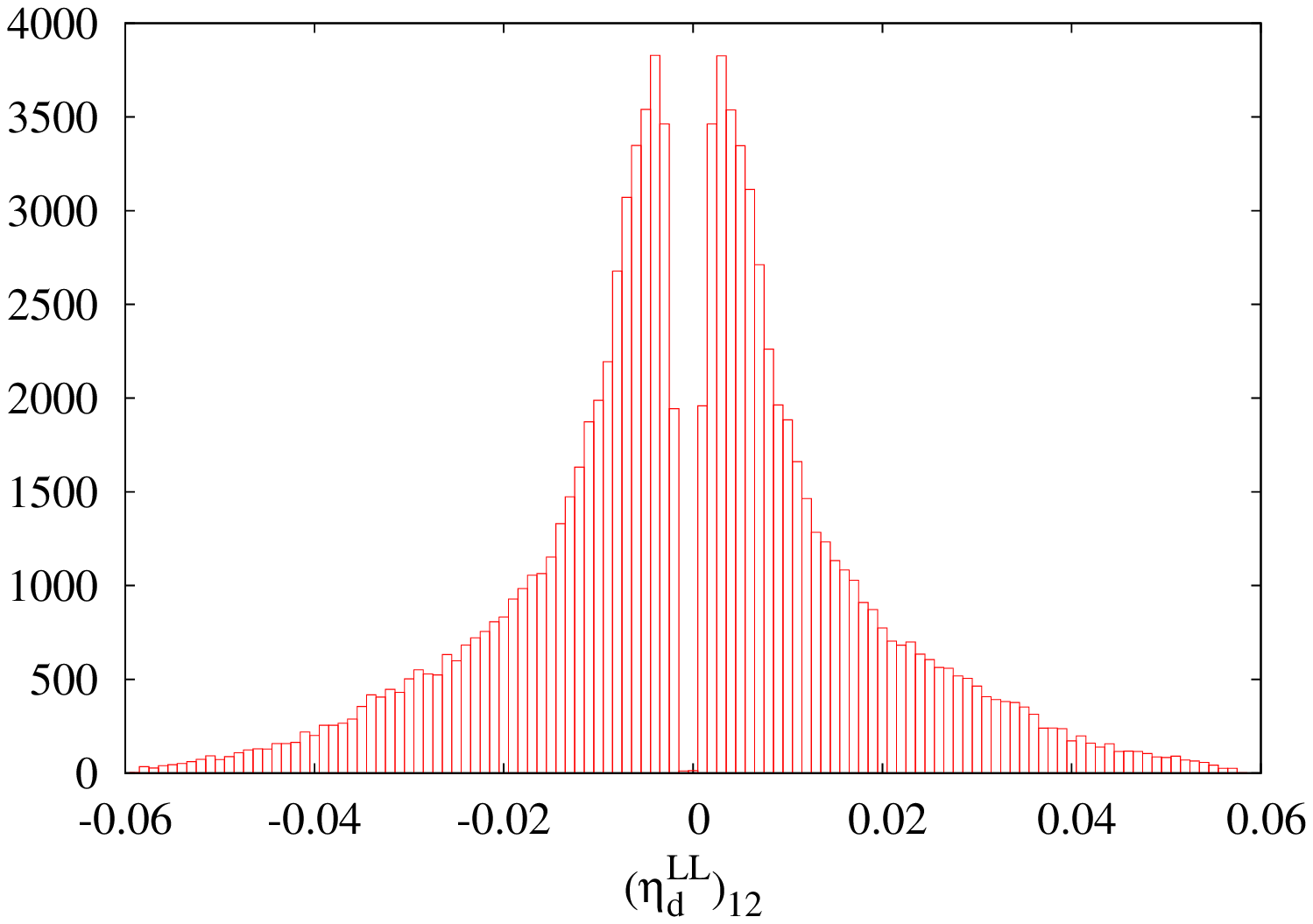}}\
\subfigure{\includegraphics[width=0.3\linewidth]{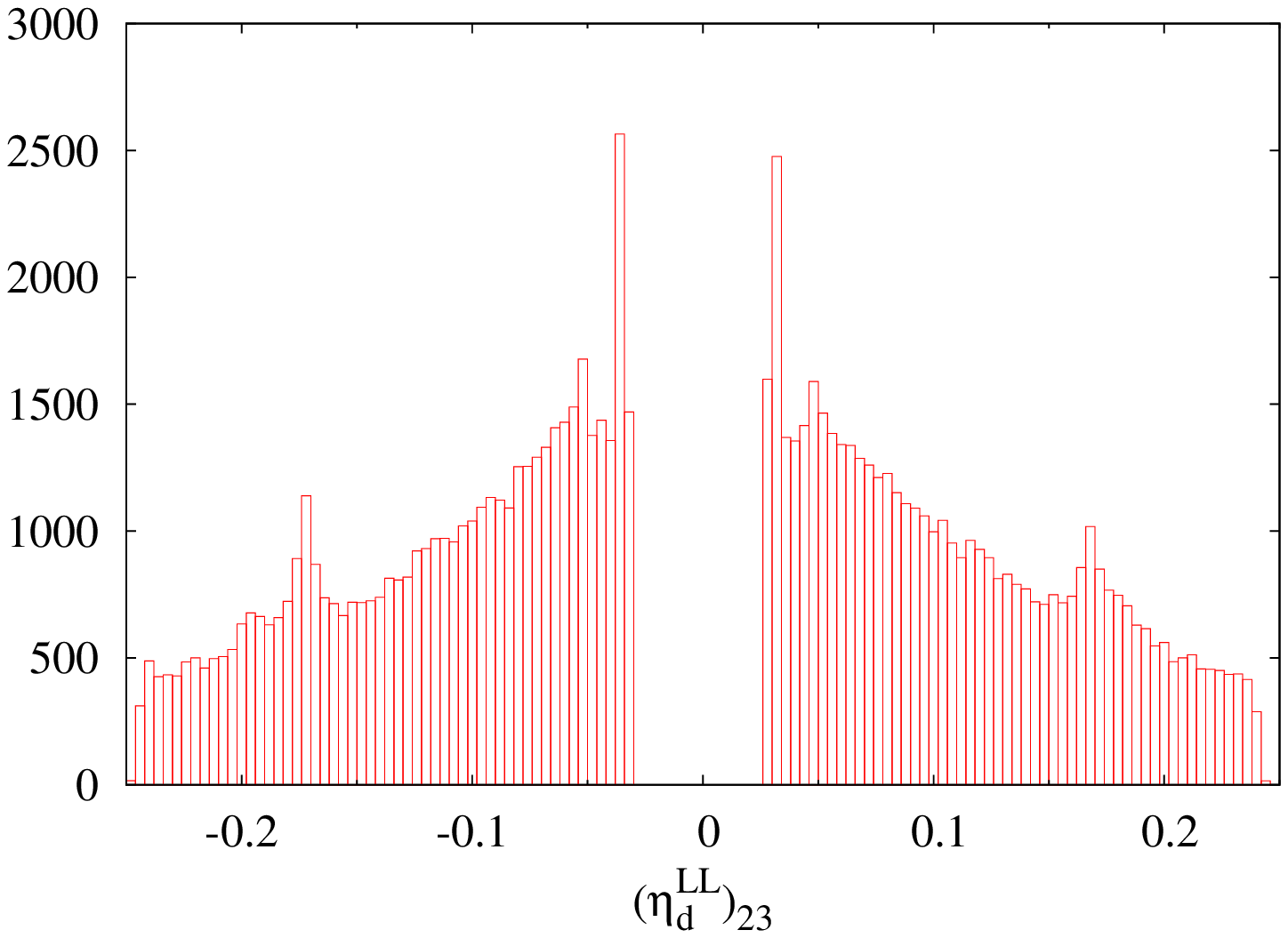}}\
\subfigure{\includegraphics[width=0.3\linewidth]{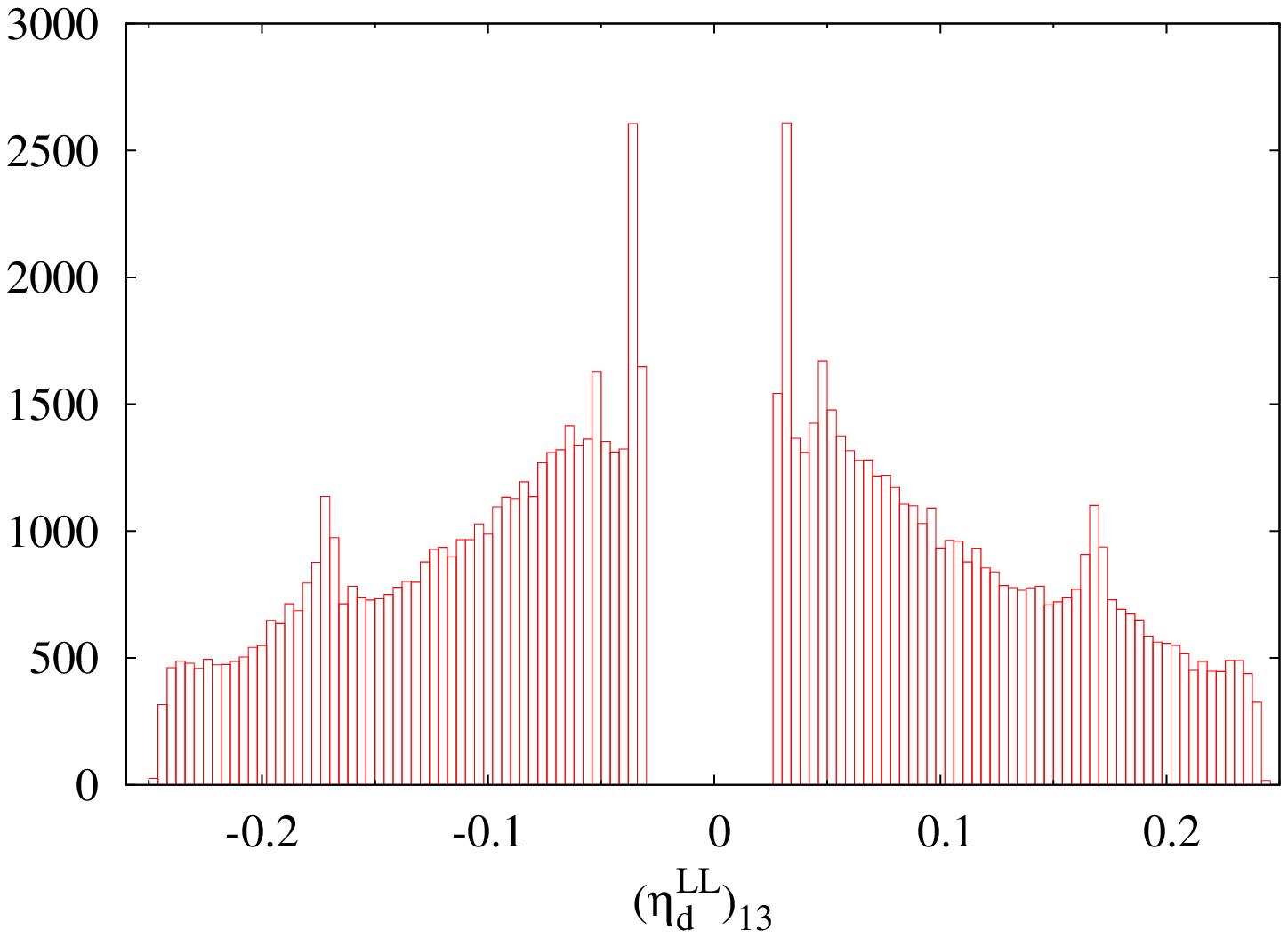}}}
\mbox{
\subfigure{\includegraphics[width=0.3\linewidth]{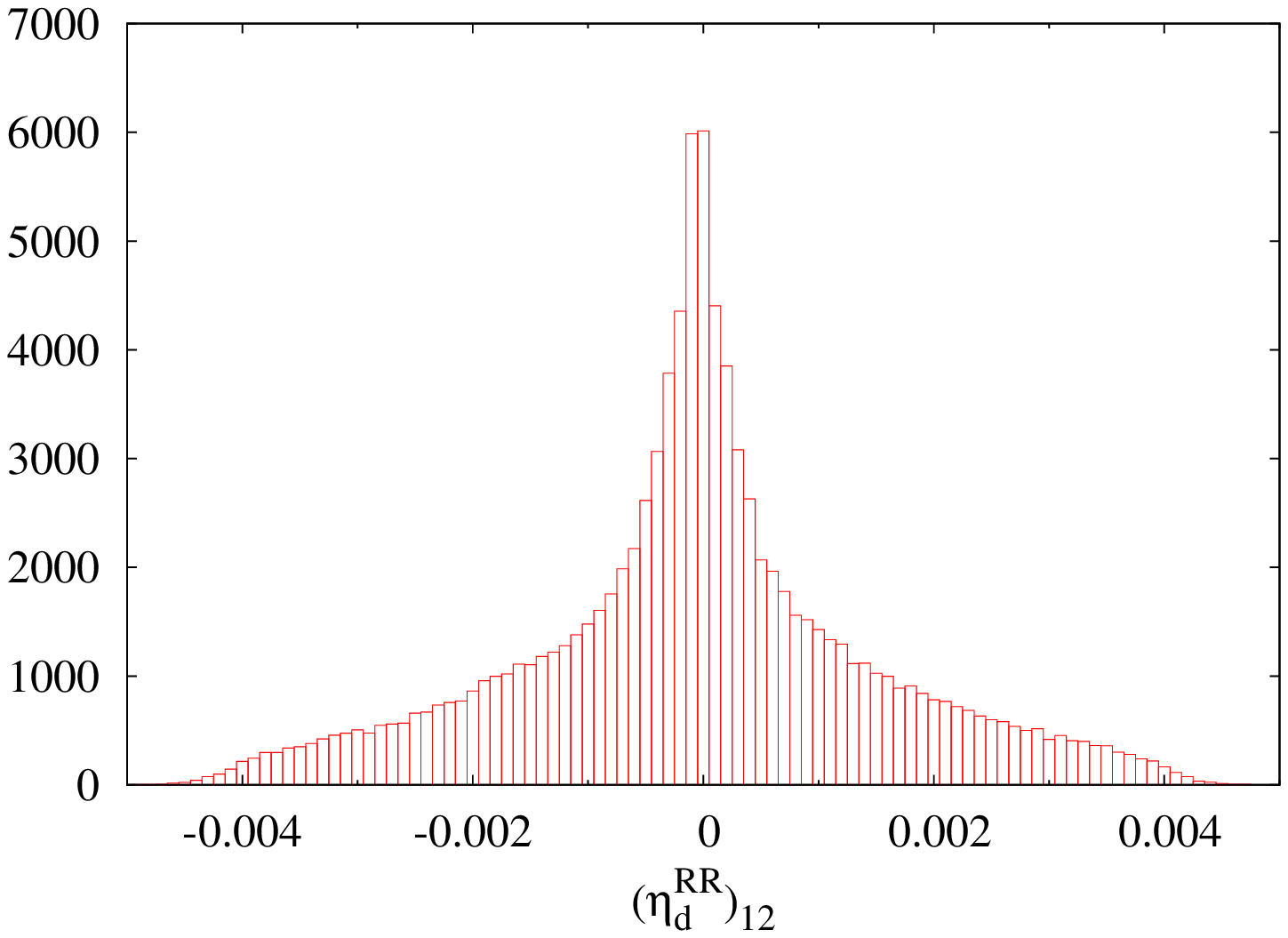}}\
\subfigure{\includegraphics[width=0.3\linewidth]{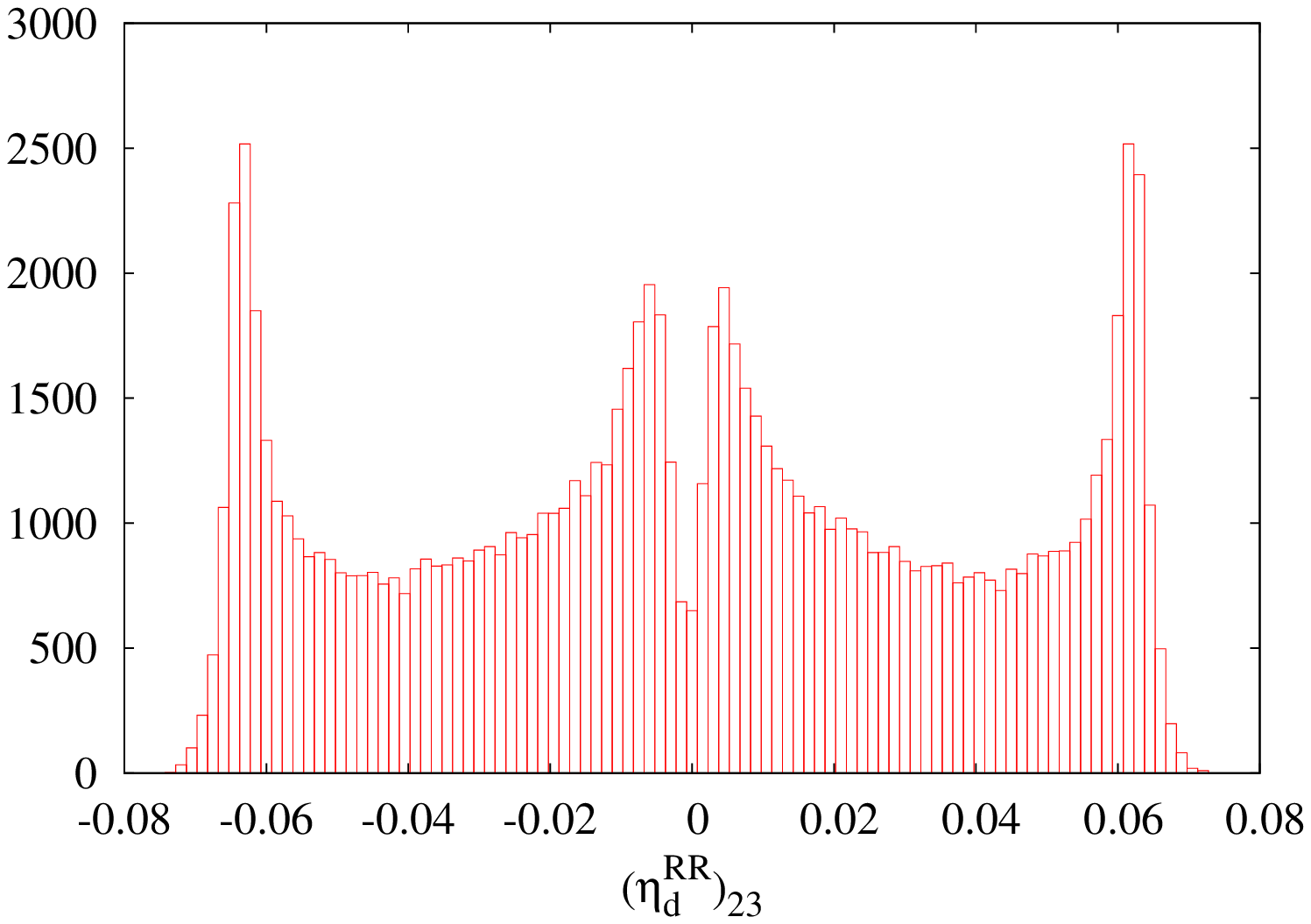}}\
\subfigure{\includegraphics[width=0.3\linewidth]{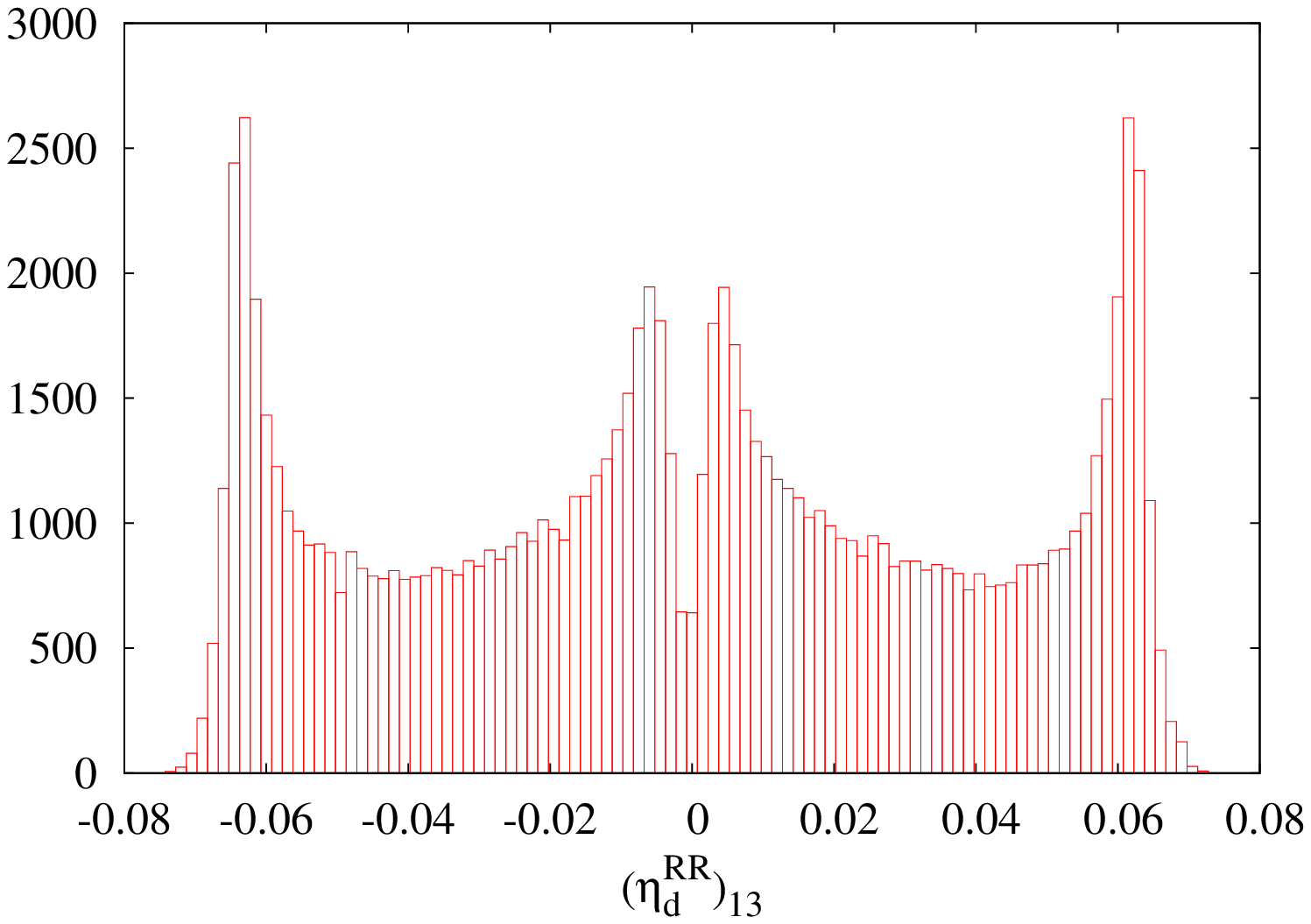}}}
\caption{Monte Carlo analysis using the model flavor texture to obtain
via the diagonalization matrices the elements
$(\eta_d^{LL,RR})_{ij}$.}
\label{fig:deltasd}
\end{center}
\end{figure}
\begin{figure}[!tp]
\begin{center}
\mbox{
\subfigure{\includegraphics[width=0.3\linewidth]{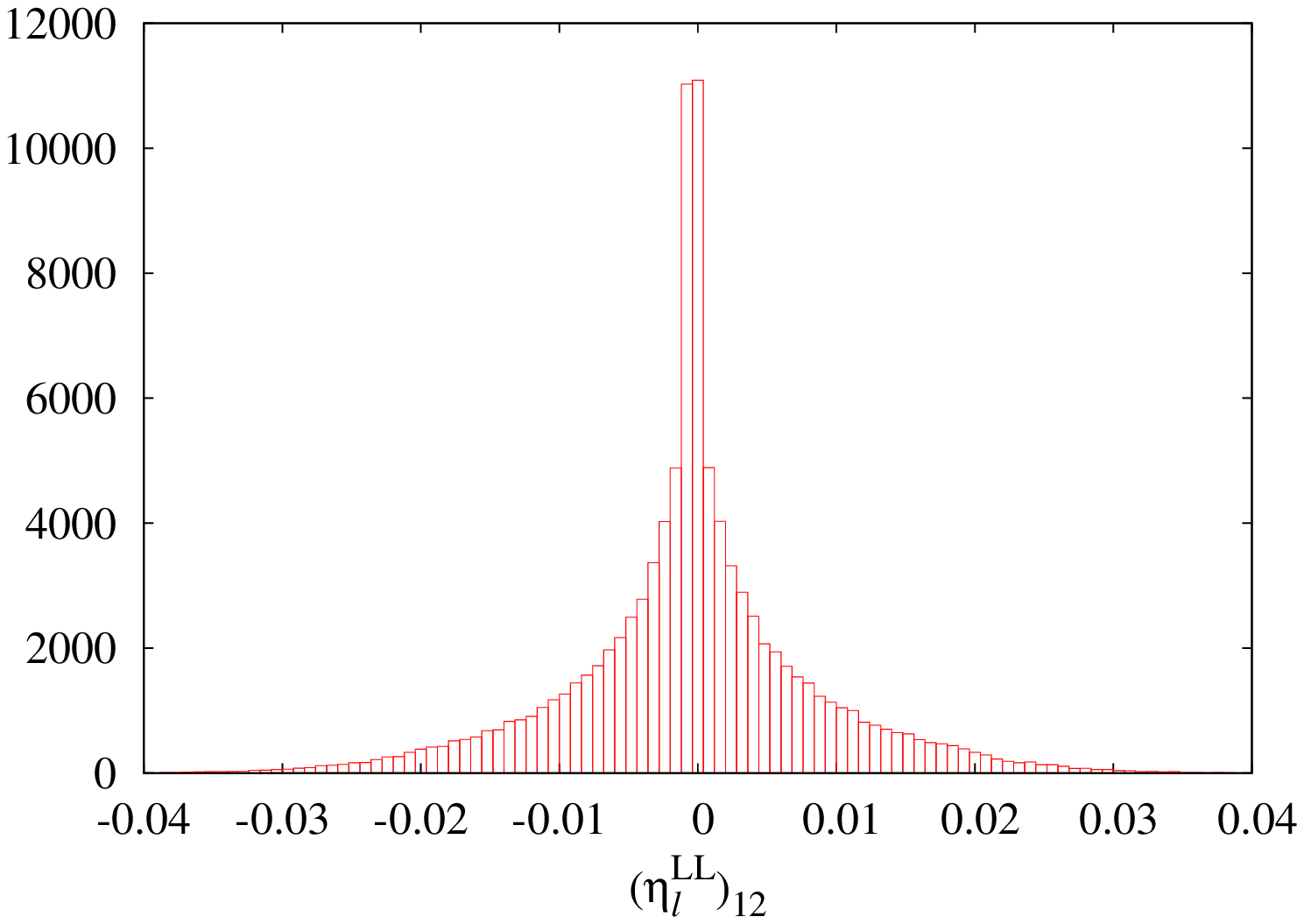}}\
\subfigure{\includegraphics[width=0.3\linewidth]{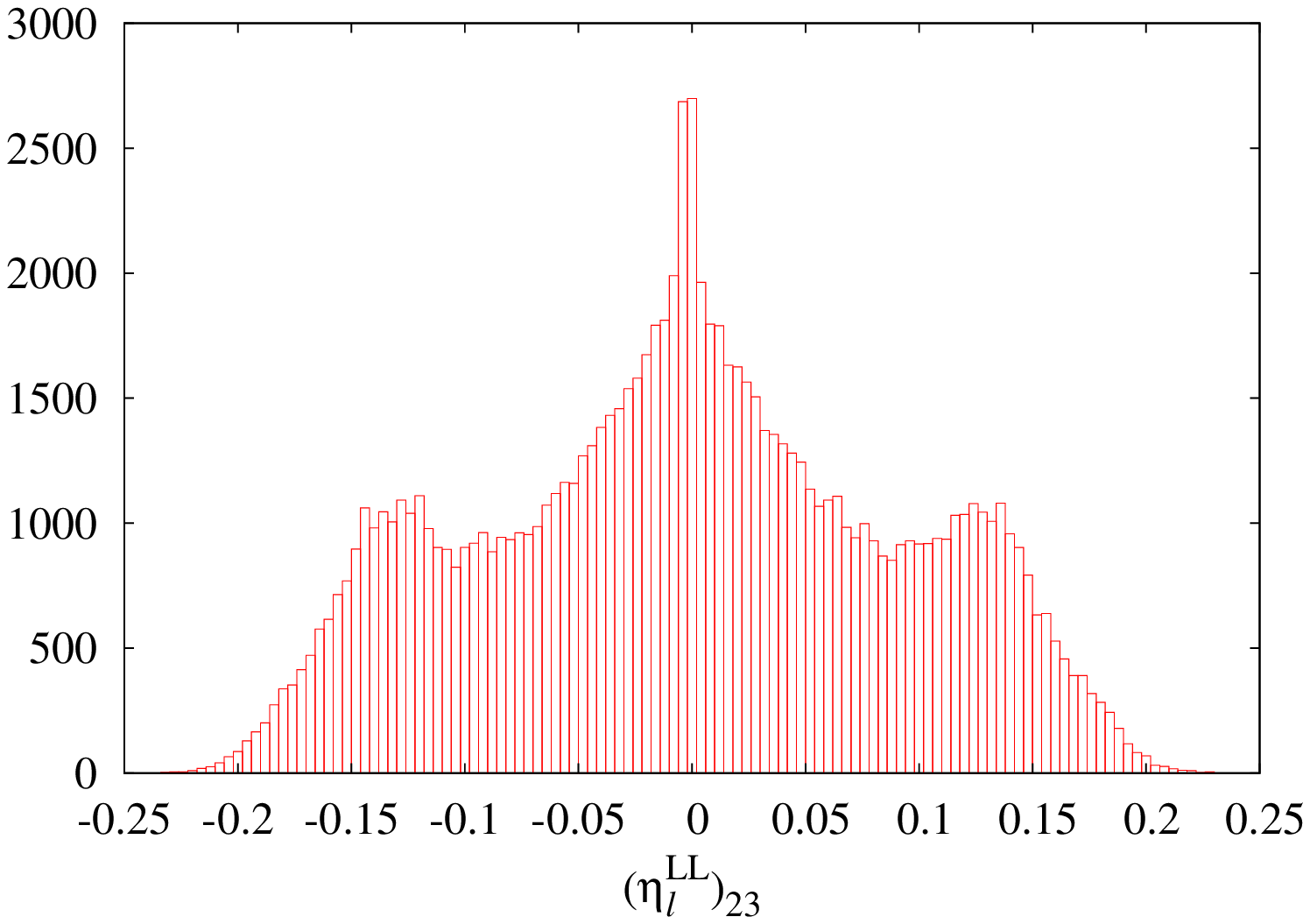}}\
\subfigure{\includegraphics[width=0.3\linewidth]{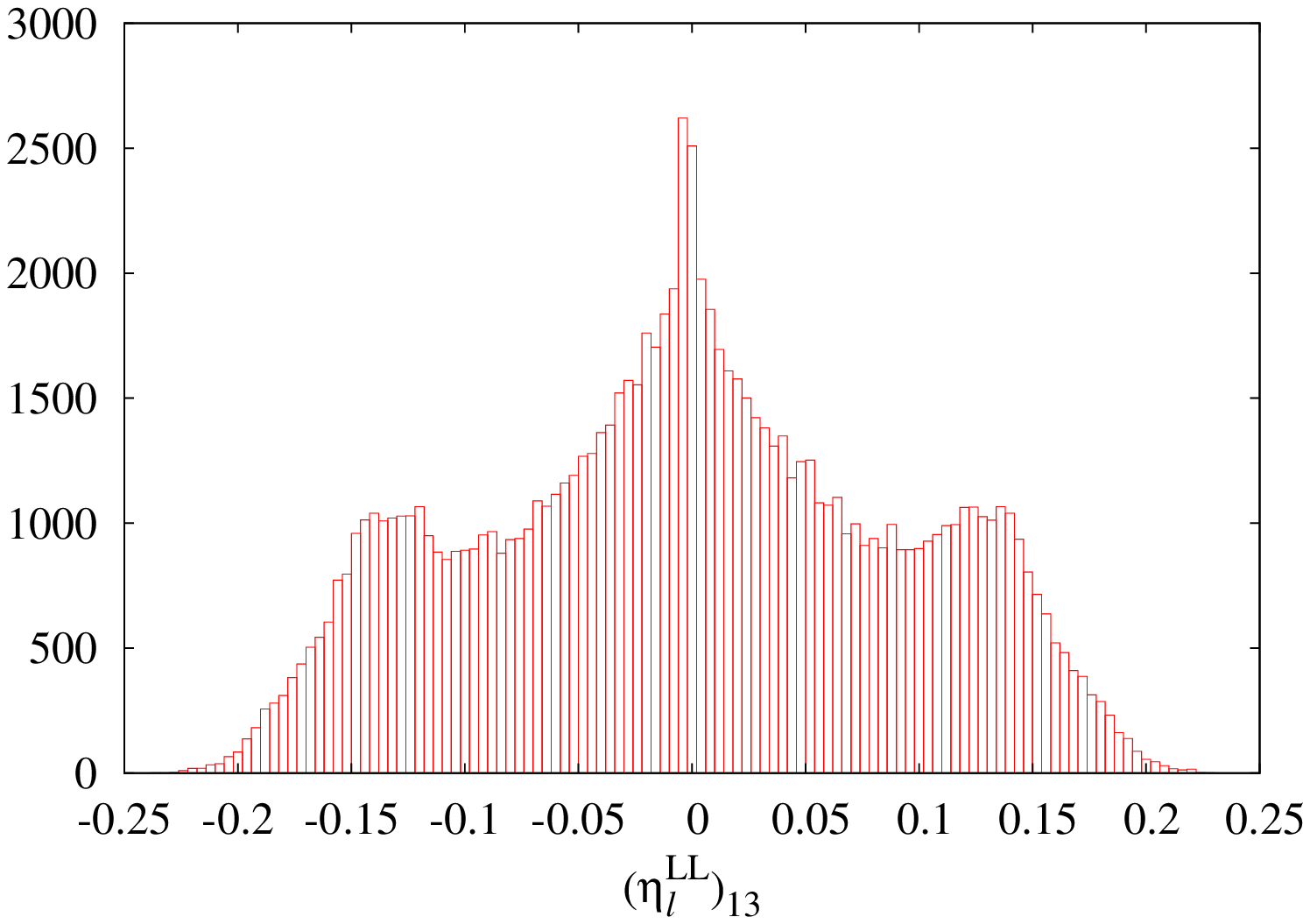}}}
\mbox{
\subfigure{\includegraphics[width=0.3\linewidth]{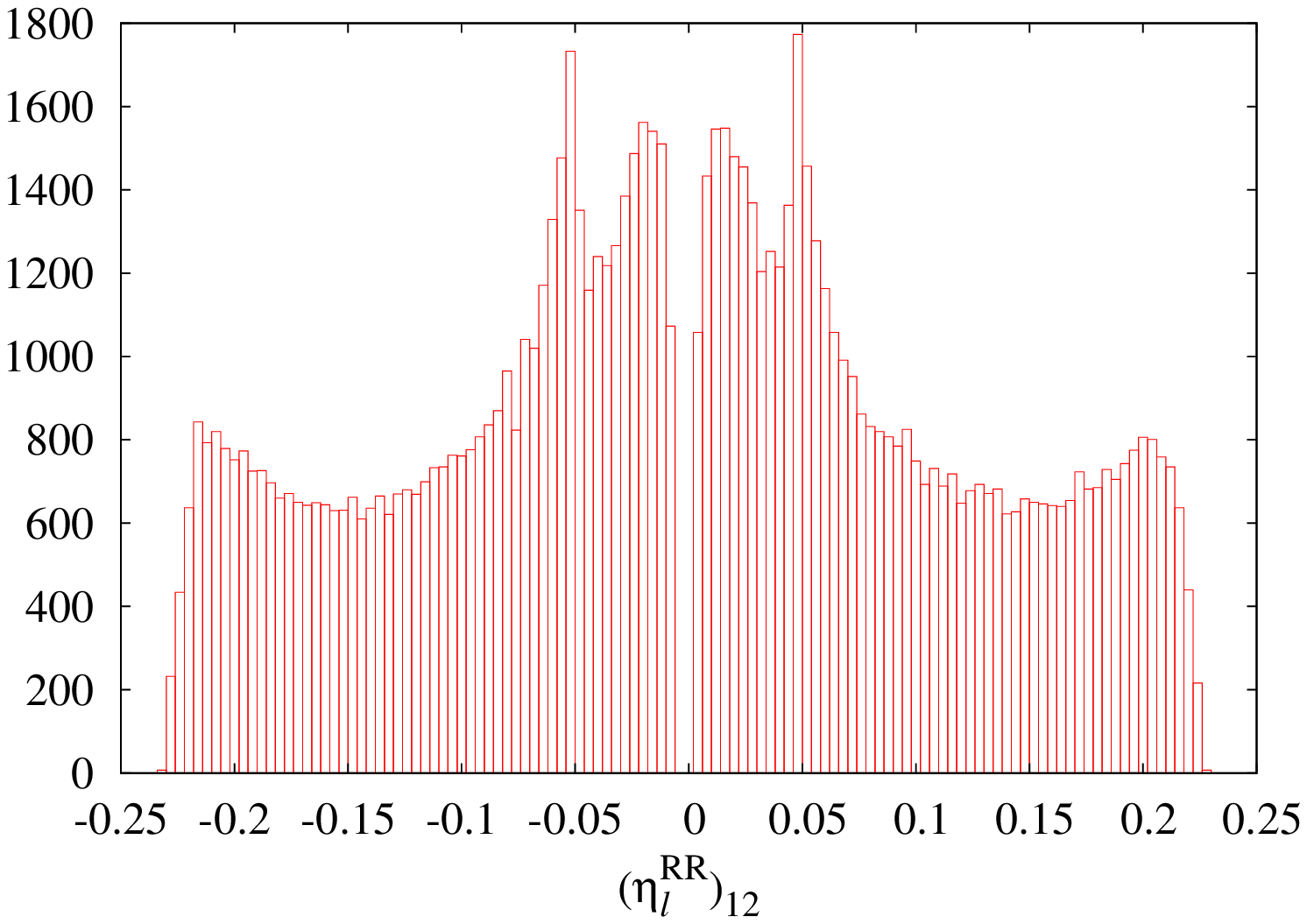}}\
\subfigure{\includegraphics[width=0.3\linewidth]{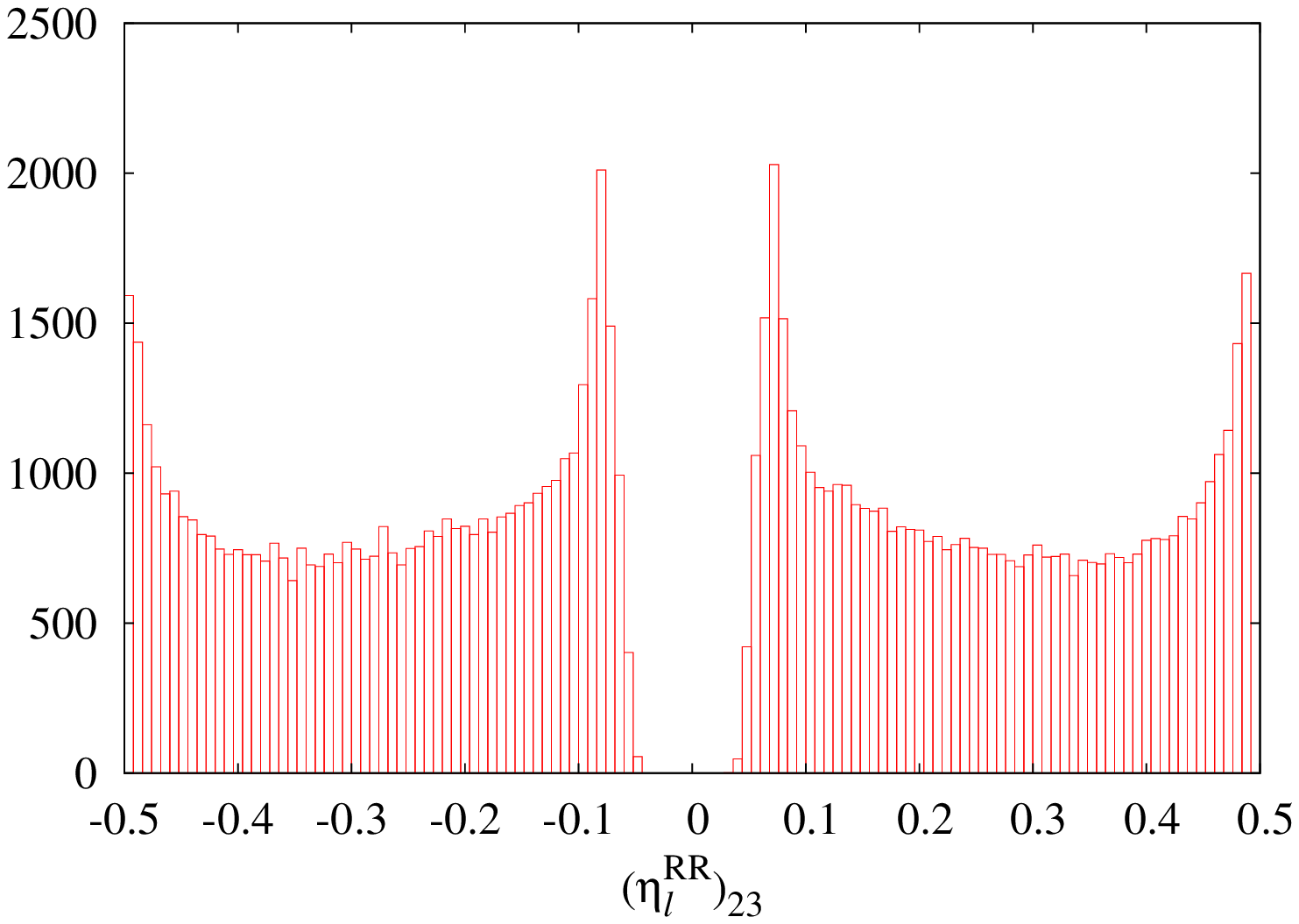}}\
\subfigure{\includegraphics[width=0.3\linewidth]{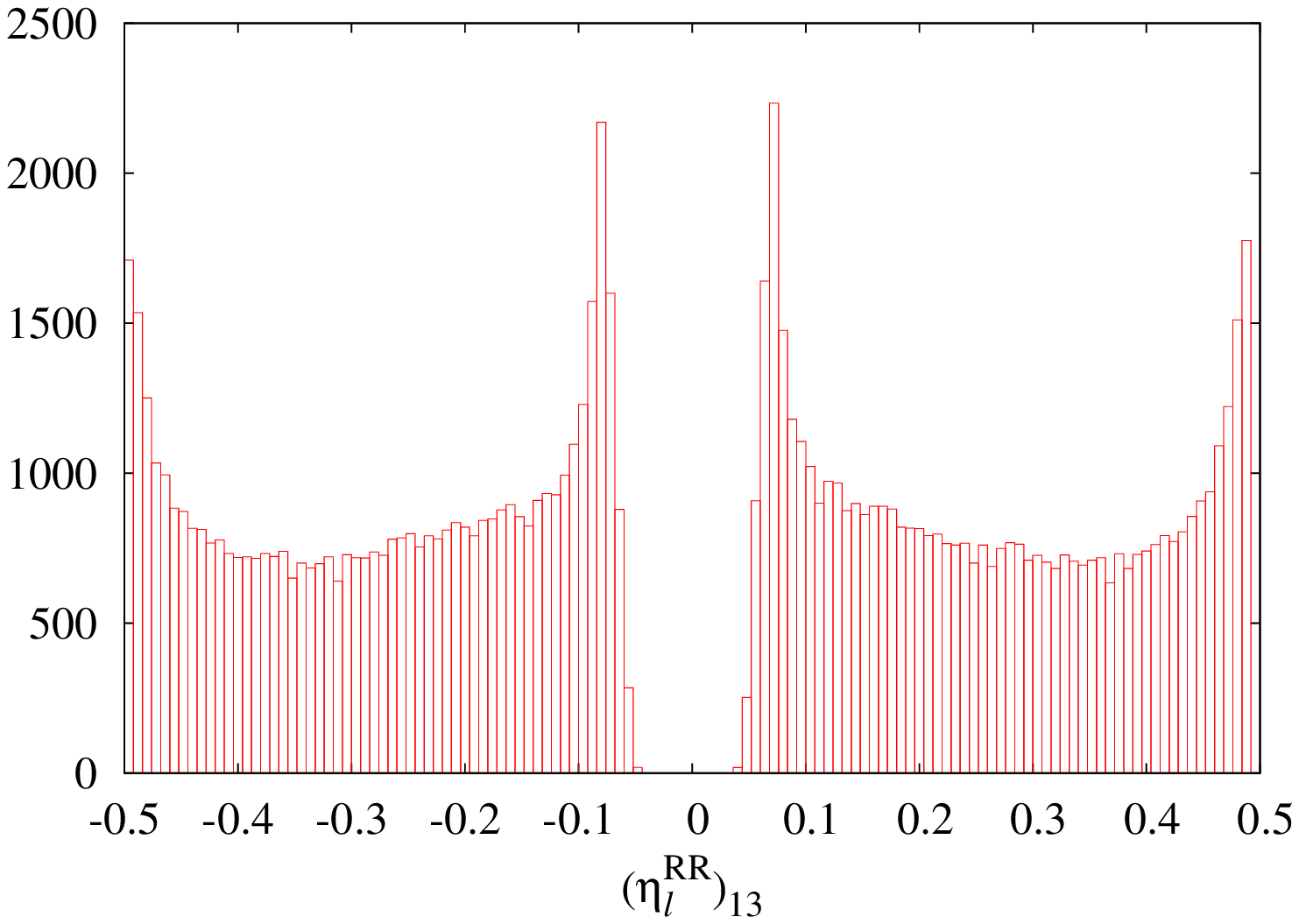}}}
\caption{Monte Carlo analysis using the model flavor texture to obtain
via the diagonalization matrices the elements
$(\eta_\ell^{LL})_{ij},(\eta_e^{RR})_{ij}$.}
\label{fig:deltase}
\end{center}
\end{figure}

Using these data, we generate maximum and mean values for the
$\eta$s, which are shown in tab.~\ref{tab:etavalues}
\begin{table}[!htp]
\begin{center}
\begin{tabular}{c|ccc}
$X$ & max$(X)$ & $\langle|X|\rangle$ & $\sqrt{\langle X^2\rangle}$\\
\hline\hline
\phantom{.}\\[-12pt]
$(\eta_u^{LL})_{12}$ & 0.0463 & 0.0133 & 0.0172\\
$(\eta_u^{LL})_{23}$ & 0.2103 & 0.1106 & 0.1254\\
$(\eta_u^{LL})_{13}$ & 0.2101 & 0.1114 & 0.1261\\
\hline
\phantom{.}\\[-12pt]
$(\eta_u^{RR})_{12}$ & 0.0462 & 0.0072 & 0.0116\\
$(\eta_u^{RR})_{23}$ & 0.2102 & 0.0787 & 0.0974\\
$(\eta_u^{RR})_{13}$ & 0.2103 & 0.0800 & 0.0986\\
\hline\hline
\phantom{.}\\[-12pt]
$(\eta_d^{LL})_{12}$ & 0.0584 & 0.0134 & 0.0175\\
$(\eta_d^{LL})_{23}$ & 0.2452 & 0.1116 & 0.1265\\
$(\eta_d^{LL})_{13}$ & 0.2455 & 0.1122 & 0.1273\\
\hline
\phantom{.}\\[-12pt]
$(\eta_d^{RR})_{12}$ & 0.0049 & 0.0011 & 0.0015\\
$(\eta_d^{RR})_{23}$ & 0.0729 & 0.0334 & 0.0397\\
$(\eta_d^{RR})_{13}$ & 0.0734 & 0.0340 & 0.0402\\
\hline\hline
\phantom{.}\\[-12pt]
$(\eta_\ell^{LL})_{12}$ & 0.0530 & 0.0060 & 0.0088\\
$(\eta_\ell^{LL})_{23}$ & 0.2314 & 0.0743 & 0.0914\\
$(\eta_\ell^{LL})_{13}$ & 0.2327 & 0.0750 & 0.0920\\
\hline
\phantom{.}\\[-12pt]
$(\eta_e^{RR})_{12}$ & 0.2291 & 0.0969 & 0.1169\\
$(\eta_e^{RR})_{23}$ & 0.4954 & 0.2645 & 0.2994\\
$(\eta_e^{RR})_{13}$ & 0.4955 & 0.2660 & 0.3009
\end{tabular}
\caption{Maximum, average and second moment average of the $\eta$
  elements, calculated by the Monte Carlo method. }
\label{tab:etavalues}
\end{center}
\end{table}

\end{document}